\documentclass[twocolumn, trackchanges]{aastex631}
\usepackage{xcolor}
\usepackage{CJK}
\usepackage{apjfonts}

\newcommand\msun{$\rm M_{\odot}$}
\usepackage{ulem}
\usepackage{amsmath}
\usepackage{cancel}
\usepackage{soul}
\usepackage{subfigure}
\newcommand{\N}[1]{\textcolor{teal}{ #1}}

\shorttitle{SN~2022crv: IIb, Or Not IIb: That is the Question}
\shortauthors{Dong et al.}
\graphicspath{{./}{figures/}}

\begin{document}

\title{Characterizing the Rapid Hydrogen Disappearance in SN2022crv: Evidence of a Continuum between Type Ib and IIb Supernova Properties}

\correspondingauthor{Yize Dong}
\email{yizdong@ucdavis.edu}

\author[0000-0002-7937-6371]{Yize Dong \begin{CJK*}{UTF8}{gbsn}(董一泽)\end{CJK*}}
\affiliation{Department of Physics and Astronomy, University of California, 1 Shields Avenue, Davis, CA 95616-5270, USA}

\author[0000-0001-8818-0795]{Stefano Valenti}
\affiliation{Department of Physics and Astronomy, University of California, 1 Shields Avenue, Davis, CA 95616-5270, USA}

\author[0000-0002-5221-7557]{Chris Ashall}
\affiliation{Department of Physics, Virginia Tech, Blacksburg, VA 24061, USA}

\author[0000-0003-2544-4516]{Marc Williamson}
\affiliation{Department of Physics, New York University, New York, NY 10003, USA}

\author[0000-0003-4102-380X]{David J.\ Sand}
\affiliation{Steward Observatory, University of Arizona, 933 North Cherry Avenue, Rm. N204, Tucson, AZ 85721-0065, USA}

\author[0000-0001-9038-9950]{Schuyler D.\ Van Dyk}
\affiliation{Caltech/IPAC, Mailcode 100-22, Pasadena, CA 91125, USA}

\author[0000-0003-3460-0103]{Alexei V. Filippenko}
\affiliation{Department of Astronomy, University of California, Berkeley, CA 94720-3411, USA}

\author[0000-0001-8738-6011]{Saurabh W.\ Jha}
\affiliation{Department of Physics and Astronomy, Rutgers, the State University of New Jersey,\\136 Frelinghuysen Road, Piscataway, NJ 08854-8019, USA}

\author[0000-0001-9589-3793]{Michael Lundquist}
\affiliation{W.~M.~Keck Observatory, 65-1120 M\=amalahoa Highway, Kamuela, HI 96743-8431, USA}

\author[0000-0001-7132-0333]{Maryam Modjaz}
\affiliation{Department of Astronomy, University of Virginia, Charlottesville, VA 22904, USA}

\author[0000-0003-0123-0062]{Jennifer E.\ Andrews}
\affiliation{Gemini Observatory, 670 North A`ohoku Place, Hilo, HI 96720-2700, USA}

\author[0000-0001-5754-4007]{Jacob E.\ Jencson}
\affiliation{Steward Observatory, University of Arizona, 933 North Cherry Avenue, Rm. N204, Tucson, AZ 85721-0065, USA}

\author[0000-0002-0832-2974]{Griffin Hosseinzadeh}
\affiliation{Steward Observatory, University of Arizona, 933 North Cherry Avenue, Rm. N204, Tucson, AZ 85721-0065, USA}

\author[0000-0002-0744-0047]{Jeniveve Pearson}
\affiliation{Steward Observatory, University of Arizona, 933 North Cherry Avenue, Rm. N204, Tucson, AZ 85721-0065, USA}

\author[0000-0003-3108-1328]{Lindsey A.\ Kwok}
\affiliation{Department of Physics and Astronomy, Rutgers, the State University of New Jersey,\\136 Frelinghuysen Road, Piscataway, NJ 08854-8019, USA}

\author[0000-0002-1250-4690]{Teresa Boland}
\affiliation{Department of Physics and Astronomy, Rutgers, the State University of New Jersey,\\136 Frelinghuysen Road, Piscataway, NJ 08854-8019, USA}

\author[0000-0003-1039-2928]{Eric Y.\ Hsiao}
\affiliation{Department of Physics, Florida State University, 77 Chieftan Way, Tallahassee, FL 32306, USA}

\author[0000-0001-5510-2424]{Nathan Smith}
\affiliation{Steward Observatory, University of Arizona, 933 North Cherry Avenue, Rm. N204, Tucson, AZ 85721-0065, USA}

\author[0000-0002-1381-9125]{Nancy Elias-Rosa}
\affiliation{Institute of Space Sciences (ICE, CSIC), Campus UAB, Carrer de Can Magrans, s/n, E-08193 Barcelona, Spain}

\author[0000-0003-4524-6883]{Shubham Srivastav}
\affiliation{Astrophysics Research Centre, School of Mathematics and Physics, Queens University Belfast, Belfast BT7 1NN, UK}
\author[0000-0002-8229-1731]{Stephen Smartt}
\affiliation{Astrophysics Research Centre, School of Mathematics and Physics, Queens University Belfast, Belfast BT7 1NN, UK}
\author{Michael Fulton}
\affiliation{Astrophysics Research Centre, School of Mathematics and Physics, Queens University Belfast, Belfast BT7 1NN, UK}

\author[0000-0002-2636-6508]{WeiKang Zheng}
\affiliation{Department of Astronomy, University of California, Berkeley, CA 94720-3411, USA}
\author[0000-0001-5955-2502]{Thomas G. Brink}
\affiliation{Department of Astronomy,University of California, Berkeley, CA 94720-3411, USA}

\author[0000-0002-9301-5302]{Melissa Shahbandeh}
\affiliation{Department of Physics and Astronomy, Johns Hopkins University, Baltimore, MD 21218, USA}
\affiliation{Space Telescope Science Institute, 3700 San Martin Drive, Baltimore, MD 21218, USA}

\author[0000-0002-4924-444X]{K.\ Azalee Bostroem}
\altaffiliation{DiRAC Fellow}
\affiliation{Department of Astronomy, University of Washington, 3910 15th Avenue NE, Seattle, WA 98195-0002, USA}

\author[0000-0003-2744-4755]{Emily Hoang}
\affil{Department of Physics and Astronomy, University of California, 1 Shields Avenue, Davis, CA 95616-5270, USA}

\author[0000-0003-0549-3281]{Daryl Janzen}
\affiliation{Department of Physics \& Engineering Physics, University of Saskatchewan, 116 Science Place, Saskatoon, SK S7N 5E2, Canada}

\author{Darshana Mehta}
\affiliation{Department of Physics and Astronomy, University of California, 1 Shields Avenue, Davis, CA 95616-5270, USA}

\author[0000-0002-7015-3446]{Nicolas Meza}
\affiliation{Department of Physics and Astronomy, University of California, 1 Shields Avenue, Davis, CA 95616-5270, USA}

\author[0000-0002-4022-1874]{Manisha Shrestha}
\affil{Steward Observatory, University of Arizona, 933 North Cherry Avenue, Tucson, AZ 85721-0065, USA}

\author[0000-0003-2732-4956]{Samuel Wyatt}
\affiliation{Steward Observatory, University of Arizona, 933 North Cherry Avenue, Rm. N204, Tucson, AZ 85721-0065, USA}

\author[0000-0002-4449-9152]{Katie Auchettl}
\affiliation{School of Physics, The University of Melbourne, Parkville, VIC 3010, Australia}
\affiliation{Department of Astronomy and Astrophysics, University of California, Santa Cruz, CA,  95064, USA }

\author[0000-0003-4625-6629]{Christopher R. Burns}
\affiliation{Observatories of the Carnegie Institution for Science, 813 Santa Barbara St., Pasadena, CA 91101, USA}

\author[0000-0003-4914-5625]{Joseph Farah}
\affiliation{Las Cumbres Observatory, 6740 Cortona Drive, Suite 102, Goleta, CA 93117-5575, USA}
\affiliation{Department of Physics, University of California, Santa Barbara, CA 93106-9530, USA}

\author[0000-0002-1296-6887]{L. Galbany}
\affiliation{Institute of Space Sciences (ICE-CSIC), Campus UAB, Carrer de Can Magrans, s/n, E-08193 Barcelona, Spain}
\affiliation{Institut d'Estudis Espacials de Catalunya (IEEC), E-08034 Barcelona, Spain}

\author[0000-0003-0209-9246]{Estefania Padilla Gonzalez}
\affiliation{Las Cumbres Observatory, 6740 Cortona Drive, Suite 102, Goleta, CA 93117-5575, USA}
\affiliation{Department of Physics, University of California, Santa Barbara, CA 93106-9530, USA}

\author[0000-0002-6703-805X]{Joshua Haislip}
\affiliation{Department of Physics and Astronomy, University of North Carolina, 120 East Cameron Avenue, Chapel Hill, NC 27599, USA}

\author[0000-0001-9668-2920]{Jason T. Hinkle}
\affiliation{Institute for Astronomy, University of Hawaii, 2680 Woodlawn Drive, Honolulu, HI 96822, USA}

\author[0000-0003-4253-656X]{D.\ Andrew Howell}
\affiliation{Las Cumbres Observatory, 6740 Cortona Drive, Suite 102, Goleta, CA 93117-5575, USA}
\affiliation{Department of Physics, University of California, Santa Barbara, CA 93106-9530, USA}

\author[0000-0001-6069-1139]{Thomas De Jaeger}
\affiliation{CNRS/IN2P3, Sorbonne Universit\'{e}, Universit\'{e} Paris Cit\'{e}), Laboratoire de Physique Nucl\'{e}aire et de Hautes \'{E}nergies, 75005, Paris, France}

\author[0000-0003-3642-5484]{Vladimir Kouprianov}
\affiliation{Department of Physics and Astronomy, University of North Carolina, 120 East Cameron Avenue, Chapel Hill, NC 27599, USA}

\author[0000-0001-8367-7591]{Sahana Kumar}
\affil{Department of Astronomy, University of Virginia, Charlottesville, VA 22904, USA}

\author[0000-0002-3900-1452]{Jing Lu}
\affil{Department of Physics and Astronomy, Michigan State University, East Lansing, MI 48824, USA}

\author[0000-0001-5807-7893]{Curtis McCully}
\affiliation{Las Cumbres Observatory, 6740 Cortona Drive, Suite 102, Goleta, CA 93117-5575, USA}
\affiliation{Department of Physics, University of California, Santa Barbara, CA 93106-9530, USA}

\author[0000-0001-5221-0243]{Shane Moran}
\affiliation{Tuorla Observatory, Department of Physics and Astronomy, University of Turku, 20014 Turku, Finland}

\author[0000-0003-2535-3091]{Nidia Morrell}
\affiliation{Las Campanas Observatory, Carnegie Observatories, Casilla 601, La Serena, Chile}

\author[0000-0001-9570-0584]{Megan Newsome}
\affiliation{Las Cumbres Observatory, 6740 Cortona Drive, Suite 102, Goleta, CA 93117-5575, USA}
\affiliation{Department of Physics, University of California, Santa Barbara, CA 93106-9530, USA}

\author[0000-0002-7472-1279]{Craig Pellegrino}
\affiliation{Las Cumbres Observatory, 6740 Cortona Drive, Suite 102, Goleta, CA 93117-5575, USA}
\affiliation{Department of Physics, University of California, Santa Barbara, CA 93106-9530, USA}

\author[0000-0002-1633-6495]{Abigail Polin}
\affiliation{The Observatories of the Carnegie Institution for Science, 813 Santa Barbara St., Pasadena, CA 91101, USA}
\affiliation{TAPIR, Walter Burke Institute for Theoretical Physics, 350-17, Caltech, Pasadena, CA 91125, USA}

\author[0000-0002-5060-3673]{Daniel E.\ Reichart}
\affiliation{Department of Physics and Astronomy, University of North Carolina, 120 East Cameron Avenue, Chapel Hill, NC 27599, USA}

\author{B. J.\ Shappee}
\affiliation{Institute for Astronomy, University of Hawaii, 2680 Woodlawn Drive, Honolulu, HI 96822, USA}

\author[0000-0002-5571-1833]{Maximilian D. Stritzinger}
\affil{Department of Physics and Astronomy, Aarhus University, Ny Munkegade 120, DK-8000 Aarhus C, Denmark}

\author[0000-0003-0794-5982]{Giacomo Terreran}
\affiliation{Las Cumbres Observatory, 6740 Cortona Drive, Suite 102, Goleta, CA 93117-5575, USA}
\affiliation{Department of Physics, University of California, Santa Barbara, CA 93106-9530, USA}

\author[0000-0002-2471-8442]{M. A. Tucker}
\altaffiliation{CCAPP Fellow}
\affiliation{Center for Cosmology and Astroparticle Physics,
The Ohio State University, 
191 West Woodruff Ave,
Columbus, OH, USA}
\affiliation{Department of Astronomy, 
The Ohio State University, 
140 West 18th Avenue,
Columbus, OH, USA}
\affiliation{Department of Physics,
The Ohio State University,
191 West Woodruff Ave,
Columbus, OH, USA}







\begin{abstract}
We present optical and near-infrared observations of SN~2022crv, a stripped envelope supernova in NGC~3054, discovered within 12 hrs of explosion by the Distance~Less~Than~40~Mpc Survey. 
We suggest SN~2022crv is a transitional object on the continuum between SNe~Ib and SNe~IIb.
A high-velocity hydrogen feature ($\sim$$-$20,000 -- $-$16,000 $\rm km\,s^{-1}$) was conspicuous in SN~2022crv at early phases, and then quickly disappeared. We find that a hydrogen envelope of $\sim$$10^{-3}$~\msun{} can reproduce the behaviour of the hydrogen feature observed. 
The lack of early envelope cooling emission implies that SN~2022crv had a compact progenitor with an extremely low amount of hydrogen. 
A nebular spectral analysis shows that SN~2022crv is consistent with the explosion of a He star with a final mass of $\sim$4.5 -- 5.6~\msun{} evolved from a $\sim$16 -- 22~\msun{} zero-age main sequence star in a binary system with $\sim$1.0 -- 1.7~\msun{} of oxygen finally synthesized in the core. The high metallicity at the supernova site indicates the progenitor experienced strong stellar wind mass loss. In order to retain a small amount of residual hydrogen at such a high metallicity, the initial orbital separation of the binary system is likely larger than $\sim$1000~$\rm R_{\odot}$.
The near-infrared spectra of SN~2022crv show a unique absorption feature on the blue side of He I line at $\sim$1.005~$\mu$m. This is the first time such a feature has been observed in a Type Ib/IIb, and could be due to \ion{Sr}{2}. Further detailed modeling of SN~2022crv can shed light on the progenitor and the origin of the mysterious absorption feature in the near infrared. 

\end{abstract}



\section{Introduction} \label{sec:intro}
Stripped envelope supernovae (SESNe) are a subclass of core-collapse supernovae (SNe) that have partly or completely lost their progenitor envelope prior to their explosions (see \citealt{Modjaz2019NatAs...3..717M} for a recent review).
SESNe are spectroscopically classified as SNe IIb, SNe Ib and SNe Ic \citep{Harkness1987,Wheeler1990,Filippenk1988,Filippenko1997,Clocchiatti1997} depending on the presence or absence of H and He lines in the optical spectra. SNe Ib show strong He lines but not H lines, while SNe Ic show neither H nor He lines. SNe IIb show clear H lines at early phases, and then the H lines become weaker over time and the spectra would be similar to SNe Ib at late phases. The sequence of SESNe (IIb$\rightarrow$Ib$\rightarrow$Ic) is commonly believed to be a result of differing amounts of stripping of the outer envelopes of their progenitors \citep{Filippenko1997,Yoon2015,Gal-Yam2017,Hiramatsu2021ApJ...913...55H}.

These hydrogen-deficient SN progenitors 
have been suggested to either arise from massive and metal-rich stars undergoing mass loss via stellar winds \citep{Woosley1993,Woosley1995ApJ...448..315W,Woosley2002,Eldridge2004,Meynet2005,Yoon2017b} or from binary interactions \citep{Podsiadlowski1992,Woosley1995ApJ...448..315W, Wellstein1999,Eldridge2004,Fryer2007, Yoon2010, Eldridge2013, Yoon2010, Yoon2017a, Gotberg2018A&A...615A..78G}.
Due to the high observed rate of SESNe, relatively weak stellar winds, and the small number of very massive stars assuming a Salpeter initial mass function \citep{Salpeter1955ApJ...121..161S}, most SESNe progenitors likely result from binary interaction 
\citep{Smith2011}.
This has been supported by direct imaging of the progenitor \citep{Aldering1994,Maund2004,Maund2011,VanDyk2011,Eldridge2013,Fremling2014,VanDyk2014,Eldridge2015,Folatelli2016,Kilpatrick2017,Tartaglia2017, Kilpatrick2021}, X-ray/radio observations \citep[e.g.,][]{Wellons2012,Drout2016,Brethauer2022}, and relatively low ejecta mass found from SESNe light curves \citep{Drout2011, Lyman2016}. 

With increasing numbers of well-observed SESNe for each subclass, many objects are becoming difficult to classify unambiguously because they are being discovered with overlapping properties.
For instance, an absorption feature at around 6200 \AA\ has been found in some SNe Ib and could be attributed to high-velocity $\rm H\alpha$ \citep{Deng2000,Branch2002,Elmhamdi2006,Parrent2007,Stritzinger2009,James2010ApJ...718..957J,Stritzinger2020,Holmbo2023A&A...675A..83H}, indicating that these objects may still contain a small amount of hydrogen. However, the hydrogen features detected in these objects could also be due to \ion{Si}{2}~$\lambda$6355, \ion{C}{2}~$\lambda$6580 or \ion{Ne}{1}~$\lambda$6402 \citep{Deng2000,Branch2002,Hamuy2002,Tanaka2009,Stritzinger2009,Dessart2011MNRAS.414.2985D,Hachinger2012}. \cite{Folatelli2014ApJ...792....7F} identified a small sample of transitional Type Ib/c SNe that seem to shift from Type Ic to Type Ib over time. These objects show initially weak helium features with nearly constant velocities during the photospheric phase, suggesting a dense shell in the ejecta. However, the weak hydrogen features seen in these objects technically result in a peculiar Type IIb classification. \cite{Milisavljevic2013} found that SESN~2011ei showed unambiguous hydrogen features at early times but these features quickly disappeared on a timescale of one week, suggesting the progenitor retained a thin hydrogen envelope at the time of explosion. Such a transformation is much faster than those observed in typical Type IIb SNe, which usually occur on a timescale of months. This implies that some Type IIb SNe may be misclassified as Type Ib SNe if they are not caught early enough. 

All these observations point towards a continuum between SNe IIb and SNe Ib, and there even may be some hydrogen hidden in Type Ib SNe, which is also supported by theoretical studies. For instance, \cite{Yoon2010} found that many Type Ib/c SNe progenitors formed in close binary systems are expected to maintain a thin hydrogen layer during their pre-supernova stage, producing the high-velocity hydrogen features observed in SNe Ib. In a series of models with various amounts of hydrogen, \cite{Hachinger2012} found that if the hydrogen envelope mass at the time of core collapse is between about 0.025 and 0.033~\msun, the difference between Type Ib and Type IIb could be unclear.

Although observational evidence has suggested that there is likely a continuum in the amount of helium between SNe IIb and SNe Ib \citep[although see \citealt{Holmbo2023A&A...675A..83H}]{Liu2016, Fremling2018A&A...618A..37F}, whether SNe IIb and SNe Ib can be clearly differentiated observationally is still an open question. 
To better characterize the classification of different types of SESNe and thus understand their progenitors, \cite{Liu2016} performed an analyse of a sample of SESNe. They proposed that the strength of $\rm H\alpha$ (or the absorption feature at around 6200 \AA) can be used to differentiate SNe Ib and SNe IIb at all epochs. 
\cite{Prentice2017} reassessed the classification system of SESNe using the spectra of a sample of SESNe and found that there is a clear distinction between He-poor SNe (SNe Ic) and He-rich SNe (SNe Ib/IIb). They attributed the 6200 \AA\ feature in SNe Ib to $\rm H\alpha$ and further suggested that the He-rich SNe can be split into subgroups based on the profile of the $\rm H\alpha$ line. To fully utilize spectra taken of SESNe, \cite{Williamson2019} proposed a new classification technique based on a support vector machine (SVM). This technique can identify transitional SESNe, i.e., SNe that present spectral features resembling more than one SESNe subtype, and thus reflects the physical properties of their progenitors. 
These sample studies imply the presence of a gradual transition between Type Ib SNe and Type IIb SNe depending on the amount of hydrogen remaining in the progenitor envelope. 
To better understand the connections between the different SESNe types and the evolution of their progenitors, a sample of SESNe that retain a small amount of hydrogen envelope is required. 
These objects need to be discovered shortly after the explosion since only the very early spectra convey signals from the outer layer of the progenitor star.
In this paper, we present optical and infrared data of SN~2022crv, a SESN discovered within $\sim$12 hrs of explosion by the Distance Less Than 40 Mpc \citep[DLT40,][]{Tartaglia2018} survey and densely monitored for over one year.
A hydrogen feature is detected in SN~2022crv at early phases and then quickly disappear shortly after the maximum.
Detailed analyses suggest that there is very little hydrogen in the SN envelope, making the object a transitional object on the continuum between SNe Ib and SNe IIb. 

This paper is organized as follows: the observations of SN~2022crv are presented in Section \ref{sec:observations}, while the observational properties, such as the reddening, distance and explosion epoch are constrained in Section \ref{sec:obs_properties}. We describe the photometric and spectroscopic properties of SN~2022crv in Section \ref{sec:phot_evol} and \ref{sec:spec_evol}, respectively. The physical implications of the observations are discussed in Section \ref{sec:discussion}, and finally we conclude in Section \ref{sec:summary}.

\begin{figure}
\includegraphics[width=1.\linewidth]{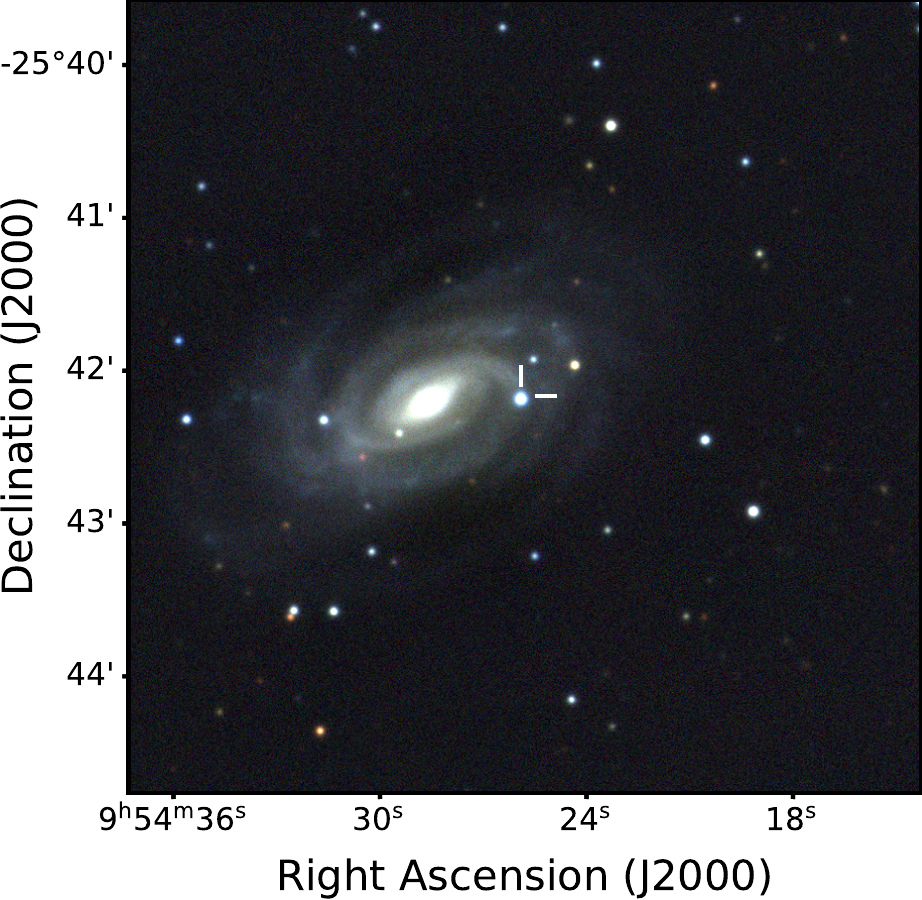}
\caption{Composite $gri$ image of SN~2022crv in NGC3054 obtained with the Las Cumbres Observatory on 2022 March 04. The position of SN~2022crv is indicated by white tick markers.
\label{fig:sn_image}}
\end{figure}

\begin{deluxetable}{cc}
\tablenum{1}
\tablecaption{Basic properties of SN~2022crv\label{tab:sn_properties}}
\tablewidth{0pt}
\tablehead{
}
\startdata
Host galaxy & NGC~3054\\
RA (2000) & 09\textsuperscript{h}54\textsuperscript{m}25\fs91\\
DEC (2000) & $-25\degr 42\arcmin 11\farcs 16$\\
Distance & $31.6_{-5.3}^{+6.4}$ Mpc\\
Distance modulus& $32.5 \pm 0.4$ mag \\
Redshift& 0.008091$\pm$0.000023\\
$E(B-V)_{\rm MW}$ & $0.0642_{-0.007}^{+0.007}$ mag*\\
$E(B-V)_{\rm host}$ & $0.146_{-0.009}^{+0.009}$ mag\\
Explosion epoch (JD) & $2459627.19 \pm 0.30$ (2022-02-16)\\
$V_{\rm max}$ (JD) & 2459645.42 $\pm 0.30$ (2022-03-06)\\
\enddata
\tablenotetext{\star}{\cite{Schlafly2011}.}
\end{deluxetable}

\begin{figure*}
\includegraphics[width=1.\linewidth]{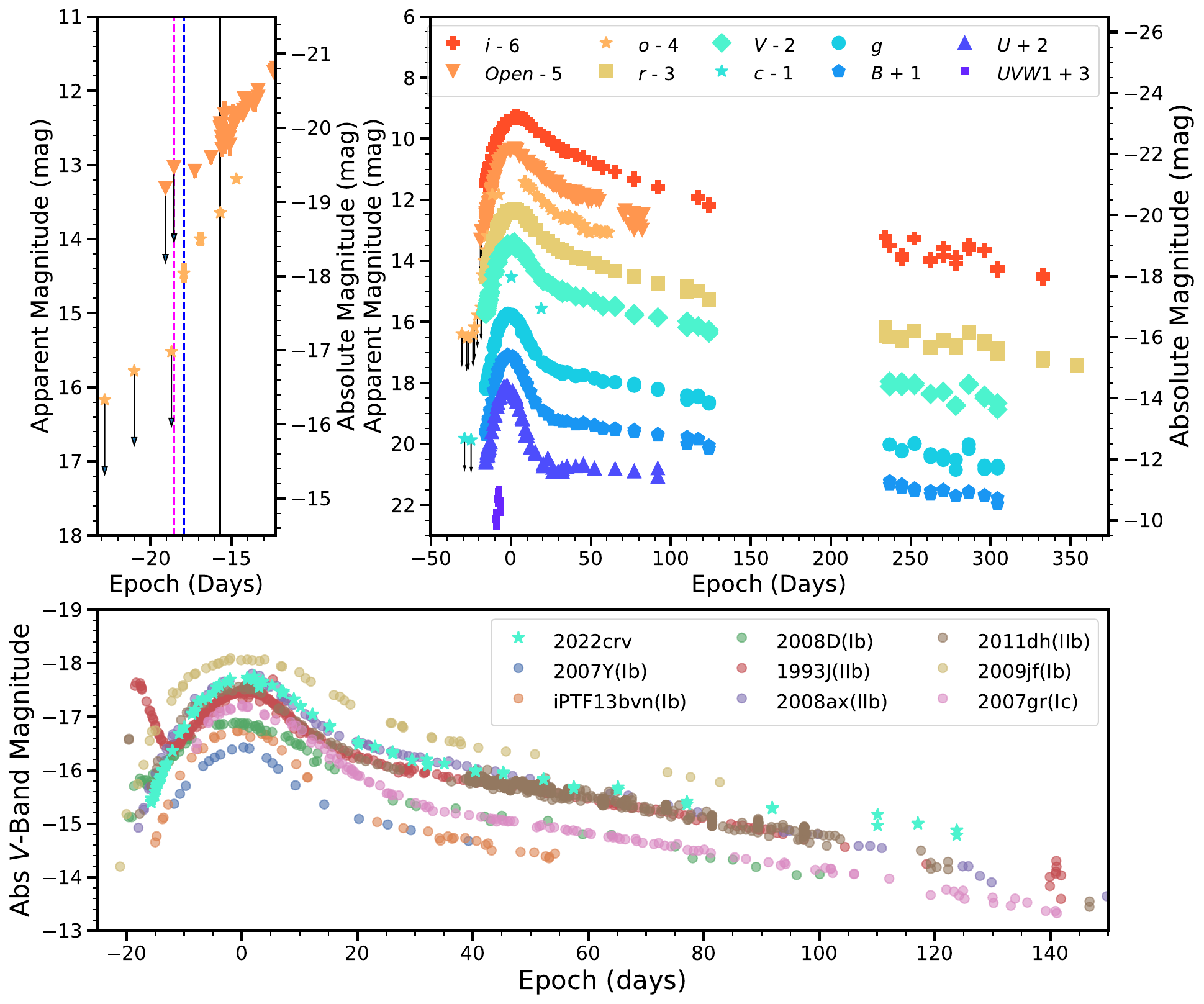}
\caption{\textit{Upper right:} Multiband light curves for SN~2022crv with respect to the $V$-band maximum (JD~2459645.42). Marker size is larger than uncertainties. The DLT40 $Open$ filter light curve is calibrated to the $r$ band. \textit{Upper left:} A zoom around a few days before and after the discovery. Only the $Open$- and $o$-band light curves are plotted here to show the constraint on the explosion epoch. The last nondetection and the earliest detection are marked by the magenta vertical dashed line and blue vertical dashed line, respectively. The black vertical line marks when the multiband followup began.
\textit{Bottom:} The $V$-band light curve of SN~2022crv compared to other SESNe. The magnitudes of other objects have been scaled to match the peak of SN~2022crv. The data used to create this figure are available.
 \label{fig:photometry}}
\end{figure*}

\section{Observations}\label{sec:observations}
SN~2022crv/DLT22d was discovered at RA(2000) $=$ 09\textsuperscript{h}54\textsuperscript{m}25\fs91, Dec(2000) $=-25\degr 42\arcmin 11\farcs 16$ 
in the nearby barred spiral (SBc) galaxy NGC~3054 (see Figure \ref{fig:sn_image}) on 2022-02-17.20 (\citealt{Dong2022TNS}, JD 2459627.696, $r$ = 18.04), during the course of the DLT40 SN search \citep{Tartaglia2018}, utilizing the 0.4-m \rm{PROMPT-MO-1} telescope \citep{Reichart2005} at the Meckering Observatory. 
An earlier detection was recovered in the $o$-band imaging taken by the Asteroid Terrestrial-Impact Last Alert System (ATLAS, \citealt{Tonry2011,Tonry2018,Smith2020}) on 2022-02-16.99. The object was initially classified as a Type Ib using a spectrum taken with GMOS on Gemini North on 2022-02-19.42 \citep{Andrews2022}. The basic properties of SN~2022crv are summarized in Table \ref{tab:sn_properties}.

\begin{deluxetable}{Ccccc} \label{tab:log_of_spec}
\tabletypesize\scriptsize
\tablecaption{ Spectroscopic observations of SN~2022crv.}
\tablewidth{0pt}
\tablehead{
\colhead{Phase (Days)$\star$} &
\colhead{MJD} &
\colhead{Telescope} &
\colhead{Instrument} & 
\colhead{Range (\AA)} 
}
\startdata 
-15.5/2.7 & 59629.43 & Gemini-N & GMOS & 3760-7030 \\
-15.5/2.8 & 59629.46 & KeckII & NIRES & 9650-24660 \\
-15.3/2.9 & 59629.61 & FTS & FLOYDS & 4850-10170 \\
-15.1/3.1 & 59629.82 & SALT & RSS & 3930-7790 \\
-13.4/4.8 & 59631.49 & FTN & FLOYDS & 3510-9990 \\
-11.5/6.7 & 59633.39 & FTN & FLOYDS & 3510-9990 \\
-10.7/7.5 & 59634.23 & SOAR & Goodman & 3780-5460 \\
-10.5/7.8 & 59634.46 & FTN & FLOYDS & 3500-9990 \\
-7.6/10.6 & 59637.34 & FTN & FLOYDS & 3510-10000 \\
-6.6/11.6 & 59638.32 & LBT & MODS & 3400-10100 \\
-6.5/11.8 & 59638.46 & FTN & FLOYDS & 3510-9990 \\
-3.5/14.7 & 59641.43 & FTS & FLOYDS & 3510-9990 \\
-0.6/17.7 & 59644.35 & FTN & FLOYDS & 3500-9990 \\
2.3/20.5 & 59647.22 & Bok & B\&C & 4010-7490 \\
2.7/21.0 & 59647.65 & FTS & FLOYDS & 3500-9990 \\
3.4/21.6 & 59648.29 & LBT & MODS & 3610-10290 \\
4.4/22.6 & 59649.28 & FTN & FLOYDS & 3500-9990 \\
4.4/22.6 & 59649.32 & UH88 & SNIFS & 3410-9090 \\
6.2/24.4 & 59651.07 & Baade & FIRE & 7910-25250 \\
6.4/24.6 & 59651.34 & FTN & FLOYDS & 3500-9990 \\
6.6/24.8 & 59651.50 & Baade & IMACS & 4230-9410 \\
7.4/25.7 & 59652.34 & KeckII & NIRES & 9650-24660 \\
8.3/26.6 & 59653.27 & KeckII & NIRES & 9660-24670 \\
10.4/28.6 & 59655.30 & FTN & FLOYDS & 3500-10000 \\
12.1/30.3 & 59657.00 & MMT & MMIRS & 9500-24290 \\
12.4/30.6 & 59657.32 & UH88 & SNIFS & 3410-9090 \\
15.6/33.8 & 59660.50 & Baade & IMACS & 4230-9410 \\
15.7/33.9 & 59660.60 & FTS & FLOYDS & 3510-9500 \\
17.3/35.6 & 59662.25 & Shane & Kast & 3620-10750 \\
18.3/36.5 & 59663.19 & LBT & MODS & 3400-9990 \\
19.2/37.5 & 59664.17 & MMT &  Binospec & 5240-6750 \\
19.3/37.6 & 59664.25 & UH88 & SNIFS & 3410-9090 \\
24.0/42.2 & 59668.90 & NOT & ALFOSC & 3410-9660 \\
24.6/42.8 & 59669.50 & Baade & IMACS & 4230-9410 \\
25.3/43.5 & 59670.21 & Shane & Kast & 3630-10740 \\
28.6/46.9 & 59673.57 & FTS & FLOYDS & 3510-9990 \\
35.4/53.6 & 59680.28 & FTN & FLOYDS & 3500-9990 \\
35.4/53.6 & 59680.29 & UH88 & SNIFS & 3410-9090 \\
36.6/54.8 & 59681.50 & Baade & IMACS & 4230-9410 \\
44.6/62.9 & 59689.54 & FTS & FLOYDS & 3500-10000 \\
48.4/66.6 & 59693.29 & UH88 & SNIFS & 3410-9090 \\
50.3/68.6 & 59695.24 & KeckI & LRIS & 3150-5640 \\
50.3/68.6 & 59695.24 & KeckI & LRIS & 5420-10300 \\
50.3/68.6 & 59695.25 & KeckI & LRIS & 5490-7140 \\
54.9/73.2 & 59699.86 & SALT & RSS & 3930-7790 \\
56.5/74.7 & 59701.42 & FTS & FLOYDS & 3500-10000 \\
58.2/76.5 & 59703.15 & LBT & MODS & 3720-10000 \\
64.4/82.6 & 59709.29 & KeckII & NIRES & 9660-24670 \\
69.3/87.6 & 59714.25 & KeckII & NIRES & 9650-24670 \\
70.3/88.6 & 59715.25 & KeckII & NIRES & 9660-24670 \\
71.5/89.7 & 59716.43 & FTS & FLOYDS & 3500-9990 \\
99.5/117.7 & 59744.41 & FTS & FLOYDS & 4010-10000 \\
102.8/121.0 & 59747.73 & SALT & RSS & 3930-7790 \\
267.4/285.6 & 59912.29 & SOAR & Goodman & 3500-7130 \\
272.3/290.6 & 59917.27 & SOAR & Goodman & 5000-9000 \\
321.0/339.2 & 59965.90 & SALT & RSS & 5480-8990 \\
353.2/371.4 & 59998.14 & Clay & LDSS3 & 4000-10000\\
355.1/373.3 & 60000.03 & GTC  & OSIRIS & 4000-10000
\enddata
\tablenotetext{\star}{Phase is measured from $\rm \textit{V}_{max}$/explosion.}
\end{deluxetable}

\begin{figure}
\includegraphics[width=1.\linewidth]{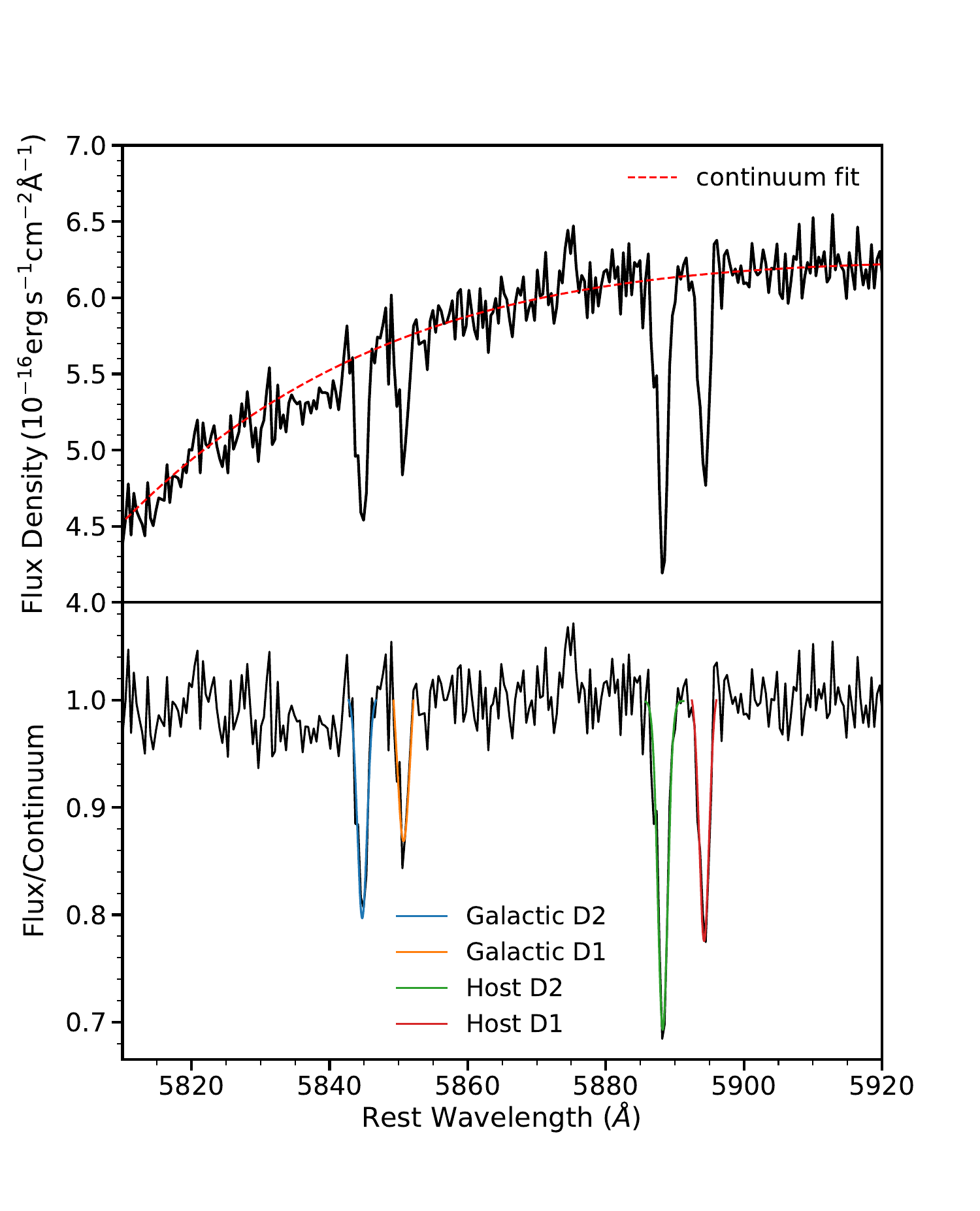}
\caption{\textit{Top:} A high-resolution spectrum taken with LRIS on 2022-04-26 (MJD~59695.25) showing the Na\,I\,D lines from the host galaxy and the Milky Way. The continuum is fitted with a 4th order polynomial function. \textit{Bottom:} The continuum-subtracted spectrum. The Na\,I\,D lines are fitted with Gaussian functions using a MCMC routine.
\label{fig:NaID}}
\end{figure}

After the discovery, the object was intensely followed by the DLT40 survey using the PROMPT5 and PROMPT-MO telescopes in the \textit{Open} filter and the ATLAS survey in the \textit{o} and \textit{c} filters. In addition, high-cadence multiband photometric observations were collected by the world-wide network of robotic telescopes at Las Cumbres Observatory \citep{Brown2013} through the Global Supernova Project. 
Ultraviolet and optical imaging was taken with the Neil Gehrels \textit{Swift} Observatory \citep{Gehrels2004} at early times.
The object was also followed by the Katzman Automatic Imaging Telescope (KAIT) as part of the Lick Observatory Supernova Search \citep[LOSS;][]{Filippenko2001}, and the 1~m Nickel telescope at Lick Observatory. \textit{B, V, R} and \textit{I} multiband images of SN~2022crv were obtained with both telescopes,
while additional \textit{clear} band (close to the \textit{R} band; see \citealt{Li2003}) images were also obtained with KAIT.
The multiband light curves are shown in Figure \ref{fig:photometry}. The reduction process of the photometric data is presented in Appendix \ref{sec:phot_reduction}.

Nineteen low-resolution optical spectra were collected with the FLOYDS spectrograph \citep{Brown2013} on the 2m Faulkes Telescopes South and North (FTS \& FTN) through the Global Supernova Project. In addition, many optical spectra were obtained with the Robert Stobie Spectrograph (RSS) on the Southern African Large Telescope \citep[SALT;][]{smith2006}, the Kast Spectrograph on the 3m Shane Telescope \citep{miller1994} at Lick Observatory, the Boller \& Chivens Spectrograph (B\&C) on the Bok Telescope \citep{green1995}, one of the Multi-Object double Spectrographs \citep[MODS1,][]{MODS} on LBT, the Binospec instrument on the MMT \citep{Binospec}, the Goodman High Throughput Spectrograph on the Southern Astrophysical Research Telescope \citep[SOAR;][]{clemens2004}, the Low-Resolution Imaging Spectrometer \citep[LRIS;][]{Oke1995} on the Keck~I telescope, the Gemini Multi-Object Spectrograph \citep[GMOS;][]{Hook2004} on the Gemini North telescope, the Andalucia Faint Object Spectrograph and Camera (ALFOSC) on the Nordic Optical Telescope (NOT) as part of the NUTS2 collaboration, the SuperNova Integral Field Spectrograph \citep[SNIFS;][]{Lantz2004} mounted at the University of Hawaii 2.2 m telescope (UH88) at Mauna Kea as part of the Spectroscopic Classification of Astronomical Transients survey \citep[SCAT;][]{Tucker2022PASP..134l4502T}, the Inamori-Magellan Areal Camera \& Spectrograph \citep[IMACS;][]{Dressler2011} on the 6.5m Magellan Baade telescope as part of the Precision Observations of Infant Supernova Explosions \citep[POISE;][]{Burns2021ATel14441....1B}, the Optical System for Imaging and low/intermediate-Resolution Integrated Spectroscopy \citep[OSIRIS;][]{Cepa10.1117/12.395520} spectrograph on the 10.4 m Gran Telescopio Canarias (GTC), and the Low-Dispersion Survey Spectrograph 3 \citep[LDSS3, which was updated from LDSS2;][]{Allington-Smith1994} on the 6.5m Magellan Clay telescope. 

Near-infrared (NIR) spectra were taken with the Near-Infrared Echellette Spectrometer \citep[NIRES;][]{Wilson2004} on the Keck~II telescope, the Folded-port InfraRed Echellette \cite[FIRE;][]{Simcoe2013} on the Magellan Baade telescope, and the MMT and Magellan Infrared Spectrograph (MMIRS) on the MMT \citep{McLeod2012}. 

A log of the spectroscopic observations is given in Table \ref{tab:log_of_spec}. The reduction process for the spectroscopic data is presented in Appendix \ref{sec:spec_reduction}.


\begin{figure*}
\centering
\includegraphics[width=.9\linewidth]{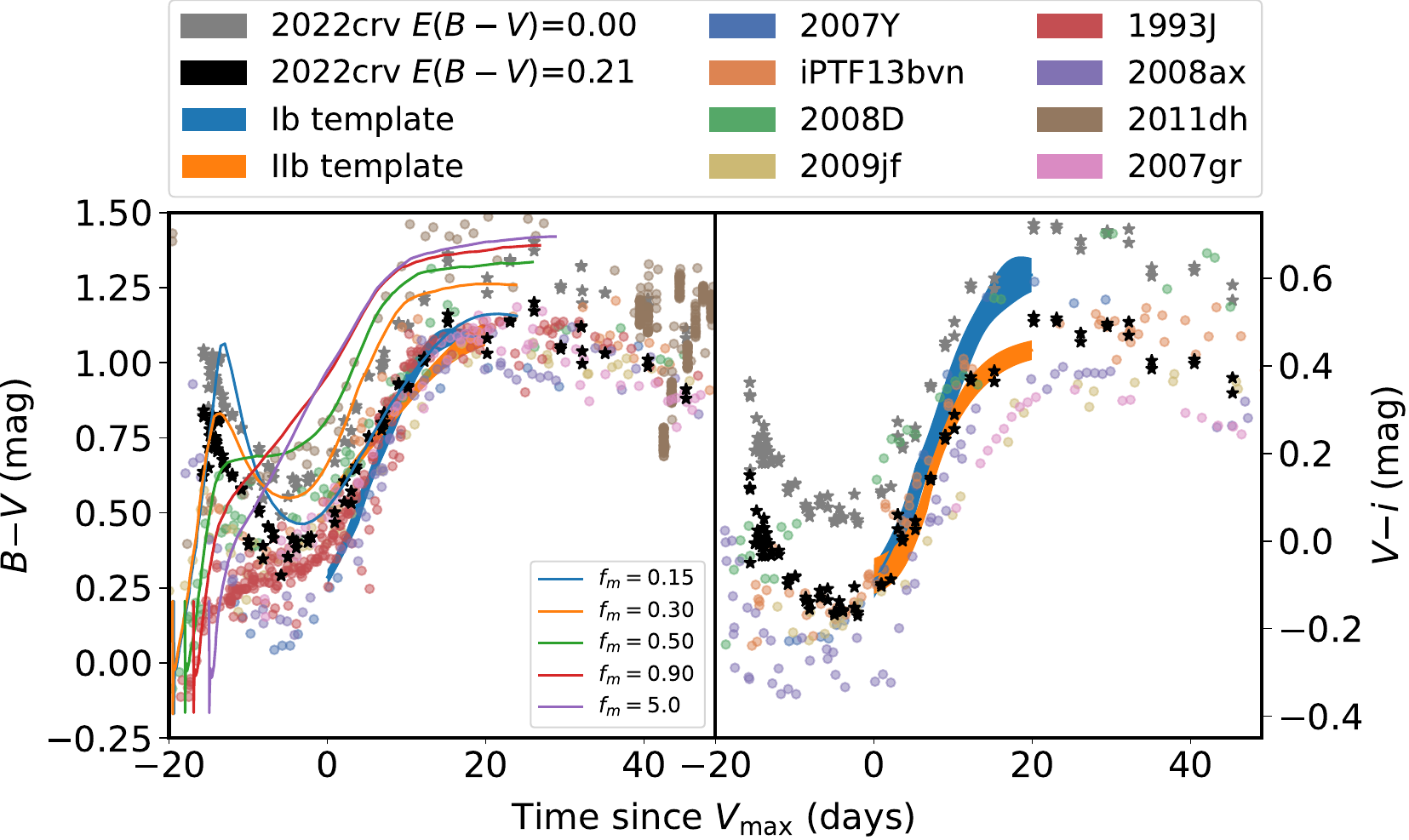}
\caption{\textit{Top:} Color evolution of SN~2022crv, after correcting for a total color excess of $E(B-V)$ = 0.21~mag. The color templates from \cite{Stritzinger2018}. A sample of Ib/IIb SNe after correcting with the published reddening estimates are also plotted. The $B-V$ color evolution is also compared with the SNe Ib color evolution models by \cite{Yoon2019} with various degrees of $\rm ^{56}Ni$ (indicated by $\rm f_m$). SN~2022crv is more consistent with the $\rm f_m$=0.15/0.3 model, suggesting a relatively weak $^{56}$Ni mixing in the ejecta (see Section \ref{sec:phot_evol}). 
\textit{Bottom:} Spectral comparison between a sample of Type Ib SNe and SN~2022crv at about 10 days after the $V$-band maximum. The transmission curve for the $B$, $V$, and $i$ bands have been plotted to indicate their wavelength ranges. The grey line shows the spectrum of 2022crv without reddening correction. An $E(B-V) = 0.21$ gives a consistent spectrum profile with other objects.
\label{fig:color_comp}}
\end{figure*}



\section{Observational Properties}\label{sec:obs_properties}
\subsection{Explosion Epoch}
By combining the early light curves from the DLT40 and ATLAS surveys, we are able to put a strong constraint on the explosion date of SN~2022crv. The SN was not detected during the DLT40 search on JD~2459626.89 (with a 3$\sigma$ limit of 18.44 mag).
On JD~2459627.49, $\sim$14.4 hours after our non-detection, the SN was detected in an $o$-band ATLAS image. The first detection and the last non-detection of SN~2022crv are highlighted in the left panel of Figure \ref{fig:photometry}. By taking the average between the first detection and the last non-detection, we constrain the explosion epoch to be JD~2459627.19$\pm$0.30, which will be adopted throughout the paper. 

\subsection{Reddening Estimation}
The equivalent width (EW) of the Na\,I\,D line is often used to estimate the SN reddening with the assumption that it is a good tracer of the amount of gas and dust along the line of sight \citep[e.g.,][]{Munari1997,Poznanski2012}. In order to measure the line-of-sight reddening towards SN~2022crv, we analyze the medium-resolution spectrum (R$\sim$4000) taken with LRIS on 2022-04-26 (Figure \ref{fig:NaID}). We apply a 4th-order polynomial to fit the continuum, and measure the EW of the Na\,I\,D lines on the continuum-subtracted spectrum (Figure \ref{fig:NaID} bottom panel). The measured EW of the host galaxy Na\,I\,D $\lambda$5890 ($\rm D_2$) and Na\,I\,D $\lambda$5896 ($\rm D_1$) are $0.544_{-0.012}^{+0.012}$~\AA\ and $0.378_{-0.011}^{+0.011}$~\AA, respectively. 
The measured EW of the Galactic Na\,I\,$\rm D_2$ and Na\,I\,$\rm D_1$ are $0.404_{-0.011}^{+0.011}$~\AA\ and $0.247_{-0.011}^{+0.011}$~\AA\ respectively. Using Eq.9 in \cite{Poznanski2012} and applying the renormalization factor of 0.86 from \cite{schlafly_blue_2010}, we found a host extinction of $E(B-V)_{\rm host} = 0.146_{-0.009}^{+0.009}$ mag with 20\% systematic uncertainty \citep{Poznanski2012}.
The Milky Way extinction is measured to be $E(B-V)_{\rm MW} = 0.070_{-0.004}^{+0.004}$ mag, consistent with the value from \citep{Schlafly2011} of $E(B-V)_{\rm MW}$ = 0.0642 (0.0007) mag. The latter will be adopted in this paper. 

For SESNe, the intrinsic color shortly after maximum light is found to be tightly distributed, both observationally \citep{Drout2011,Taddia2015,Stritzinger2018} and theoretically \citep{Dessart2016,Woosley2021}. 
By studying a sample of SESNe, \cite{Drout2011} and \cite{Stritzinger2018} found that, compared to other phases, the scatter in various color indices for each sub-type of SESNe is smaller shortly after the maximum.
Based on the modeling of a range of SESNe, \cite{Woosley2021} confirmed that there are pinches in color indices shortly after the peak. They also found that the spectra of SESNe are similar to each other approximately 10 days after the peak. In the top panel of Figure \ref{fig:color_comp}, we compare the \textit{B}$-$\textit{V} and \textit{V}$-$\textit{i} evolution of SN~2022crv with other well-studied SNe Ib/IIb, including SN~2009jf  \citep{Valenti2011,Sahu2011}, SN~2007Y \citep{Stritzinger2009}, SN~2008D \citep{Modjaz2009}, SN~1993J \citep{Filippenko1993}, and SN~2008ax \citep{Pastorello2008}. The color templates for SNe Ib and IIb from \cite{Stritzinger2018} are also plotted. We found that a total extinction of $E(B-V)$=0.21 mag gives SN~2022crv a consistent color evolution with our comparative sample of SNe Ib/IIb and templates shortly after $V_{\rm max}$. In the bottom panel of Figure \ref{fig:color_comp}, we compare the spectrum of SN~2022crv with the other SNe Ib/IIb at around 10 days after $V_{\rm max}$. The original spectrum of SN~2022crv is redder than those of our other objects, while a total extinction of $E(B-V)=\,0.21$~mag gives a spectral slope more consistent with the population.
Therefore, throughout this paper, we will adopt an $E(B-V)_{\rm tot} = 0.21$~mag, assuming a $\rm R_{V}$ = 3.1.

\subsection{Distance}
The distance of NGC~3054 listed on the NASA/IPAC Extragalactic Database (NED) ranges from 12.9 Mpc to 40.0 Mpc. SN~2006T \citep[Type IIb;][]{Monard2006} also exploded in NGC~3054, and \cite{Lyman2016} used a distance of 32.9 Mpc for SN~2006T, while \cite{Taddia2018} adopted a distance of 31.6 Mpc. To be consistent with the distance of SN~2006T used by previous works, we assume a distance of 31.6$^{+6.4}_{-5.3}$ Mpc (a distance modulus of 32.5$\pm$0.4 mag) based on the Tully-Fisher distance \citep{Tully2009}.


\subsection{Host Properties} \label{sec:host_properties}
SN2022crv is at a projected offset of 36\farcs 4 (5.6$^{+1.1}_{-0.9}$ kpc) from the host galaxy NGC~3054. To estimate the metallicity at the SN position, we measured the flux of the strong ionized gas emission lines (H$\beta$, \ion{O}{3}]~$\lambda$5007, [\ion{N}{3}]~$\lambda$6584, H$\alpha$, [\ion{S}{2}]~$\lambda\lambda$6717, 6731) from the spectrum taken 371.5 days after the SN explosion. The continuum is removed by fitting a linear component around the narrow emission lines. By using the strong-line diagnostics presented in \cite{Curti2020MNRAS.491..944C}, the weighted average oxygen abundance at the SN site is measured to be 12 + log(O/H) = 8.83$\pm$0.08
and the results for each indicator calibration are shown in Table \ref{tab:metal}. Assuming a solar oxygen abundance of 12 + log(O/H) = 8.69 \citep{Allende_Prieto2001ApJ...556L..63A} as well as a solar metallicity ($\rm Z_{\odot}$) of 0.0134 \citep{Asplund2009ARA&A..47..481A}, 12 + log(O/H) = 8.83 is equivalent to a metallicity of Z $\simeq$ 1.4\,$\rm Z_{\odot} \simeq 0.019$. 
The $\rm O_{3}N_{2}$ calibration indicator in \cite{Pettini2004MNRAS.348L..59P} gives 12 + log(O/H) = 8.88$\pm$0.14, consistent with our measurements above.
Comparing to all the SESNe in the PMAS/PPak Integral-field Supernova Hosts Compilation (PISCO) sample \citep{Galbany2018ApJ...855..107G}, SN~2022crv has one of the most metal-rich environments among the Type Ib/IIb SNe in the sample. Given the high metallicity, the progenitor star of SN~2022crv likely experienced strong wind mass loss \citep{Vink2001A&A...369..574V, Crowther2007ARA&A..45..177C, Mokiem2007A&A...473..603M}. 
Just before completing our paper, there is another paper on SN~2022crv coming out \citep{gangopadhyay2023bridging}, and they found a high progenitor mass-loss rate based on the radio light curve, consistent with what we suggested here.
The implication of the high mass loss rate will be further discussed in Section \ref{sec:O_mass_progenitor}.

\begin{deluxetable}{Cccc} \label{tab:metal}
\tabletypesize\scriptsize
\tablecaption{The flux of ionized gas emission lines at the SN position measured from the spectrum taken on 2023-03-23 (371 days after the explosion). The oxygen abundance values are calculated based on the strong-line diagnostics presented in \cite{Curti2020MNRAS.491..944C}. \label{tab:meta}}
\tablewidth{0pt}
\tablehead{
\colhead{Indicator} &
\colhead{Line Ratio} &
\colhead{Value} &
\colhead{12+log[O/H]} 
}
\startdata 
R3&[\ion{O}{3}]~$\lambda$5007/H$\beta$& 0.13&8.84$\pm$0.07 \\
N2&[\ion{N}{3}]~$\lambda$6584/H$\alpha$& 0.39&8.75$\pm$0.10 \\
S2&[\ion{S}{2}]~$\lambda\lambda$6717, 6731/H$\alpha$& 0.18&8.87$\pm$0.06 \\
RS_{32}&[\ion{O}{3}]~$\lambda$5007/H$\beta$ + [\ion{S}{2}]~$\lambda$6717, 6731/H$\alpha$& 0.31&8.85$\pm$0.08 \\
O_{3}S_{2}&([\ion{O}{3}]~$\lambda$5007/H$\beta$) / ([\ion{S}{2}]~$\lambda\lambda$6717, 6731/H$\alpha$)& 0.72&8.76$\pm$0.11 \\
O_{3}N_{2}&([\ion{O}{3}]~$\lambda$5007/H$\beta$) / ([\ion{N}{2}]~$\lambda$6584/H$\alpha$)& 0.34&8.84$\pm$0.09 \\
\hline
Weighted\ Average &&& 8.83$\pm$0.08 \\
\enddata
\end{deluxetable}

\begin{figure}
\includegraphics[width=1.\linewidth]{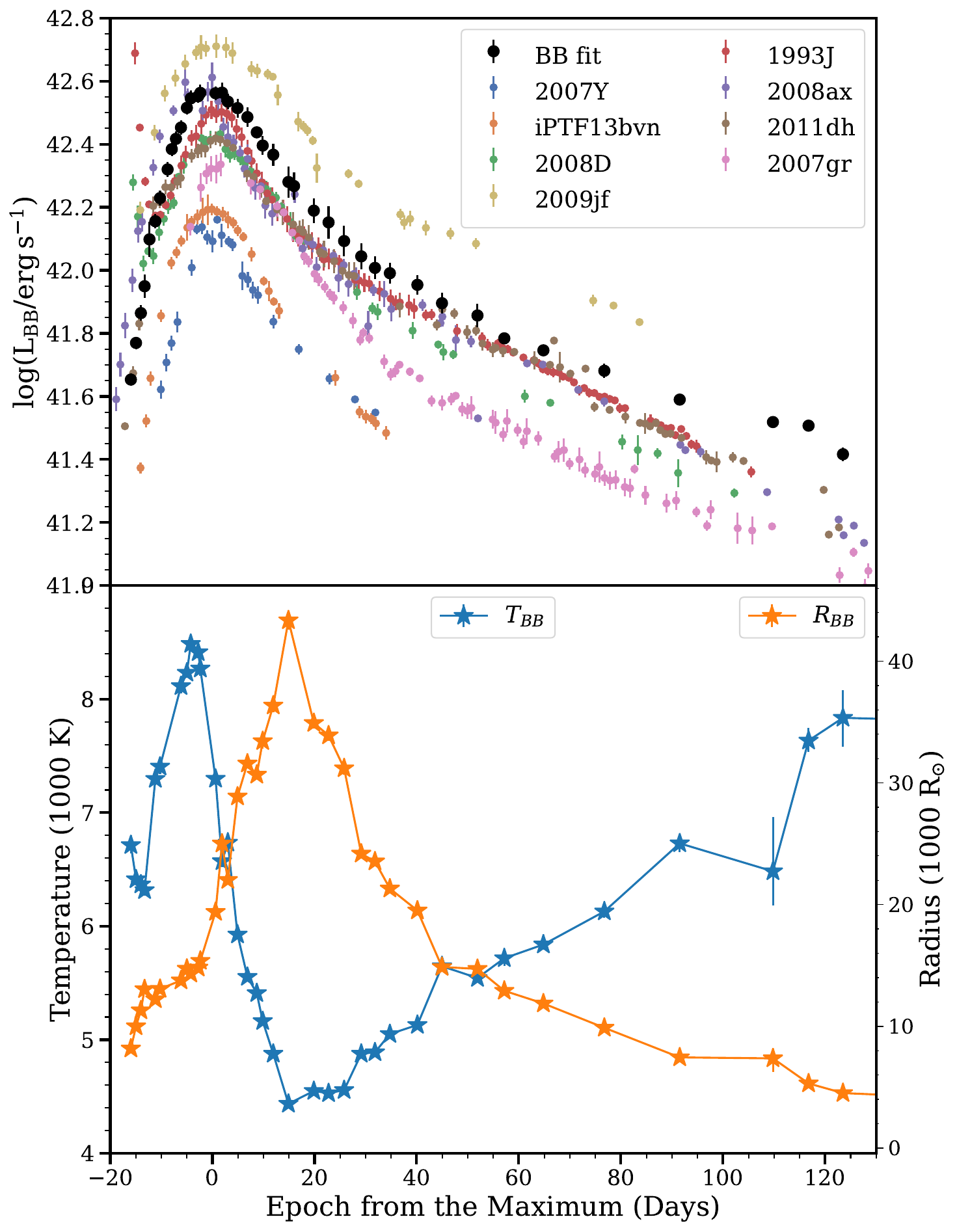}
\caption{\textit{Top:} Bolometric light curve of SN~2022crv compared with other Type Ib/IIb SNe. The unfilled points represent the bolometric light curve from the blackbody fit. 
\textit{Bottom:} The evolution of temperature and radius of SN~2022crv derived from the blackbody fit. 
\label{fig:bolo_lc}}
\end{figure}

\begin{figure*}
\centering
\includegraphics[width=0.9\linewidth]{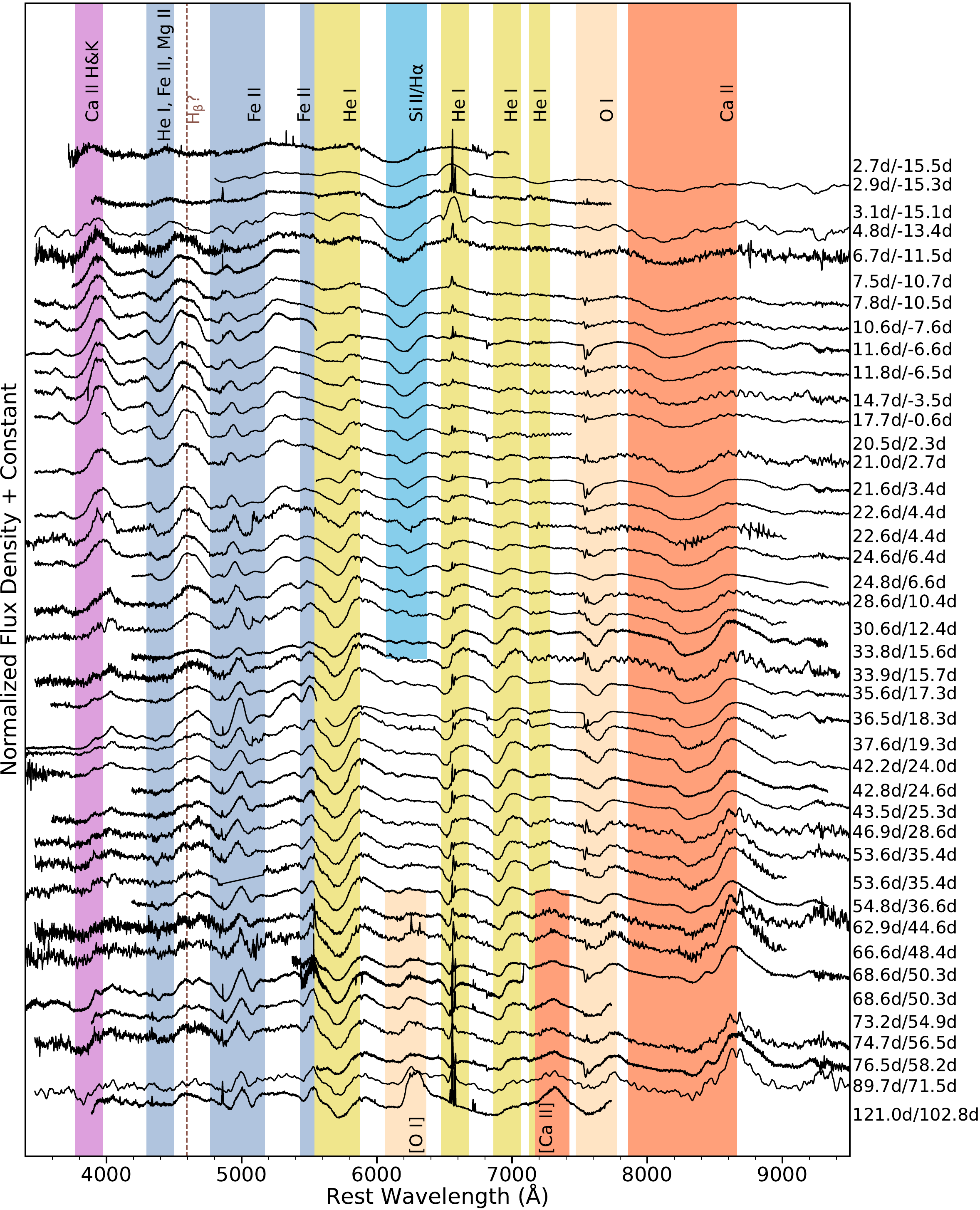}
\caption{The optical spectroscopic evolution of SN~2022crv from the photospheric phase to the early nebular phase. The phase is measured from the explosion/$V$-band maximum.
\label{fig:spec_evol_1}}
\end{figure*}

\begin{figure*}[h!]
\includegraphics[width=1.\linewidth]{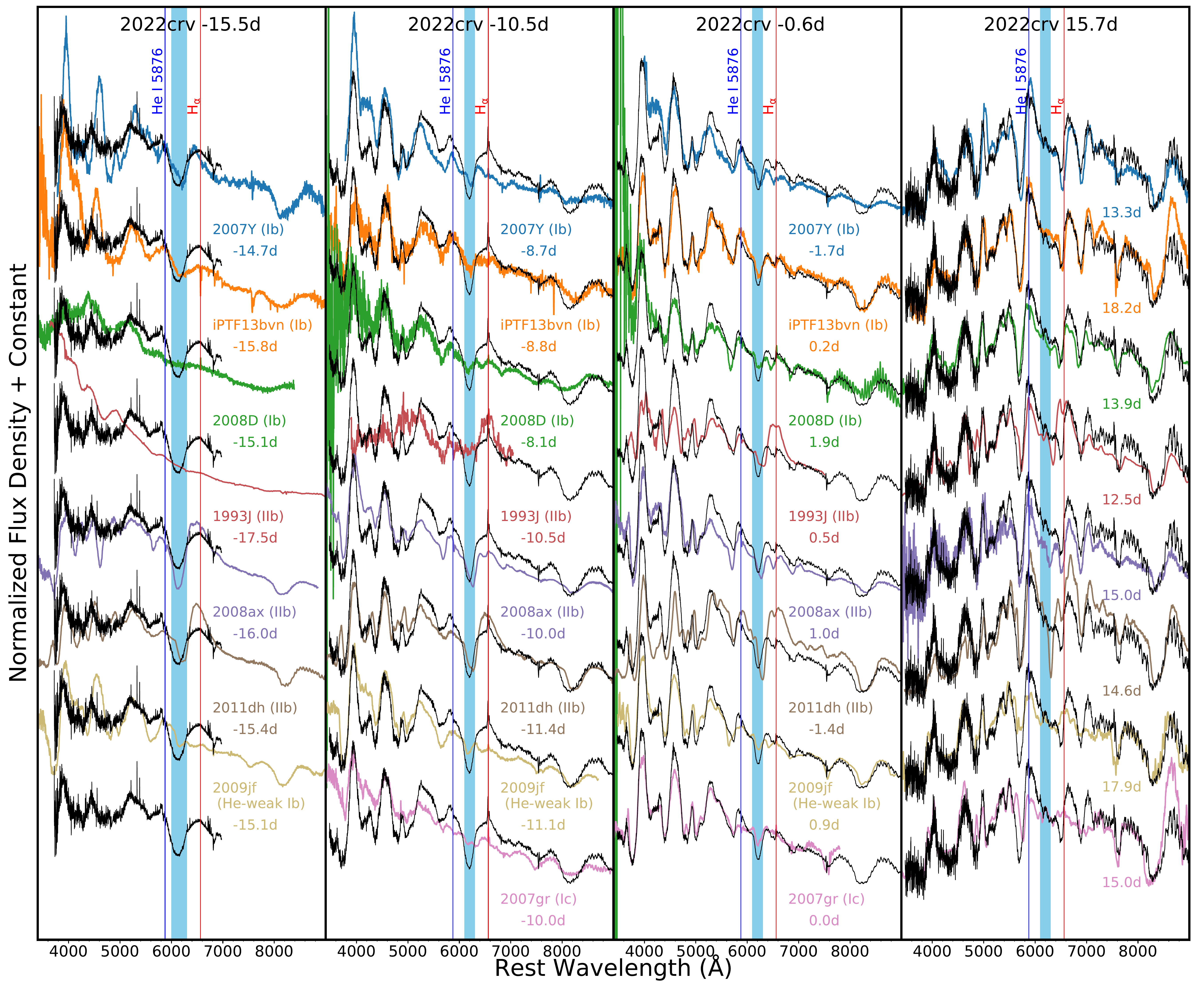}
\caption{Optical spectral comparison of SN~2022crv to other well-studied SESNe, including SNe Ib, SN~2007Y, iPTF13bvn, SN~2008D; SNe IIb, SN~1993J, SN~2008ax, SN~2011dh; and SNe Ic, SN~2007gr.  The blue shaded areas mark the 6200 \AA\ features.
\label{fig:spec_comp}}
\end{figure*}

\section{photometric evolution}\label{sec:phot_evol}
\subsection{Light Curve Evolution}
The multiband light curves of SN~2022crv are shown in Figure \ref{fig:photometry}. SN~2022crv reaches a \textit{V}-band maximum of $M_{V}$~=~$-$17.7$\pm$0.4 mag on JD~2459645.42, $\sim$18.2 days after the date of explosion. Around 60 days after $V_{\rm max}$, the light curves show a linear decline. The $V$-band decline rate is about 1.4~$\rm mag\,(100d)^{-1}$, faster than the expected radioactive decay rate of $\rm ^{56}Co\rightarrow{}^{56}Fe$ [0.98~$\rm mag\,(100d)^{-1}$] \citep{Nadyozhin1994}, which is consistent with other SNe Ib/IIb. In the bottom panel of Figure \ref{fig:photometry}, we compare the $V$-band light curve of SN~2022crv with those of other well-studied SESNe (SNe Ib: SN~2007Y \citep{Stritzinger2009}, iPTF13bvn \citep{Fremling2014}, SN~2008D \citep{Modjaz2009}, SN~2009jf \citep{Valenti2011}; SNe IIb: SN~1993J \citep{Filippenko1993}, SN~2008ax \citep{Pastorello2008}, SN~2011dh \citep{Ergon2014A&A...562A..17E}; and SN Ic: SN~2007gr \citep{Hunter2009A&A...508..371H}). These objects are selected because they have well-sampled light curves and spectral sequences.

The apparent magnitudes of other SESNe in the sample have been shifted to match the peak of SN~2022crv. SN~2022crv does not show an initial peak due to shock cooling like SN~1993J does and has a faster rise than SN~2008D. Overall, the light curve shape of SN~2022crv is similar to SN~2009jf, SN~2007Y and SN~2008ax. 

\citet{Chevalier2010ApJ...711L..40C} suggested that Type IIb SNe can be divided into two categories according to whether the progenitor is compact or extended. If the amount of hydrogen is low enough, the compact Type IIb would merge into Type Ib.
Due to the lack of cooling envelope emission, SN~2022crv likely has a compact progenitor, which will be further discussed in Section \ref{sec:hydro_envelope}. 


The $B-V$ and $V-i$ color evolution of SN~2022crv is compared to our sample of SNe Ib/IIb in Figure \ref{fig:color_comp}. The color evolution of SN~2022crv after $V_{max}$ is similar to those of other SESNe in comparison. After correcting for reddening, the $B-V$ color of SN~2022crv shows a rapid initial rise and reaches a peak of $\sim$0.8 mag. The color then evolves toward the blue down to $\sim$0.4 mag. After maximum light, the $B-V$ color increases and reaches a peak of $\sim$1.1~mag before entering the nebular phase. The $V-i$ color of SN~2022crv shows a similar trend. 
Through hydrodynamical simulations, \cite{Yoon2019} found that the early color evolution of SNe Ib/c is strongly affected by the $\rm ^{56}Ni$ mixing level in the SN ejecta. The $\rm ^{56}Ni$ mixing level is characterized by $\rm f_{m}$ in \cite{Yoon2019}, with a larger value of $\rm f_{m}$ representing a more mixing.
The $B-V$ evolution of SN~2022crv matches with the $\rm f_{m} = 0.15$ or the $\rm f_{m} = 0.3$ model presented in \cite{Yoon2019} (Figure \ref{fig:color_comp}), implying weak $\rm ^{56}Ni$ mixing in SN~2022crv.

\subsection{Bolometric Light Curve}
\label{sec:bolo}
In this section, we build the bolometric light curve of SN~2022crv, which will be used to determine the physical parameters of explosion in Section \ref{sec:arnett_model}.

Due to the lack of photometric coverage in the UV and NIR, we calculated the bolometric correction based on the $B-i$ color evolution and applied it to the $B$-band light curve, as described in \cite{Lyman2014,Lyman2016}. The bolometric magnitudes are converted to bolometric luminosity assuming $M_{Bol,\,\odot}$ = 4.74 and $L_{Bol,\, \odot}$ = 3.9$\times 10^{33}$~$\rm erg\,s^{-1}$. The bolometric light curve of SN~2022crv is shown in Figure \ref{fig:bolo_lc} along with those of a sample of Type Ib/IIb SNe. 
The peak bolometric luminosity of SN~2022crv is 3.39$\rm \times10^{42} erg\,s^{-1}$ and is almost as bright as SN~2009jf, which puts SN~2022crv on the brighter end among the Type Ib/IIb SNe \citep{Lyman2016, Taddia2018, Prentice2019MNRAS.485.1559P}.
As a sanity check, we also calculated the bolometric light curve by fitting the reddening-corrected SED of SN~2022crv with a blackbody spectrum at each epoch using a Markov Chain Monte Carlo (MCMC) routine in the Light Curve Fitting package \citep{hosseinzadeh_light_2020}. The blackbody bolometric light curve is plotted in Figure \ref{fig:bolo_lc} and is consistent with the bolometric light curve derived from bolometric correction. 

From the blackbody fits, we also derived the radius and temperature evolution of SN~2022crv (bottom panel of Figure \ref{fig:bolo_lc}). The temperature increases after a rapid initial drop at early phases, and reaches a peak of $\sim8000\,$K several days before $V_{max}$, then it rapidly decreases until reaching a minimum of $\sim4500\,$K before the object settles into the nebular phase. The behaviour of the temperature evolution is consistent with the color evolution shown in Figure \ref{fig:color_comp}. The radius of SN~2022crv continuously increases until it peaks at around 20 days after $V_{\rm max}$. The radius and temperature evolution of SN~2022crv is similar to other SESNe \citep{Taddia2018}.

\begin{deluxetable*}{cccccccc}
\tabletypesize\scriptsize
\tablecaption{Position of the minimum of the absorption component of the P-cygni profile (\AA). \label{tab:velo1}}
\tablewidth{0pt}
\tablehead{
\colhead{Epoch} &
\colhead{MJD} &
\colhead{$\rm H{\alpha}$/\ion{Si}{2}~$\lambda$6355} &
\colhead{\ion{He}{1}~$\lambda$5876} &
\colhead{\ion{He}{1}~$\lambda$6678} &
\colhead{\ion{Fe}{2}~$\lambda$4561} &
\colhead{\ion{Fe}{2}~$\lambda$5018} &
\colhead{\ion{Fe}{2}~$\lambda$5169} 
}
\startdata 
2.7&59629.4&6172.5&5623.7&$\cdots$ &$\cdots$ &$\cdots$ &$\cdots$  \\
2.9&59629.6&6183.5&5590.9&$\cdots$ &$\cdots$ &$\cdots$ &$\cdots$  \\
3.1&59629.8&6177.0&5638.5&$\cdots$ &$\cdots$ &$\cdots$ &$\cdots$  \\
4.8&59631.5&6218.0&5678.4&$\cdots$ &$\cdots$ &$\cdots$ &$\cdots$  \\
6.7&59633.4&6237.1&5709.2& &4378.0&4814.7&4997.8 \\
7.5&59634.2&$\cdots$ &$\cdots$ & $\cdots$&4397.2&4842.4&5014.6 \\
7.8&59634.5&6240.0&5657.1& &4400.2&4843.7&5014.8 \\
10.6&59637.3&6246.9&5725.6&6500.2&4418.3&4856.7&5020.4 \\
11.6&59638.3&$\cdots$ &$\cdots$ & $\cdots$&4419.6&4859.1&5025.3 \\
11.6&59638.3&6252.2&$\cdots$ &6503.2&$\cdots$ & $\cdots$& $\cdots$ \\
11.8&59638.5&6247.9&5730.6&6513.6&4417.4&4857.0&5029.8 \\
14.7&59641.4&6253.6&5769.5&6516.0&4428.1&4863.0&5041.3 \\
17.7&59644.3&6261.7&5771.3&6556.0&4436.3&4870.9&5049.0 \\
21.0&59647.6&6271.4&5762.9&6562.6& $\cdots$& $\cdots$&$\cdots$  \\
21.6&59648.3& $\cdots$&$\cdots$ &$\cdots$ &4438.9&4886.3&5060.6 \\
21.6&59648.3&6279.7&5767.4&6580.3& $\cdots$&$\cdots$&$\cdots$  \\
22.6&59649.3&6280.6&5767.6&6568.9& $\cdots$&$\cdots$&$\cdots$  \\
22.6&59649.3&6290.1&5774.3&6580.0&4440.8&4878.9&5077.2 \\
24.6&59651.3&6288.1&5760.5&6567.7&$\cdots$ & $\cdots$& $\cdots$ \\
24.8&59651.5&6289.8&5759.7&6573.8&4438.1&4904.1&5073.3 \\
28.6&59655.3&6303.8&5745.4&6560.9& $\cdots$&4920.3&5088.0 \\
30.6&59657.3&6320.6&5744.5&6563.6& $\cdots$&4927.3&5093.5 \\
\enddata
\tablenotetext{\star}{Epoch is measured from the explosion (JD~2459627.19).}
\end{deluxetable*}

\begin{deluxetable}{cccccccc}
\tabletypesize\scriptsize
\tablecaption{The position of the flux minimum of the absorption component of the P cygni profile (\AA). \label{tab:velo2}}
\tablewidth{0pt}
\tablehead{
\colhead{Epoch} &
\colhead{MJD} &
\colhead{\ion{He}{1}~$\lambda$5876} &
\colhead{\ion{He}{1}~$\lambda$6678} &
\colhead{\ion{Fe}{2}~$\lambda$5018} &
\colhead{\ion{Fe}{2}~$\lambda$5169} 
}
\startdata 
33.8&59660.5& 5734.7&6549.4& 4917.8&5099.7 \\
33.9&59660.6& 5738.8&6555.8& 4918.7&5105.2 \\
35.6&59662.2& 5737.2&6556.6& $\cdots$ &5110.8 \\
37.5&59664.2& 5743.4& $\cdots$ &$\cdots$ & $\cdots$ \\
36.5&59663.2& 5745.1&6562.7& $\cdots$ & $\cdots$ \\
36.5&59663.2& $\cdots$ & $\cdots$&  $\cdots$&5108.1 \\
37.6&59664.2& 5745.6&6566.0 &$\cdots$&5102.7 \\
42.2&59668.9& 5737.3&6560.8 &$\cdots$&5111.2 \\
42.8&59669.5& 5739.7&6565.4 &$\cdots$&5114.7 \\
43.5&59670.2& 5738.8&6565.8 &$\cdots$&5114.1 \\
46.9&59673.6& 5738.4&6567.1 &$\cdots$&5089.6 \\
53.6&59680.3& 5743.9&6571.8 &$\cdots$&5095.0 \\
53.6&59680.3& 5752.6&6582.9 &$\cdots$& $\cdots$ \\
54.8&59681.5& 5744.9&6576.5 &$\cdots$&5116.4 \\
62.9&59689.5& 5753.6&6577.4 &$\cdots$&5115.5 \\
66.6&59693.3& 5763.9&6590.2 &$\cdots$&5117.6 \\
68.6&59695.2& $\cdots$ & $\cdots$& $\cdots$ &5119.2 \\
68.6&59695.2& 5757.8&6589.7 & $\cdots$& $\cdots$ \\
68.6&59695.2& 5756.4&6594.2 & $\cdots$& $\cdots$ \\
73.2&59699.9& 5756.1&$\cdots$ & $\cdots$ &5118.7 \\
74.7&59701.4& 5755.1&6578.9&$\cdots$ & 5097.0 \\
76.5&59703.2& $\cdots$ & $\cdots$& $\cdots$& 5119.2 \\
76.5&59703.2& 5755.0& $\cdots$ &$\cdots$ & $\cdots$ \\
89.7&59716.4& 5765.0&6579.4& $\cdots$ & $\cdots$ \\
121.0&59747.7 &5770.3& $\cdots$& $\cdots$& 5119.7 \\
\enddata
\tablenotetext{\star}{Epoch is measured from the explosion (JD~2459627.19).}
\end{deluxetable}

\begin{figure}
\includegraphics[width=1.\linewidth]{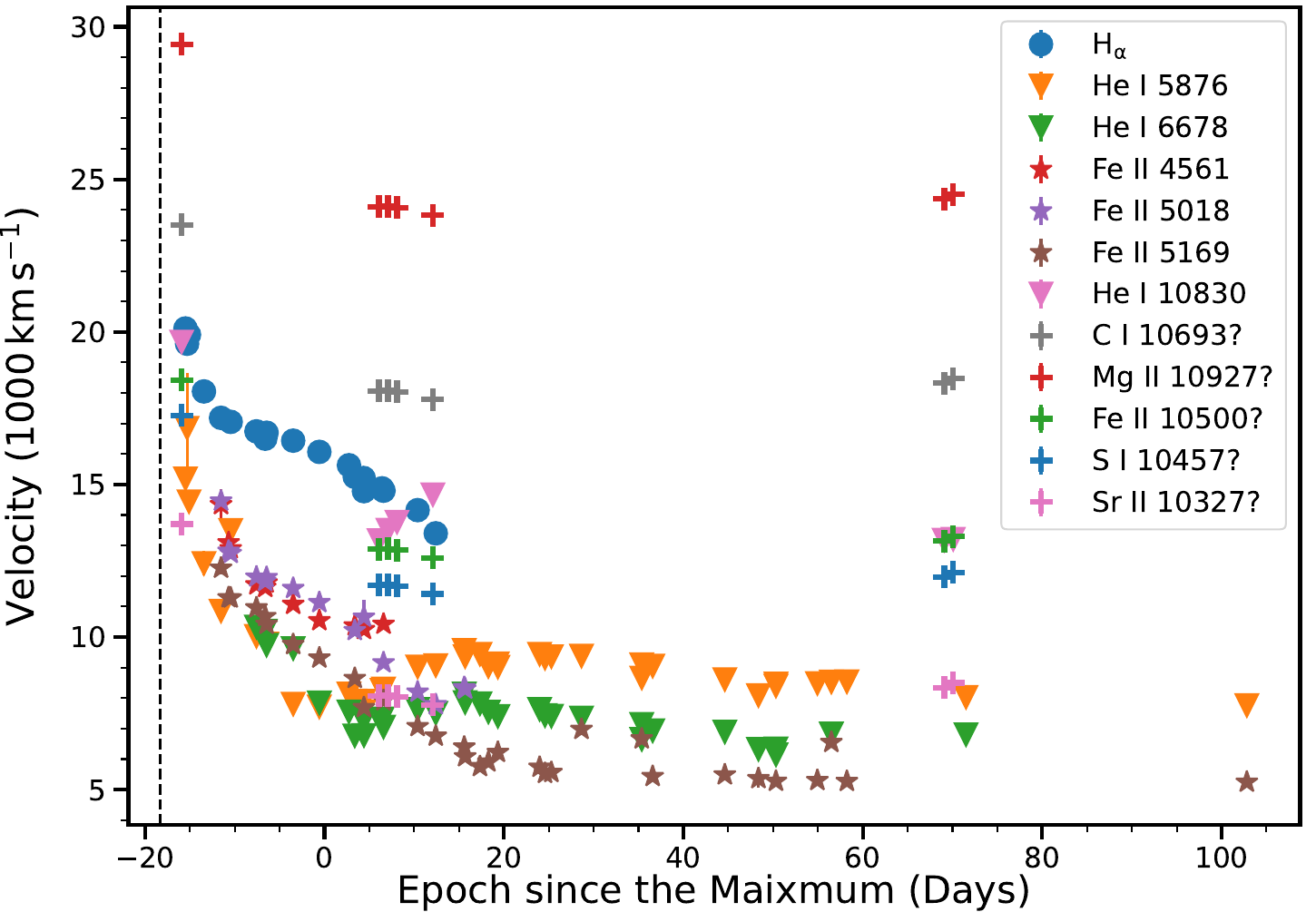}
\caption{Velocity evolution of $\rm H\alpha$, He I, Fe II and feature A (see Section \ref{sec:nir_evol}). The epoch of explosion is marked with the vertical dashed line. The velocity of feature A is calculated assuming it is from \ion{C}{1}~$\lambda$1.0693~$\mu$m, \ion{Mg}{1}~$\lambda$1.0927~$\mu$m, \ion{Fe}{2}~$\lambda$1.05~$\mu$m, \ion{S}{1}~$\lambda$1.0457~$\mu$m, and \ion{Sr}{1}~$\lambda$1.0327~$\mu$m, respectively, and is shown with the plus symbol. The velocity of $\rm H{\alpha}$ shows a rapid drop after the $V_{max}$. As discussed in Section \ref{sec:classification}, the 6200 \AA\ feature is likely dominated by Si~II after $V_{\text{max}}$, so this drop does not reliably reflect the velocity evolution of the $\rm H{\alpha}$ line.
\label{fig:line_velo_evol}}
\end{figure}

\begin{figure}
\includegraphics[width=1.\linewidth]
{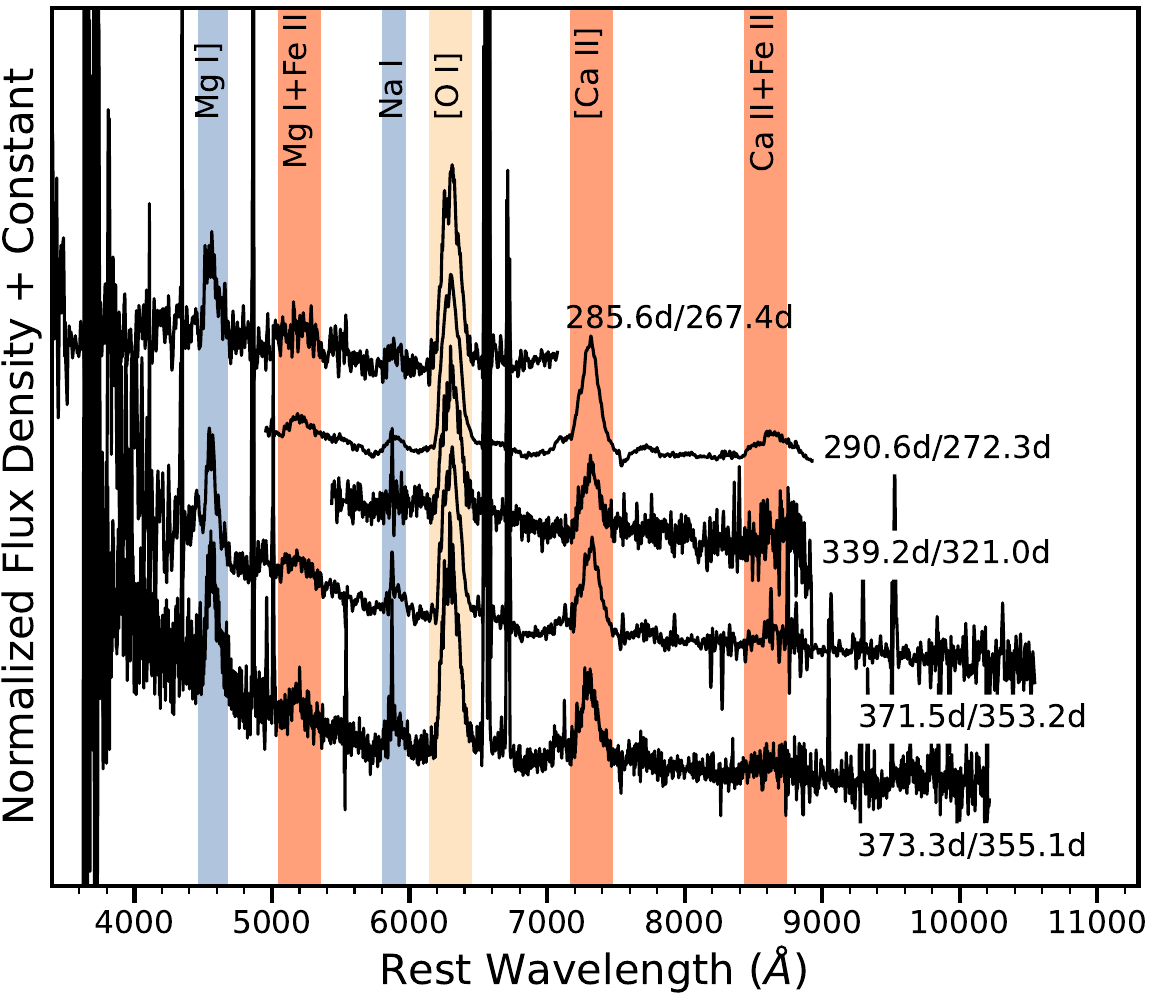}
\caption{The nebular phase spectroscopic evolution of SN~2022crv. Phase is measured from explosion/$V$-band maximum. All the spectra shown here are smoothed with a second order Savitzky-Golay filter with a window size of 11 \AA.
\label{fig:spec_evol_2}}
\end{figure}

\begin{figure}
\includegraphics[width=1.\linewidth]{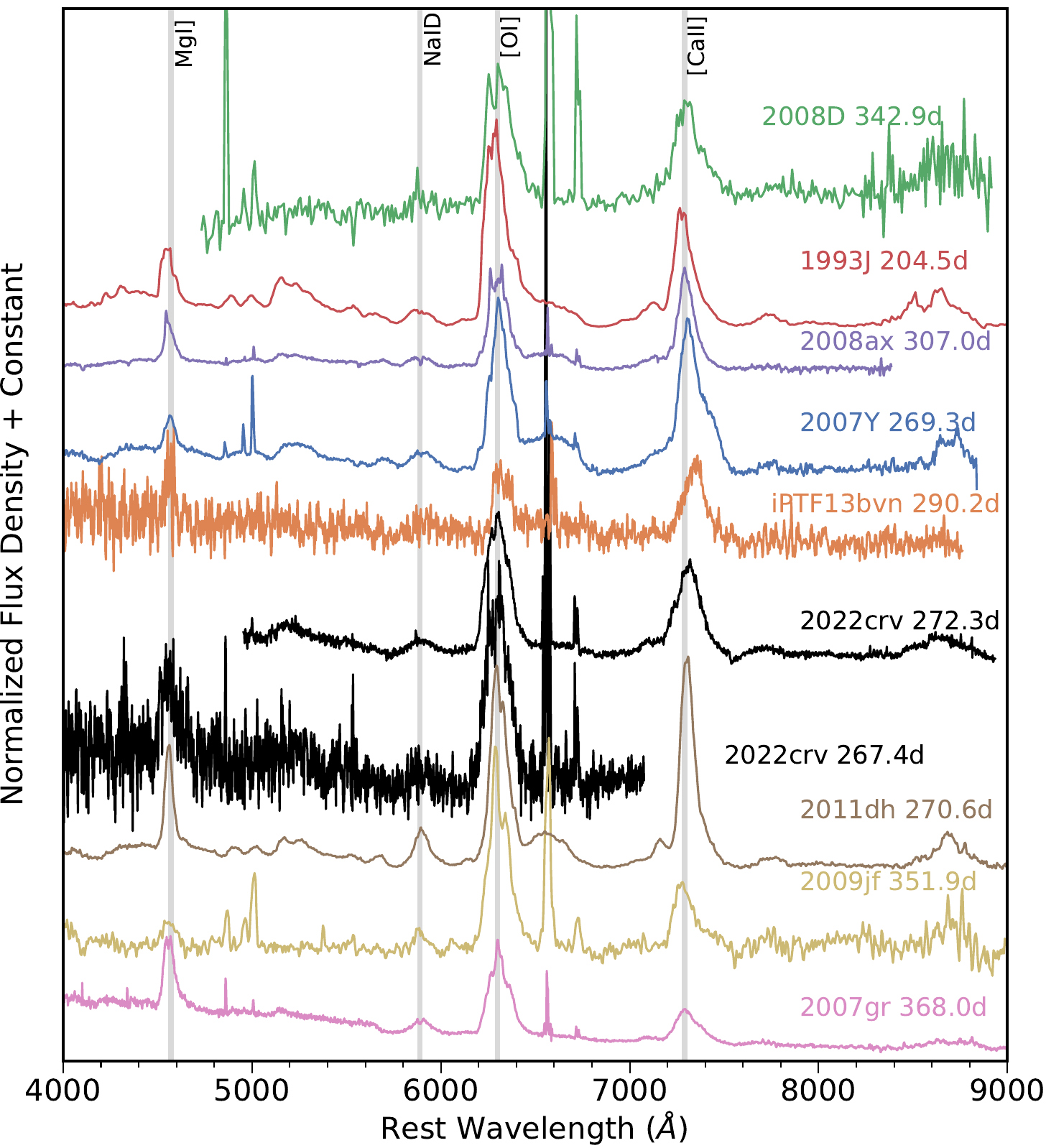}
\caption{Nebular spectra of SN~2022crv compared to other SESNe. The phase is measured from the $V$-band maximum.
\label{fig:nebular_spec_com}}
\end{figure}

\begin{figure}
\includegraphics[width=1.\linewidth]{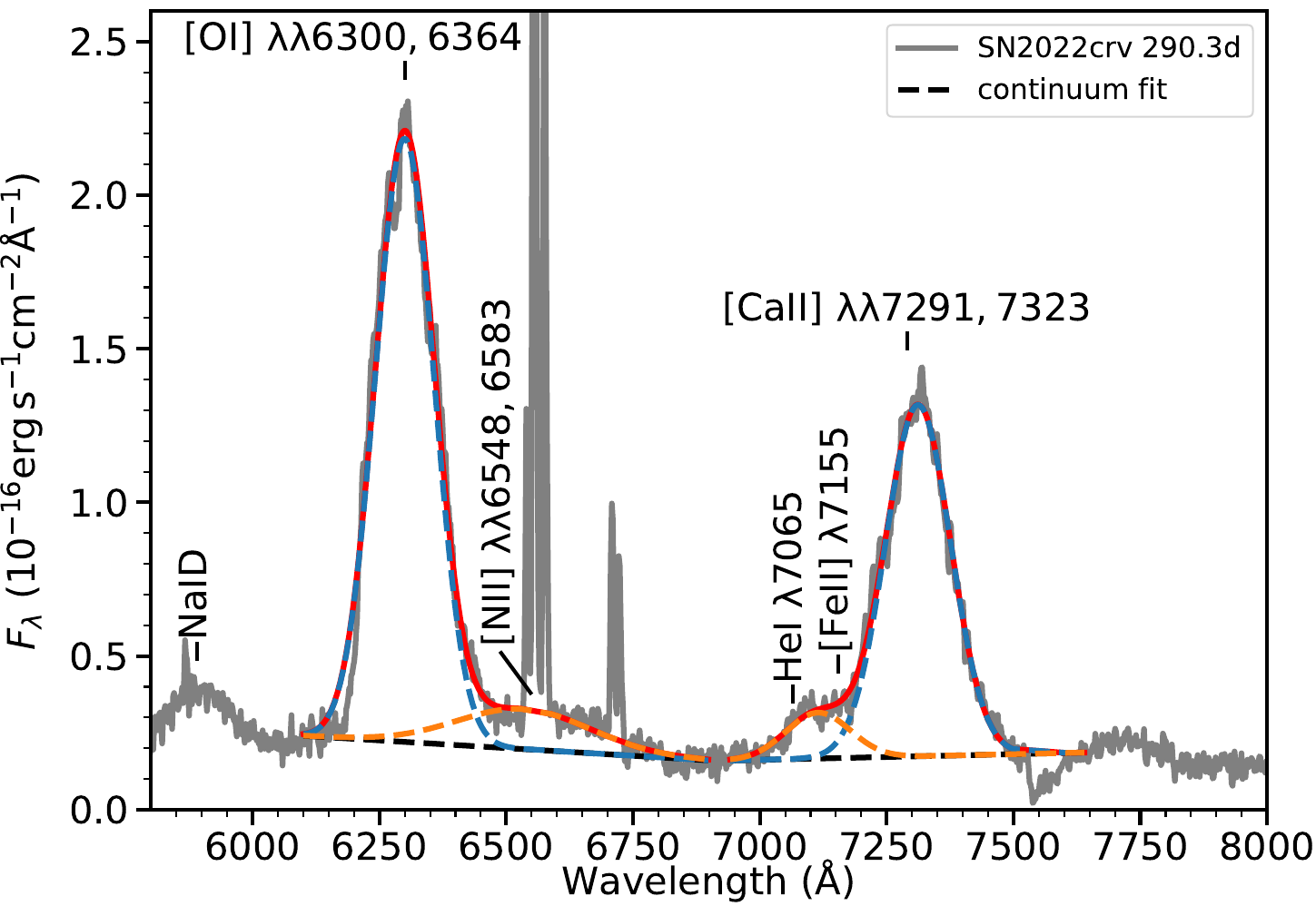}
\caption{Line identification of the nebular spectrum taken on 2022-12-04. A double Gaussian function is fitted around 6300 \AA\ and 7300 \AA, respectively, to deblend the [\ion{O}{1}] and [\ion{Ca}{2}] from other lines. The continuum is defined by a straight line connecting the two local minima.
\label{fig:nebular_OI_CaII}}
\end{figure}

\begin{figure*}
\includegraphics[width=1.\linewidth]{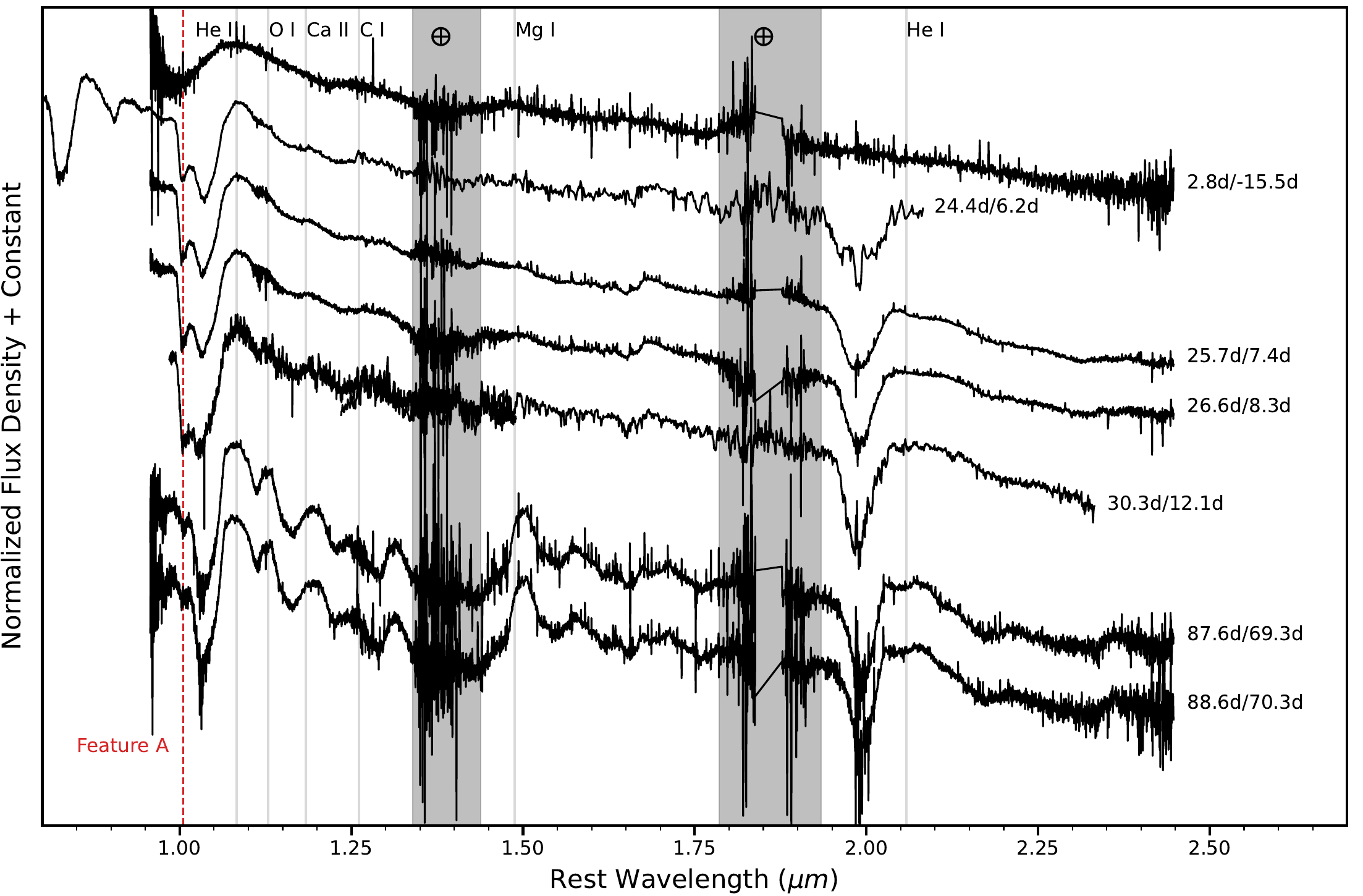}
\caption{NIR spectra of SN~2022crv. The phase is measured from the explosion/$V$-band maximum. The red dashed line marks the extra absorption feature on the blue side of \ion{He}{1}~$\lambda$1.083~$\mu$m. The high telluric absorption regions are marked with grey bands.
\label{fig:spec_nir}}
\end{figure*}

\begin{figure*}
\centering
\includegraphics[width=0.8\linewidth]{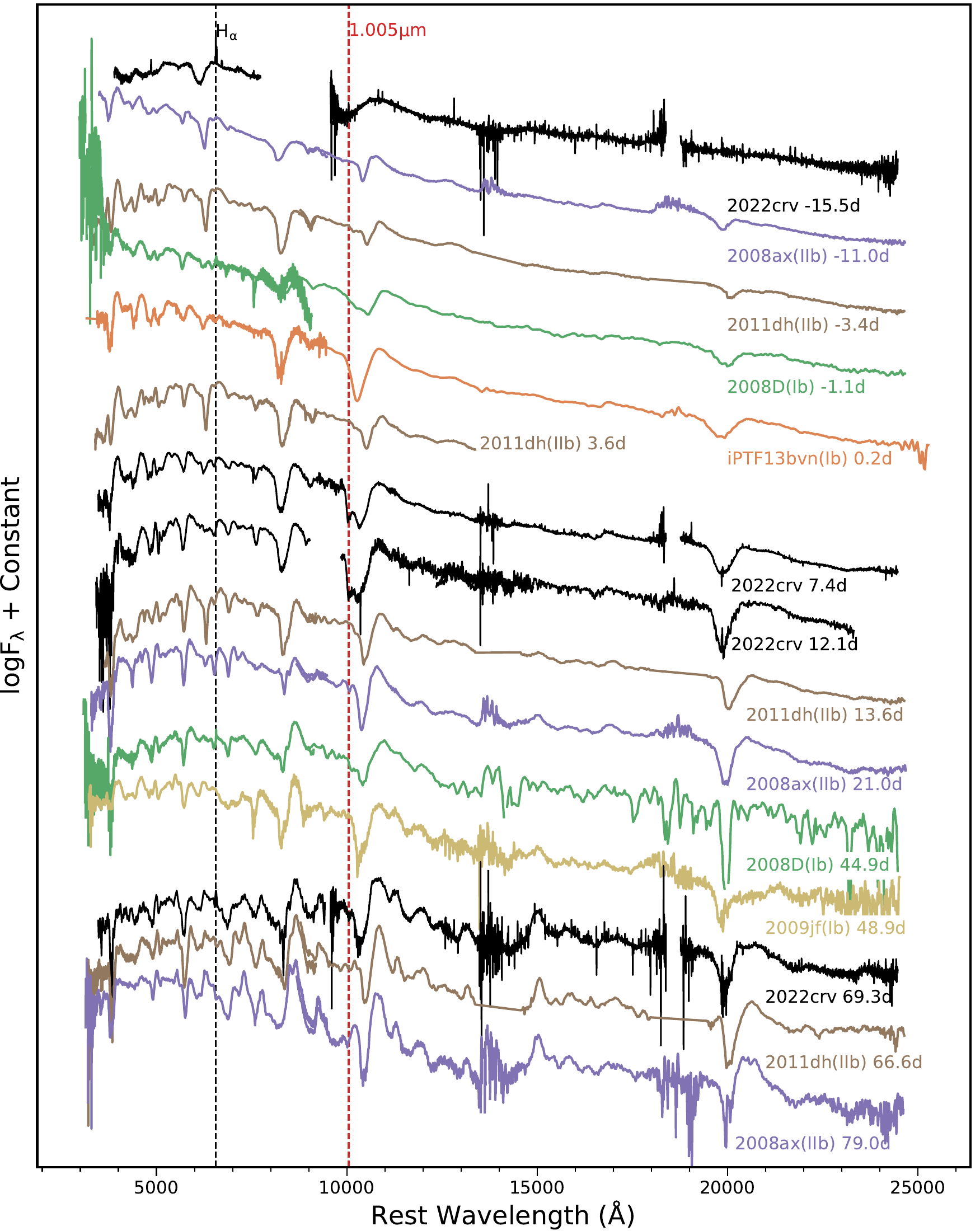}
\caption{Comparison of combined optical and NIR spectra of SN~2022crv with those of other SNe Ib/IIb. The position of $\rm H{\alpha}$ (black dashed line) and 1.005~$\mu$m (red dashed line) are marked.
\label{fig:nir_spec_comp}}
\end{figure*}

\begin{figure}
\includegraphics[width=1.\linewidth]{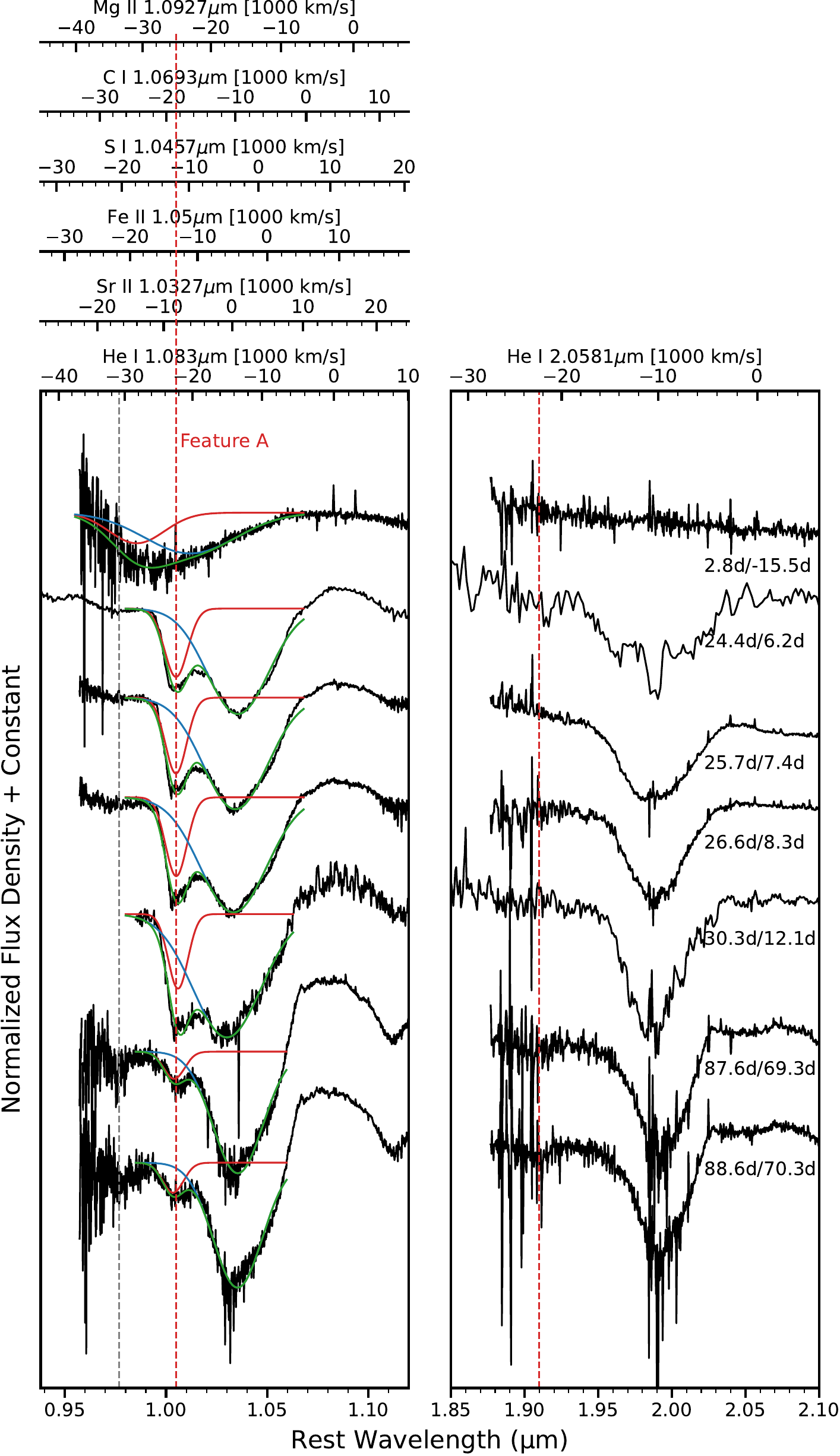}
\caption{Spectral evolution of SN~2022crv in the 1 $\mu$m and 2 $\mu$m regions. The phase if measured from the explosion/$V$-band maximum. Feature A can be fit with a two-Gaussian function.
The velocity scales of \ion{C}{1}~$\lambda$1.0693~$\mu$m, \ion{Mg}{1}~$\lambda$1.0927~$\mu$m, high velocity (HV) \ion{He}{1}~$\lambda$1.083~$\mu$m, \ion{Fe}{2}~$\lambda$1.05~$\mu$m, \ion{S}{1}~$\lambda$1.0457~$\mu$m, and \ion{Sr}{2}~$\lambda$1.0327~$\mu$m are shown in the top x-axis. In the left panel, the red dashed line roughly marks the position of feature A at 1.005~$\mu$m, corresponding to $\sim$23280~km~s$^{-1}$ assuming feature A is from \ion{He}{1}~$\lambda$1.083~$\mu$m. The red dashed in the right panel also has a velocity of $\sim$23280km/s with respect to \ion{He}{1}~$\lambda$20581. The grey dashed line marks the possible identification of \ion{Sr}{2}~$\lambda$1.0036~$\mu$m, at a velocity of $\sim$8000~km~s$^{-1}$.
\label{fig:spec_nir_velo}}
\end{figure}

\section{Spectroscopic Evolution}
\label{sec:spec_evol}
\subsection{Evolution of Optical Spectra From Photospheric Phase To Early Nebular Phase}
The optical spectra from the photospheric to early nebular phase are shown in Figure~\ref{fig:spec_evol_1}. The early spectra show a prominent absorption line at $\sim$6200 \AA, which could be due to Si II or high velocity H$\alpha$. We also note that a small notch appears at 4595 \AA\ at early phases (marked in Figure~\ref{fig:spec_evol_1}) and could arise from $\rm H{\beta}$ at a similar velocity to $\rm H{\alpha}$.
We will discuss the $\sim$6200 \AA\ absorption line further in section \ref{sec:hydro_envelope}.  \ion{He}{1}~$\lambda$5876 appears in the first spectrum and gets stronger over time. \ion{He}{1}~$\lambda$6678, \ion{He}{1}~$\lambda$7065 and \ion{He}{1}~$\lambda$7281 can be seen after $\sim -$6~d. The \ion{Fe}{2}~$\lambda$4561, \ion{Fe}{2}~$\lambda$5018 and \ion{Fe}{2}~$\lambda$5169 lines, which are good tracers of photospheric velocity, can be identified after $\sim -$11~d. Other typical lines such as Ca II H\&K $\lambda\lambda3934,3969$, \ion{Ca}{2}~$\lambda\lambda$8498, 8542, 8662 and \ion{O}{1}~$\lambda$7774 are present and are strong before the object is well into the nebular phase. The \ion{[Ca]}{2}~$\lambda\lambda$7291,7323 and [\ion{O}{1}]~$\lambda\lambda$6300,6364 lines start to emerge after day 35 and dominate the late-phase spectra. The evolution of the velocities of \ion{He}{1}~$\lambda$5876, \ion{He}{1}~$\lambda$6678, \ion{Fe}{2}~$\lambda$4561, \ion{Fe}{2}~$\lambda$5018 and \ion{Fe}{2}~$\lambda$5169 are shown in Figure~\ref{fig:line_velo_evol}. The velocity evolution of the 6200~\AA\ line is also shown in Figure~\ref{fig:line_velo_evol}, assuming it is from H$\alpha$. The position of the flux minima of these lines are listed in Tables \ref{tab:velo1} and \ref{tab:velo2}.

In Figure \ref{fig:spec_comp}, we compare the optical spectra of SN~2022crv at day $-$15.5, day $-$10.5, day $-$0.6 and day +15.7 with those of other SESNe at similar epochs, including SNe Ib: SN~2007Y \citep{Stritzinger2009}, iPTF13bvn \citep{Fremling2014}, SN~2008D \citep{Modjaz2009}; SNe IIb: SN~1993J \citep{Filippenko1993}, SN~2008ax \citep{Pastorello2008}, SN~2011dh \citep{Ergon2014A&A...562A..17E}; and SN Ic: SN~2007gr \citep{Hunter2009A&A...508..371H}. At early phases, SN~2022crv is more similar to the SNe IIb in our comparison sample; but after maximum light, SN~2022crv is almost identical to the SNe Ib sample. 
The 6200 \AA\ feature in SN~2022crv completely disappeared around 15d after $V_{\rm max}$, while the $\rm H\alpha$ line in SNe IIb is still strong at similar phases. This will be discussed further in Section \ref{sec:hydro_envelope}.

\subsection{Nebular Spectra}
\label{sec:nebular_spec}
The nebular spectra are shown in Figure \ref{fig:spec_evol_2}. In Figure \ref{fig:nebular_spec_com}, we compare the nebular spectra of SN~2022crv with those of other Type Ib/IIb SNe at similar epochs, and find that in agreement with the sample, SN~2022crv is dominated by strong [\ion{O}{1}]~$\lambda\lambda$6300, 6364 and [\ion{Ca}{2}]~$\lambda\lambda$7291, 7323. In addition, weak \ion{Mg}{1}]~$\lambda$4571 and Na\,I\,D doublet can also be seen in the spectra. The hydrogen emission is non existent or very weak in the nebular spectra, further supporting that SN~2022crv has a compact progenitor \citep{Chevalier2010ApJ...711L..40C}.

The [\ion{O}{1}] and [\ion{Ca}{2}] lines are often used to constrain the progenitor of SESNe, so a detailed analysis is presented in Figure \ref{fig:nebular_OI_CaII}. The [\ion{O}{1}] doublet is slightly blended with [\ion{N}{2}]~$\lambda\lambda$6548, 6583, and the [\ion{Ca}{2}] line is blended with \ion{He}{1}~$\lambda$7065 and [\ion{Fe}{2}]~$\lambda$7155. In order to deblend [\ion{O}{1}] and [\ion{Ca}{2}] from other lines, we fit 2 Gaussians around [\ion{O}{1}] and [\ion{Ca}{2}], respectively. The continuum is defined by a line connecting the two local minima. The nebular spectrum has been scaled to the $r-$ and $i-$band photometry. The fits are shown in Figure \ref{fig:nebular_OI_CaII}. The flux of the [\ion{O}{1}] at day 290.3 is measured to be 2.9$\times 10^{-14}$~$\rm erg\,s^{-1}\,cm^{-2}$, and the [\ion{O}{1}]/[\ion{Ca}{2}] ratio is found to be 1.5. At day 371.3, the [\ion{O}{1}]/[\ion{Ca}{2}] ratio is found to be 1.6, consistent with the value we got at 290.3 d. 
In section \ref{sec:O_mass_progenitor}, these measurements will be used to constrain the progenitor properties of SN~2022crv.

\subsection{Evolution of NIR Spectra} \label{sec:nir_evol}
Figure \ref{fig:spec_nir} shows the spectroscopic evolution of the NIR spectra of SN~2022crv. The first spectrum was taken only $\sim$2.8 days after explosion, and it shows a very high-velocity \ion{He}{1}~$\lambda$1.083~$\mu$m line ($\sim$20000 $\rm km\,s^{-1}$). In Figure \ref{fig:nir_spec_comp}, we compare the combined optical and NIR spectrum of SN~2022crv with other SESNe at various epochs, including SNe Ib: SN~2008D \citep{Modjaz2009}, SN~2009jf \citep{Valenti2011}, LSQ13abf \cite{Stritzinger2020}; and SNe IIb: SN~2008ax \citep{Pastorello2008}, SN~2011dh \citep{Ergon2014A&A...562A..17E}.

In general, the NIR line evolution of SN~2022crv is consistent with other Type Ib/IIb. However, at around 1.005~$\mu$m, there is an extra absorption feature on the blue side of \ion{He}{1}~$\lambda$1.083~$\mu$m, hereafter feature A. 
A similar feature is likely present in SN~2008D and SN~2008ax, but it is only at late times and not as strong as the feature in SN~2022crv.
To our best knowledge, it is the first time that such a strong feature is observed in a SESN.
In order to isolate feature A from the \ion{He}{1}~$\lambda$1.083~$\mu$m line, we fit two Gaussian functions around this region (see the left panel of Figure \ref{fig:spec_nir_velo}). 
For the origin of feature A, we consider six possibilities: \ion{C}{1}~$\lambda$1.0693~$\mu$m, \ion{Mg}{1}~$\lambda$1.0927~$\mu$m, high velocity (HV) \ion{He}{1}~$\lambda$1.083~$\mu$m, \ion{Fe}{2}~$\lambda$1.05~$\mu$m, \ion{S}{1}~$\lambda$1.0457~$\mu$m, and \ion{Sr}{2}~$\lambda$1.0327~$\mu$m. The velocity of feature A is indicated in Figure \ref{fig:spec_nir_velo} as well as in Figure \ref{fig:line_velo_evol} assuming this line is from the \ion{C}{1}, \ion{Mg}{1}, \ion{Fe}{2}, \ion{S}{1}, and \ion{Sr}{2} respectively. 

The derived velocities of \ion{C}{1}~$\lambda$1.0693~$\mu$m and \ion{Mg}{1}~$\lambda$1.0927~$\mu$m are higher than the velocity of H$\alpha$, so these two lines can be ruled out unless there is C and Mg present at a higher velocity than the hydrogen envelope. In addition, since there is no clear evidence of any blue component around the \ion{He}{1}~$\lambda$20581 (see the right panel of Figure \ref{fig:spec_nir_velo}), it is unlikely this extra absorption line is from HV \ion{He}{1}~$\lambda$1.083~$\mu$m. 

\ion{S}{1}~$\lambda$1.0457~$\mu$m, \ion{Fe}{2}~$\lambda$1.05~$\mu$m, and \ion{Sr}{2}~$\lambda$1.0327~$\mu$m give reasonable velocities. However, \ion{S}{1}~$\lambda$1.0457~$\mu$m is usually seen in He-poor SNe \citep{Teffs2020,Shahbandeh2022}, so it is unclear if such a strong \ion{S}{1}~$\lambda$1.0457~$\mu$m can be present in a He-strong SN. 
We could not identify other lines from \ion{S}{1} in the NIR spectra, probably because other \ion{S}{1} lines are rather weak.
The \ion{Fe}{2}~$\lambda$1.05~$\mu$m line is commonly seen in SNe Ia but not in CCSNe. If feature A is from \ion{Fe}{2}~$\lambda$1.05~$\mu$m, the derived velocity is lower than the velocity of $\rm H{\alpha}$ but larger than those of He and Fe lines in the optical, indicating that this line is formed in the outer layers of the ejecta. For SESNe, some degree of mixing is required to explain the observed properties \citep{Shigeyama1990,Woosley1997,Dessart2012,Yoon2019}, so it is possible that the \ion{Fe}{2}~$\lambda$1.05~$\mu$m line is formed by material mixed into the outer layer. In addition, the ejecta of SESNe can be aspherical \citep[e.g.,][]{Taubenberger2009,Milisavljevic2010,Fang2022ApJ...928..151F}. In this case, \ion{Fe}{2}~$\lambda$1.05~$\mu$m lines could be formed at higher velocities. However, no other \ion{Fe}{2} lines are identified in the NIR spectra, making this possibility less realistic. 
If feature A is from \ion{Sr}{2}~$\lambda$1.0327~$\mu$m, its velocity could be similar to the velocity of Fe lines in optical. \ion{Sr}{2}~$\lambda$1.0327~$\mu$m is common in Type II SNe \citep{Davis2019ApJ...887....4D}, so it is possible this line can be seen in a SESN given that they are all core collapses from massive stars. A possible absorption line from \ion{Sr}{2}~$\lambda$1.0036~$\mu$m at the same velocity ($\sim$8000km\,s$^{-1}$) is likely present in the NIR spectra, which is marked using a grey dashed line in Figure \ref{fig:spec_nir_velo}. This makes \ion{Sr}{2}~$\lambda$1.0327~$\mu$m the most plausible explanation of feature A.

In conclusion, feature A observed in SN~2022crv is likely due to \ion{Sr}{2}~$\lambda$1.0327~$\mu$m, but we could not fully exclude 
\ion{Fe}{2}~$\lambda$1.05~$\mu$m and \ion{S}{1}~$\lambda$1.0457~$\mu$m. Further detailed hydrodynamic modeling is needed to investigate the origin of feature A.


\begin{figure*}
\includegraphics[width=1.\linewidth]{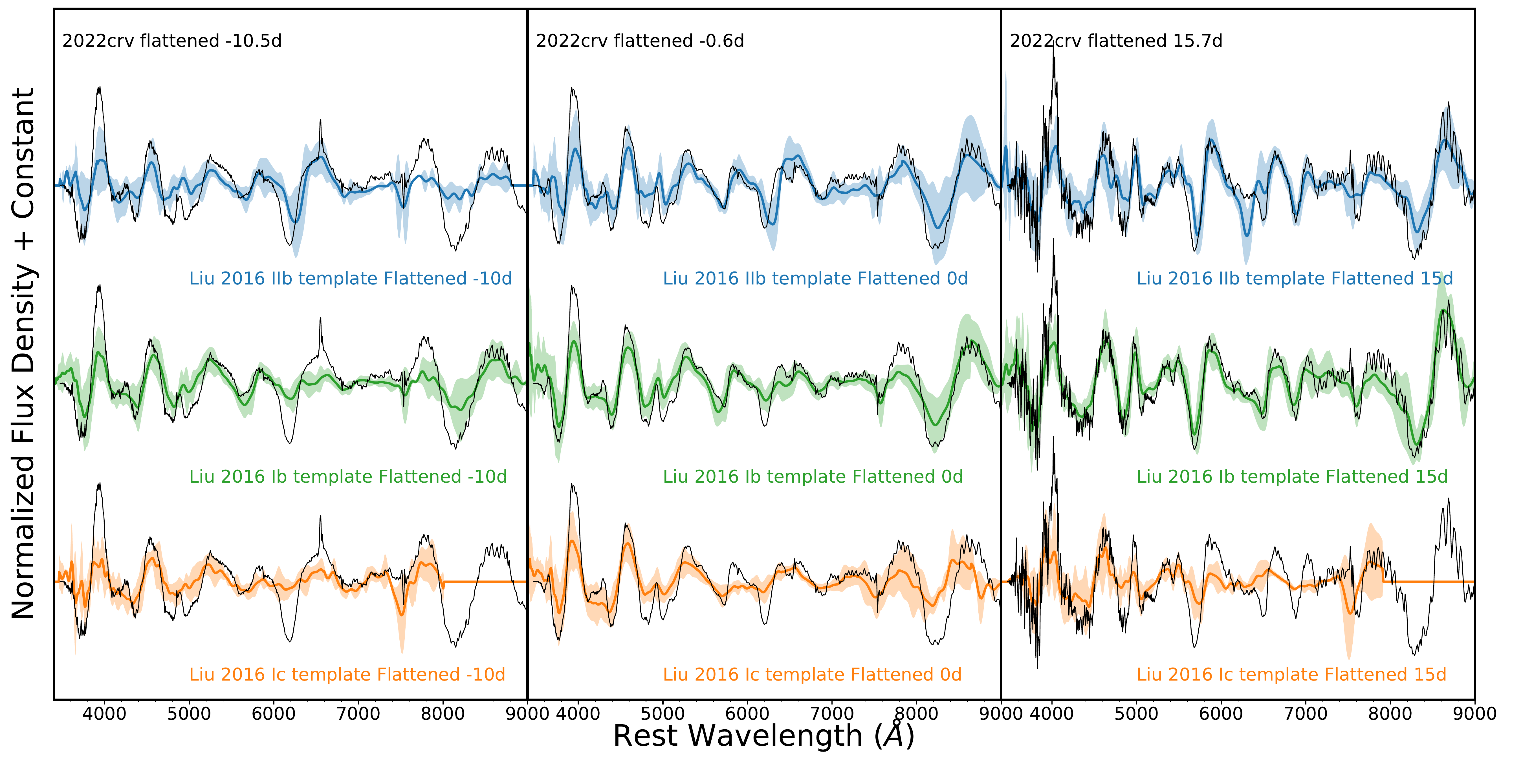}
\caption{Observed spectra of SN~2022crv compared to the mean spectra (the solid lines) and the standard deviations (the shaded regions) of SN IIb, Ib and Ic from \cite{Liu2016}. The observed spectra have been flattened using {\sc SNID}. 
\label{fig:spec_temp_comp}}
\end{figure*}

\begin{figure}
\includegraphics[width=1.\linewidth]{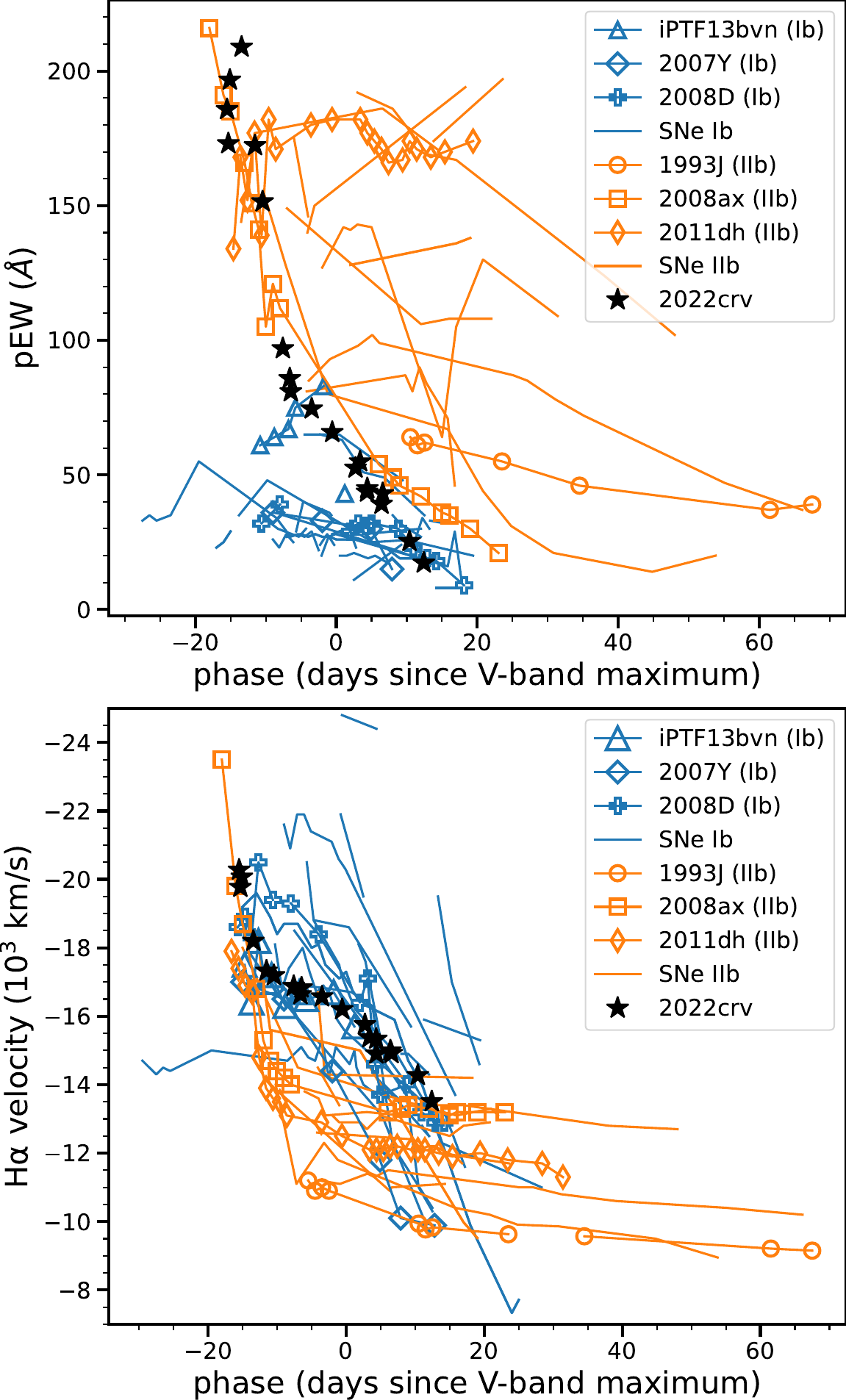}
\caption{\textit{Upper:} the pEW evolution of the 6200\AA\ feature of SN~2022crv compared to other SESNe. The 6200 \AA\ feature in SN~2022crv shows a large pEW at the beginning, similar to other SNe IIb. Then it quickly decreases until the 6200 \AA\ feature disappears at around 15 days after $V_{\rm max}$. \textit{Bottom:} Velocity evolution of the 6200 \AA\ feature in SN~2022crv assuming it is from $\rm H{\alpha}$. The velocity of the 6200 \AA\ feature in SN~2022crv is generally larger than those in SNe IIb but similar to those in SNe Ib.
\label{fig:velo_pew_com}}
\end{figure}

\begin{figure}
    \includegraphics[width=1.\linewidth]{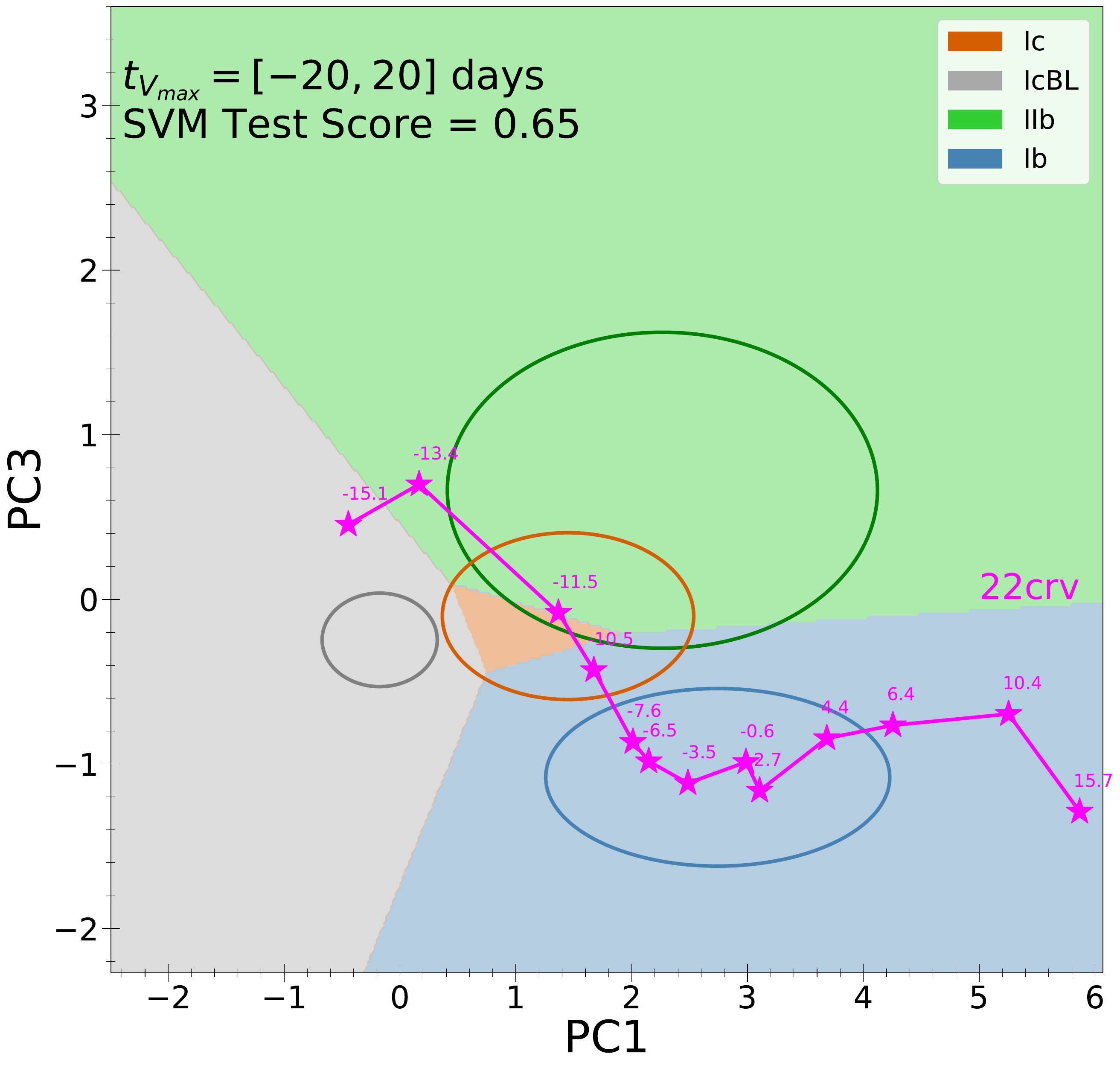}
    \caption{A SVM time dependent spectral classification of SN 2022crv (Williamson \& Modjaz 2023, in prep) based on the methods presented in \citet{Williamson2019}. 
    SN 2022crv is classified as a Type IIb SN before around -10 days from V-band maximum.  After that, SN 2022crv is more similar to a Type Ib. The trajectory in principal component analysis (PCA) phase space as 2022crv evolves (magenta) clearly shows the type transition occurring in the spectra. 
    }
    \label{fig:pca_classification}
\end{figure}

\begin{figure*}
\includegraphics[width=1.\linewidth]{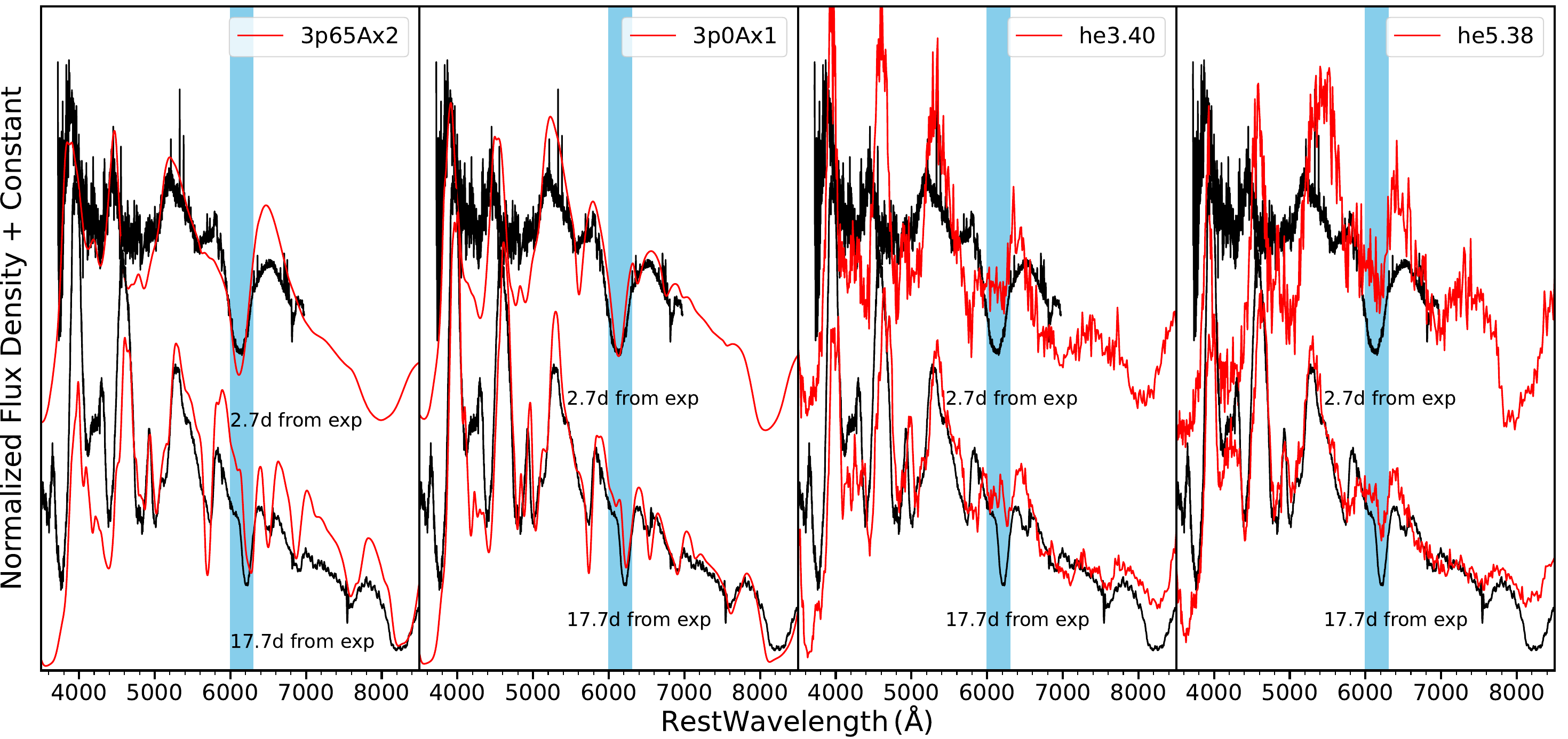}
\caption{Comparison of the optical spectra of SN~2022crv 2.7d and 17.7d from the explosion with models from \cite{Dessart2016} and \cite{Woosley2021} at similar epochs. The 6200 \AA\ feature is marked by the blue shading. The He3.40 and the He5.38 models are hydrogen-free models, while the 3p65Ax2 and the 3p0Ax1 models retain a small amount of hydrogen.
\label{fig:spec_model_comp}}
\end{figure*}

\section{Discussion}
\label{sec:discussion}

\subsection{Classification}\label{sec:classification}

Although there are clear definitions for each subtype of SESNe, actual classification for individual objects can be nontrivial. This is because the classification for some objects can be time dependent \cite[e.g.,][]{Milisavljevic2013,Folatelli2014ApJ...792....7F,Williamson2019,Holmbo2023A&A...675A..83H}. In addition, there is likely a continuum between the subtypes (although see \citealt{Holmbo2023A&A...675A..83H}). For instance, it has been suggested there is likely a gradual transition from Type IIb to Type Ib depending on the amount of residual hydrogen in the envelope \citep[e.g.,][]{Prentice2017}.
Furthermore, hydrogen has been suspected to be present in many Type Ib SNe, suggesting these objects may still maintain low-mass hydrogen envelopes \citep[e.g.,][]{ Deng2000,Branch2002,Elmhamdi2006,Parrent2007,James2010ApJ...718..957J}.
This breaks the Type Ib definition of no hydrogen, and makes the line between Type Ib and Type IIb more vague.
In recent years, many efforts have been made to improve the classification system of SESNe \citep{Liu2016, Prentice2017, Williamson2019, Holmbo2023A&A...675A..83H}. 
These studies assume there are hydrogen features in both Type Ib and Type IIb, and the difference between them is the evolution and the strength of the $\rm H\alpha$ line.
In this section, we will discuss the classification of SN~2022crv and show that this object seems to be an outlier of some classification schemes mentioned above.

SN~2022crv was initially classified as a Type Ib SN \citep{Andrews2022}. However, as we discussed in section~\ref{sec:spec_evol}, a strong absorption line is visible around 6200~\AA\  during early phases and could be related to $\rm H\alpha$, making the classification of SN~2022crv uncertain. 
Shortly after the maximum, this line disappears and the spectral evolution of SN~2022crv is almost identical to other SNe Ib. 
A small notch at around 4595 \AA\ is observed at early phases in SN~2022crv (see Figure \ref{fig:spec_evol_1}) and could be due to $\rm H\beta$.
To get a better sense of where SN~2022crv stands among SESNe, we compare the observed spectra with the mean spectral templates of SNe IIb, Ib and Ic from \cite{Liu2016} in Figure \ref{fig:spec_temp_comp}. 
The observed spectra shown here have been flattened using {\sc SNID} following the procedure outlined in \cite{Blondin2007}. At about day $-$10 and day 0, the 6200 \AA\ absorption lines in SN~2022crv are as strong as those in the mean spectra of SNe IIb.
However, the 6200 \AA\ absorption line in SN~2022crv is at a higher velocity than those in the mean spectra of SNe IIb, and at a similar velocity to those in the mean spectra of SNe Ib. At about day 15, the spectrum of SN~2022crv is almost identical to the mean spectra of SNe Ib, while the mean spectra of SNe IIb still show a strong 6200 \AA\ absorption line.

If the 6200 \AA\ line is indeed attributed to $\rm H\alpha$, SN~2022crv should be classified as a Type IIb. However, the high-velocity and the rapid disappearance of $\rm H\alpha$ complicates the classification.
\cite{Liu2016} found that the $\rm H\alpha$ velocities in SNe Ib are systematically larger than in SNe IIb, and the pEW values in SNe Ib are smaller than in SNe IIb, consistent with the consensus that SNe IIb have more hydrogen in their envelopes. They thus proposed that the pEW can be used to differentiate between SNe Ib and SNe IIb at all epochs.
That being said, if an object is classified as a Type Ib/Type IIb using the $\rm H\alpha$ pEW, the classification will not change over time since Type IIb always maintain a larger $\rm H\alpha$ pEW.
This result has been further supported by \cite{Holmbo2023A&A...675A..83H} based on the spectroscopic analysis of a large sample of SESNe.
In Figure \ref{fig:velo_pew_com}, we compare the velocity and pEW evolution of the 6200 \AA\ absorption line in SN~2022crv assuming it is from $\rm H\alpha$ with those of objects presented in \cite{Liu2016}. SN~2022crv shows a high pEW at early phases similar to other SNe IIb, followed by a rapid decline, and has a similar pEW value to other SNe Ib about 10 days after the $V$-band maximum.
The velocity of SN~2022crv is generally higher than SNe IIb and its evolution is more similar to those of SNe Ib. This pEW transition from SNe IIb to SNe Ib seems to be an outlier of the classification scheme using the pEW of $\rm H{\alpha}$ proposed by \cite{Liu2016}.

To quantitatively illustrate the complications involved in this classification, we applied the principal component and SVM classification method described in \cite{Williamson2019}. The code has been updated to allow a time-dependent classification of SESNe (Williamson \& Modjaz 2023, in prep). The result is shown in Figure \ref{fig:pca_classification}. Initially, SN~2022crv is located in the Type Ic-BL (broad line) region. This is likely due to the spectrum at day $-$15.1 which is dominated by the strong high-velocity $\rm H{\alpha}$, which can be misidentified as the broad \ion{Si}{2}~$\lambda$6355 line in SNe Ic-BL. At roughly day $-$13 to day $-$10, SN~2022crv is more similar to SNe IIb. After day $-$10, it is more consistent with SNe Ib. 

\cite{Prentice2017} proposed to use the strength and ratio between absorption and emission of $\rm H{\alpha}$ to classify He-rich SNe.
They suggested that He-rich SESNe can be split into four groups, IIb, IIb(I), Ib(II) and Ib, from Hydrogen rich to Hydrogen poor. 
Following the method described in \cite{Prentice2017}, the ratio between absorption and emission of $\rm H{\alpha}$ in SN~2022crv before peak is measured to be $\sim$0.3, and the average EW before the maximum is about 73~\AA. These values make SN~2022crv marginally fall into the IIb(I) category. The ratio between absorption and emission of $\rm H{\alpha}$ is a good probe of the extent of the hydrogen envelope, and a larger ratio indicates a more extended hydrogen envelope \citep{Prentice2017}. 
A classification of IIb(I) implies that the amount of hydrogen in SN~2022crv is larger than normal SNe Ib but smaller than SNe IIb, suggesting SN~2022crv is an intermediate object between these two classes.

In summary, SN~2022crv shares similarities with both Type IIb and Type Ib. The $\rm H{\alpha}$ velocity evolution of SN~2022crv is consistent with those of SNe Ib. The pEW evolution of SN~2022crv gradually transitions from SNe IIb to SNe Ib, a behaviour not observed in the sample studied by \cite{Liu2016}. The amount of hydrogen in SN~2022crv is likely between those in SNe Ib and SNe IIb, suggesting SN~2022crv is a representative transitional object on the continuum between SNe IIb and SNe Ib.

\subsection{Hydrogen Envelope} \label{sec:hydro_envelope}
Quantifying how much hydrogen is still present in the progenitor right before the explosion is important for understanding the mass loss history and thus the progenitor system.
In this section, we will use both a semi-analytic method and hydrodynamic modeling to constrain the mass of hydrogen retained in the progenitor right before the explosion.

The velocity of $\rm H{\alpha}$ in SN~2022crv is generally higher than those of SNe IIb.
The higher velocity of $\rm H{\alpha}$ could be a result of higher explosion energy or smaller hydrogen mass. Assuming that the He I line appears when the surrounding hydrogen envelope becomes optically thin and SN expansion is homologous, \cite{Marion2014} proposed a method to roughly estimate the mass of the hydrogen envelope: $M_{\rm H} \propto (v_{\rm H} \times t_{\rm He})^{2}$, where $v_{\rm H}$ is the velocity of the outer edge of the hydrogen envelope and $t_{\rm He}$ is the time when the He line is first observed, and the constant of proportionality is empirically calculated by scaling to a reference supernova. For SN~2022crv, $v_{\rm H}$ is about 30,000 km\,s$^{-1}$. The He line in SN~2022crv can be clearly seen in the first spectrum, so we can limit $t_{\rm He} <$  2.7d. Using SN~1993J and SN~2011dh as references, we find $\rm M_{H}(2022crv) < 0.14\times M_{H}(2011dh)$ and $\rm M_{H}(2022crv) < 0.05\times M_{H}(1993J)$. For SN~2011dh and SN~1993J, a recent study by \cite{Gilkis2022} estimated hydrogen masses of $\sim$0.035 \msun{} and $\sim$0.1 \msun{}, respectively. Therefore, the hydrogen envelope mass in SN~2022crv can be constrained to be $\rm  M_{H}(2022crv) < 5\times 10^{-3}$ \msun{}. This small hydrogen mass is consistent with the higher velocity and the quick disappearance of the H$\alpha$ in SN~2022crv. However, this seems contrary to the large pEW at early phase.

In order to investigate if a small hydrogen envelope can produce the evolution of the H$\alpha$ line in SN~2022crv, we compare the observed spectra of SN~2022crv with the SESNe models in \cite{Dessart2015,Dessart2016} (3p65Ax2 and 3p0Ax1) and \cite{Woosley2021} (He3.40 and He5.38) (see Figure \ref{fig:spec_model_comp}). The He3.40 and the He5.38 model spectra are produced based on models presented in \cite{Woosley2019} and \cite{Ertl2020}. These models do not include any hydrogen. The 3p65Ax2 and the 3p0Ax1 model spectra are based on models presented in \cite{Yoon2010}. The hydrogen masses in the 3p65Ax2 and 3p0Ax1 models are $4.72\times10^{-3}$ \msun{} and $7.92\times10^{-4}$ \msun{}, respectively.

Both the 3p65Ax2 and 3p0Ax1 models nicely reproduce the early-time broad absorption line at 6200 \AA\ and other main features in the observed spectra (Figure \ref{fig:spec_model_comp}), while the 3p0Ax1 model gives a better fit.
The 6200 \AA\ feature in these two models has been attributed to $\rm H{\alpha}$ at early phase and Si II before the maximum light \citep{Dessart2015}. 
This is also likely what happened in SN~2022crv. At early time, the broad absorption line at 6200 \AA\ is mainly from hydrogen. As the object evolves, the 6200 \AA\ feature becomes narrower and more dominated by \ion{Si}{2}.
The 6200 \AA\ feature is missing in the hydrogen-free He3.40 and He5.38 models at 2.7d. At 17.7 d after the explosion, these models show weak absorption lines at around 6200 \AA\ which are identified as Si II in \cite{Woosley2021}. We note that the models we use here are not specifically made for SN~2022crv, so the hydrogen mass we get here can be only treated as an order of magnitude estimation. For example, the models from \cite{Dessart2016} have peak luminosities fainter than SN~2022crv, this could be due to a lower explosion energy or lower nickel mass synthesized in the model. Detailed modeling would be required for future studies to better constrain the mass of hydrogen left in the progenitor envelope of SN~2022crv. 

The early light curve of SN~2022crv does not show signs of cooling envelope emission, and the nebular spectra also lack hydrogen emission, consistent with the compact progenitor scenario proposed by \cite{Chevalier2010ApJ...711L..40C}.
Therefore, SN~2022crv likely has a compact progenitor with extremely low amount of hydrogen.

In conclusion, the hydrogen envelope in SN~2022crv is likely on the order of $10^{-3}$ \msun{}. From the model comparison, we found that the 6200\AA\ line is likely a mixture of $\rm H{\alpha}$ and Si II. At early phases (within 15 days after the explosion), the 6200\AA\ line is likely dominated by hydrogen. After that, the 6200\AA\ is likely mainly due to Si II.

\subsection{Explosion Properties}

\begin{figure}
\includegraphics[width=1.\linewidth]{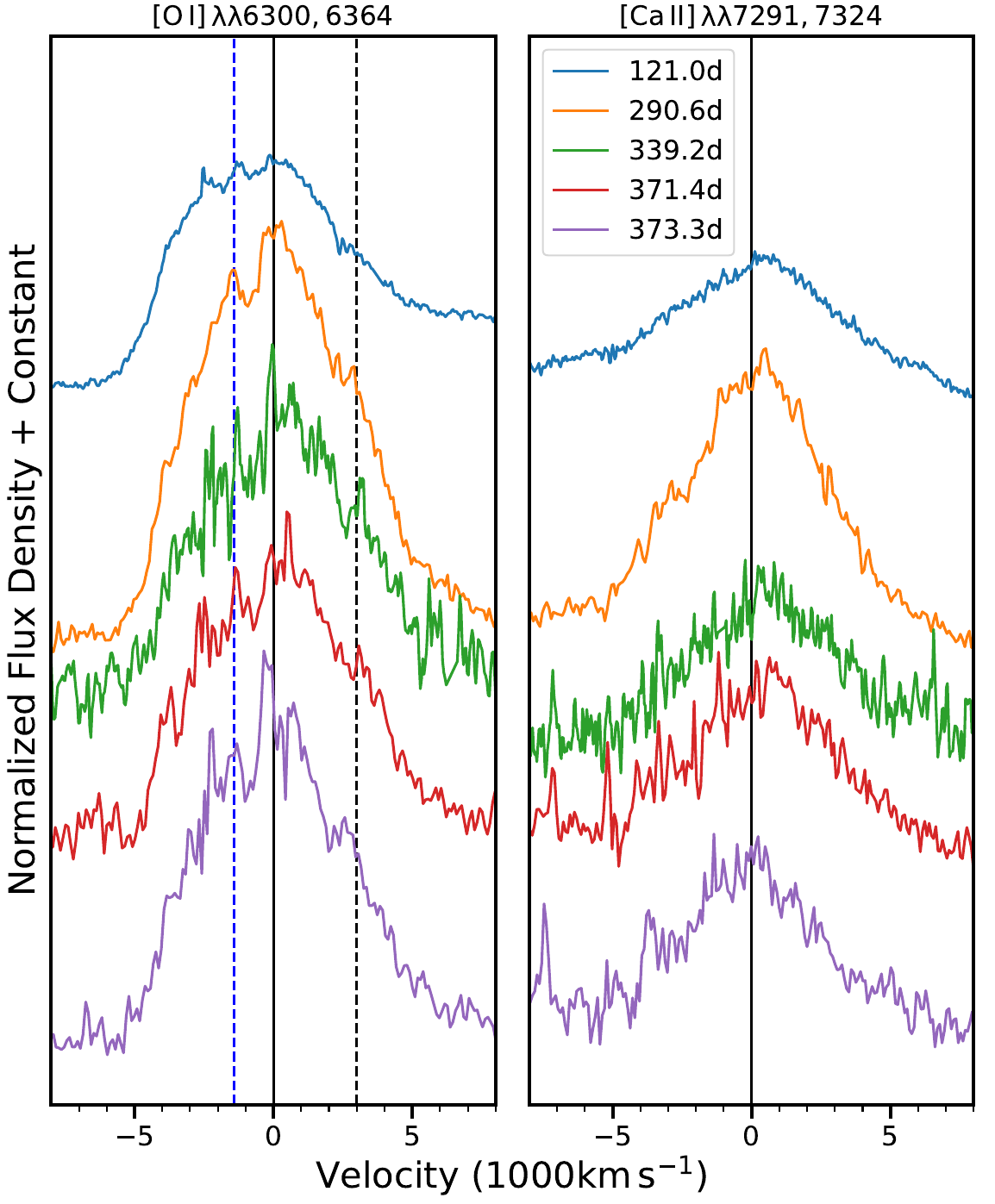}
\caption{Late-time profile of [\ion{O}{1}]~$\lambda\lambda$6300, 6364 and [\ion{Ca}{2}]~$\lambda\lambda$7291, 7323. The phase is measured from the explosion. The zero points for [\ion{O}{1}] and [\ion{Ca}{2}] are 6300~\AA\ and 7307.5~\AA\, respectively, and are marked by the solid vertical lines in each panel. The black vertical dashed line in the left panel marks 6364~\AA. The blue vertical dashed line marks the blue-shifted peak at around 6270~\AA.
\label{fig:OI_profile}}
\end{figure}

\subsubsection{Explosion Geometry}
At late times, when the SNe ejecta become optically thin, spectra can provide useful information about the inner structure of the SNe. For SESNe, the line profile of [\ion{O}{1}]~$\lambda\lambda$6300,6364 is often used to study the geometry of the explosion \citep{Mazzali2005,Maeda2008,Modjaz2008,Taubenberger2009,Milisavljevic2010,Fang2022ApJ...928..151F}.
The [\ion{O}{1}] line in SN~2022crv shows an asymmetric profile (see Figure \ref{fig:OI_profile}), with a prominent peak at around 6300 \AA\ and a blue-shifted peak at around 6270 \AA\ ($-$1430 $\rm km\,s^{-1}$). The separation of the two peaks is 30 \AA\, smaller than the separation of [\ion{O}{1}]~$\lambda\lambda$6300,6364 doublet.
An asymmetric profile could originate from the oxygen-rich material with a torus-like structure \citep{Mazzali2005, Maeda2008, Valenti2008}. However, in this model, the [\ion{O}{1}] line would display either a double-peaked profile if viewed from a direction close to the plane of the torus or a single-peaked profile if viewed from a direction perpendicular to the torus. Such a scenario is hard to explain in the case of SN~2022crv due to the lack of a redshifted emission line. If this asymmetric profile is indeed from a torus-like structure, then the emission from the rear side is likely scattered by the ejecta or absorbed by dust \citep{Milisavljevic2010}. For SN~2022crv, a clear sign of CO formation is detected about 100 days after the explosion (Rho et al., in prep), implying dust could form at sufficiently late phases.

\cite{Maurer2010} suggested that the double-peaked profile observed in the [\ion{O}{1}] line can be caused by the high-velocity $\rm H{\alpha}$ absorption. Given that a high-velocity feature is detected at early phases for SN~2022crv, this scenario can not be ruled out. If the trough at around 6285 \AA\ is due to $\rm H{\alpha}$ absorption, the corresponding velocity would be $\sim$12,700~$\rm km\,s^{-1}$. There is no clear evidence of $\rm H{\beta}$ found in the nebular spectra of SN~2022crv, likely due to the low signal-to-noise ratio of the blue portion of the spectra. If $\rm H{\alpha}$ is responsible for the asymmetric profile of the [\ion{O}{1}] line, the ejecta of SN~2022crv would be almost spherically symmetric. 

\begin{deluxetable}{cccccc}
\tablecaption{A summary of the nebular spectral models which were compared to SN~2022crv. \label{tab:nebular_model}}
\tablewidth{0pt}
\tablehead{
\colhead{Model} \vspace{-0.2cm} &
 \colhead{$M_{\rm ZAMS}$} &
\colhead{$M_{\rm preSN}$} &
\colhead{$M_{\rm He}$} &
\colhead{$M_{\rm O}$} &
\colhead{Reference\tablenotemark{$\star$}}
 \\
\colhead{} &
\colhead{(\msun{})} &
\colhead{(\msun{})} &
\colhead{(\msun{})} &
\colhead{(\msun{})} &
\colhead{}
}
\startdata 
Dessart He5.0&20.8&3.81&5.0&0.592&a\\
Dessart He6.0&23.3&4.44&6.0&0.974&a\\
Dessart He8.0&27.9&5.63&8.0&1.71&a\\
Jerkstrand m12C&12&3.12&&0.3&b\\
Jerkstrand m13G&13&3.52&&0.52&b\\
Jerkstrand m17A&17&5.02&&1.3&b\\
\enddata
\tablenotetext{\star}{(a) \cite{Woosley2019,Ertl2020,Dessart2021}; (b) \cite{Woosley2007,Jerkstrand2015}.}
\end{deluxetable}

\subsubsection{Oxygen Mass And Progenitor Mass} \label{sec:O_mass_progenitor}
Theoretical studies have shown that the initial progenitor mass of a SN strongly correlates with the oxygen mass in the SN ejecta \citep{Woosley1995ApJS..101..181W,Jerkstrand2015,Dessart2021,Dessart2023arXiv230612092D}. 
The flux of the [\ion{O}{1}]~$\lambda\lambda$6300, 6364 emission line during the nebular phase has been demonstrated to be a powerful diagnostic tool to constrain the oxygen mass \citep{Uomoto1986, Jerkstrand2012, Dessart2021}. In addition, the line ratio of [\ion{O}{1}]~$\lambda\lambda$6300, 6364/[\ion{Ca}{2}]~$\lambda\lambda$7291, 7323 can be an indicator of the progenitor mass of an SESN, since synthesized Ca is not sensitive to the main sequence mass of the progenitor \citep{Nomoto2006}.

\cite{Uomoto1986} showed that the minimal oxygen mass in the SN ejecta can be calculated with:
\begin{equation}
    \frac{M_{oxygen}}{\rm M_{\odot}} = 10^{8}\times (D/{\rm Mpc})^{2}\times F([OI])\times \exp(2280{\rm K}/T),
\end{equation}
where D is the distance of the SN in Mpc, F([\ion{O}{1}]) is the flux of the [\ion{O}{1}] line in $\rm erg\,s^{-1}\,cm^{-2}$, and $T$ is the temperature of the oxygen-emitting gas in $\rm K$. $T$ can be estimated by using the [\ion{O}{1}]~$\lambda$5577/[\ion{O}{1}]~$\lambda\lambda$6300, 6364 flux ratio. However, the [\ion{O}{1}]~$\lambda$5577 line in SN~2022crv is not clearly detected, indicating a low temperature. The [\ion{O}{1}]~$\lambda$5577 was also not detected for SN~1990I, and \cite{Elmhamdi2004} put a lower limit on the [\ion{O}{1}]~$\lambda$5577/[\ion{O}{1}]~$\lambda\lambda$6300, 6364 flux ratio by using a temperature of 3200 -- 3500~K for the line-emitting region of SN~1990I. \cite{Sahu2011} used a temperature of 4000~K for SN~2009jf since the [\ion{O}{1}]~$\lambda$5577 line was not visible in the nebular spectra of SN~2009jf. For SN~2022crv, we assume the temperature of the line-emitting region is $\sim$3200~K -- 4000~K, which results in an oxygen mass of $\sim$0.9 -- 3.5~\msun{}.

In Section \ref{sec:nebular_spec}, the [\ion{O}{1}]/[\ion{Ca}{2}] ratio for SN~2022crv is measured to be $\sim$1.5. \cite{Kuncarayakti2015} measured the [\ion{O}{1}]/[\ion{Ca}{2}] ratio for a group of CCSNe, and they found that there is a natural spread for SNe Ib/c. 
This observed spread is likely an indication of two progenitor populations of SESNe: a single massive Wolf-Rayet star and a low-mass star in a binary system.
A [\ion{O}{1}]/[\ion{Ca}{2}] ratio of 1.5 indicates that the progenitor of SN~2022crv is more likely a less massive star.
At solar metallicity, the minimum zero-age main sequence (ZAMS) mass ($\rm M_{ZAMS}$) for a single star to lose its hydrogen envelope via stellar winds and become a Wolf-Rayet star is about 25 -- 35 \msun{} \citep{Crowther2007ARA&A..45..177C, Sukhbold2016ApJ...821...38S, Ertl2020}.
As we discussed in Section \ref{sec:host_properties}, the metallicity at the location of SN~2022crv is only slightly larger than solar metallicity.
This implies the progenitor was likely in a binary system with a $\rm M_{ZAMS}$ less than about 25 -- 35 \msun{}, and at least a part of the hydrogen envelope was stripped during the binary interaction.

\begin{figure*}
\includegraphics[width=1.\linewidth]{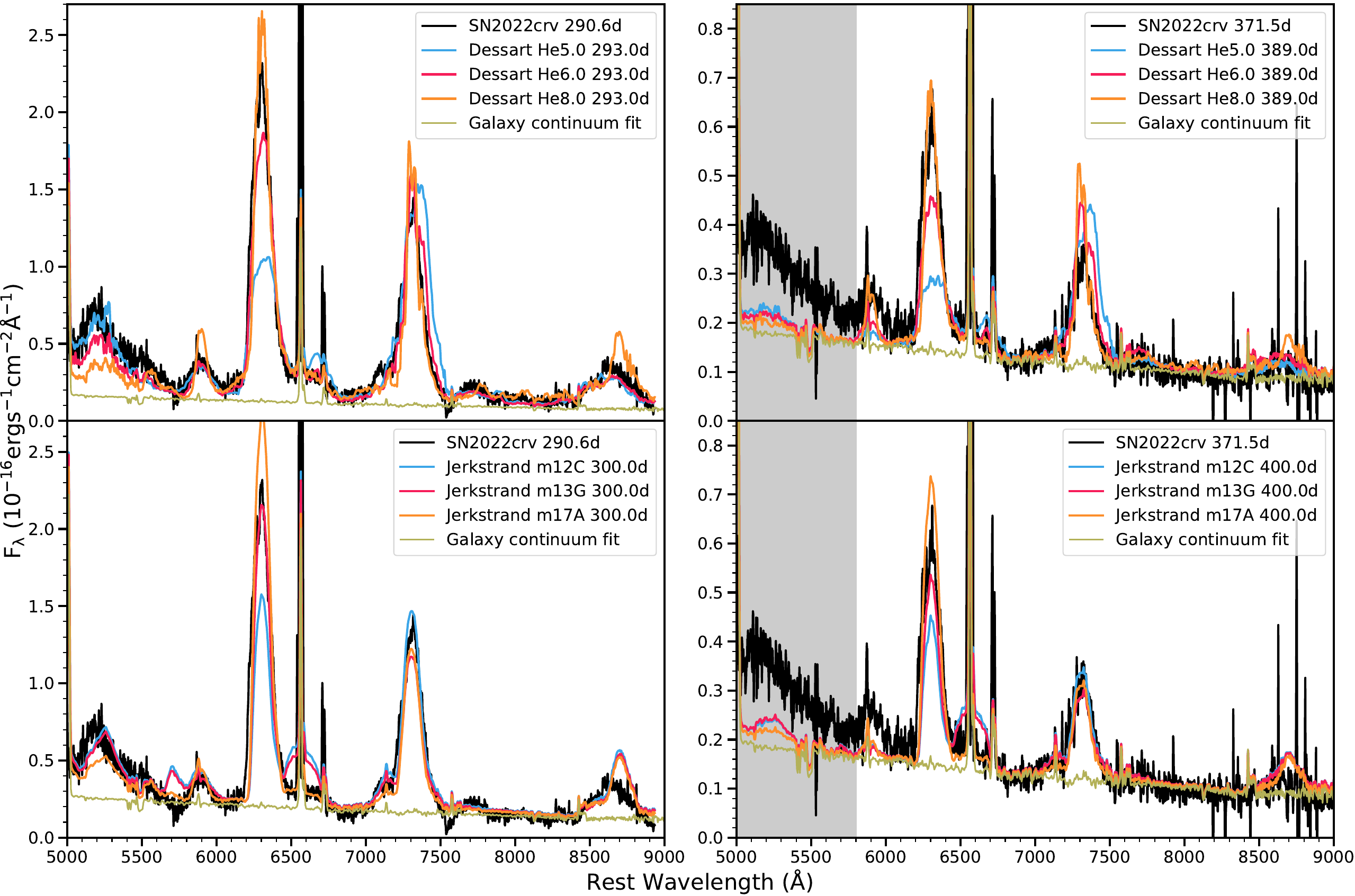}
\caption{Comparison of the nebular spectra of SN~2022crv at 290.6d and 371.5d post-explosion with models from \cite{Jerkstrand2015} and \cite{Dessart2021}. The spectrum is contaminated by the host galaxy, so we fit them with a combination of a star-forming galaxy spectrum and the model SN spectra. For the spectrum at day 371.5, there is a strong background contamination at less than $\sim$5800 \AA\ (marked with grey shading), which makes the observed spectrum clearly depart from the star-forming galaxy spectrum.
\label{fig:nebular_model_comp}}
\end{figure*}

\begin{figure}
\includegraphics[width=1.\linewidth]{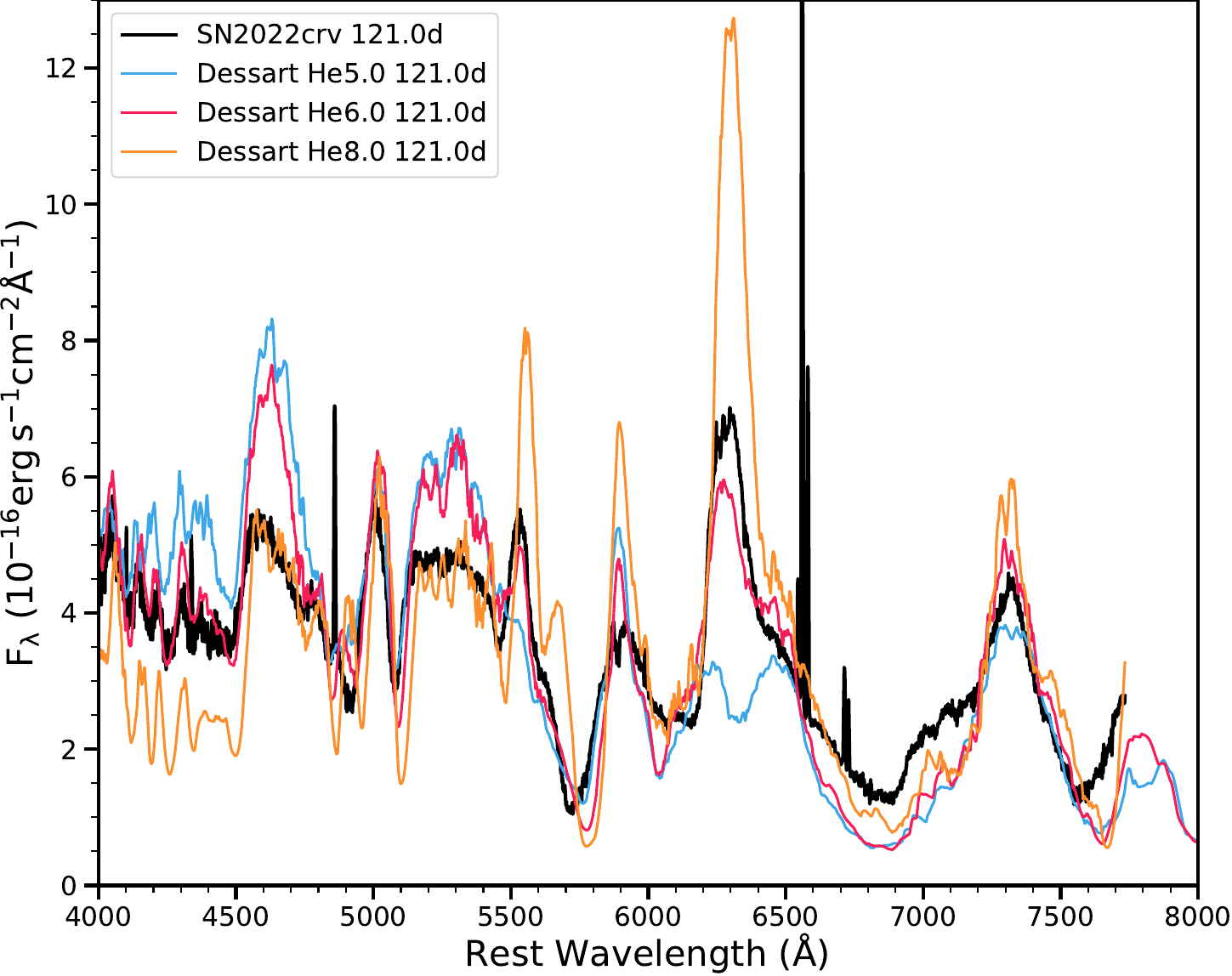}
\caption{Comparison of the spectrum of SN~2022crv at 121d after the explosion with models from \cite{Dessart2023arXiv230612092D}.
\label{fig:nebular_model_comp_121d}}
\end{figure}

To further constrain the progenitor properties of SN~2022crv, we compare the observed spectrum with theoretical models from \cite{Jerkstrand2015}, \cite{Dessart2021}, and \cite{Dessart2023arXiv230612092D}. \cite{Jerkstrand2015} took the single-star Type II models from \cite{Woosley2007} and artificially removed most of the hydrogen envelope, and then produced spectra for SNe IIb. \cite{Dessart2021, Dessart2023arXiv230612092D} modelled nebular spectra of SNe Ib/c based on models from \cite{Woosley2019} and \cite{Ertl2020}, which also include the effects of wind loss from the He star after the H envelope is fully stripped via the binary interaction. 
\cite{Dessart2023arXiv230612092D} found that the spectral features of SESNe are mainly dependent on their presupernova masses ($\rm M_{preSN}$) and the oxygen yields,  
so these models can provide constraints on the oxygen mass and $\rm M_{preSN}$ of SN~2022crv.


The nebular spectra of SN~2022crv at day 290.6 and day 371.5 are likely contaminated by the host background. 
Therefore, we fit the observed spectrum with a linear combination of the SN spectrum and a star-forming galaxy template spectrum from \cite{Kinney1996ApJ...467...38K} as the first-order approximation of the host background. As the SN gets fainter, this contamination becomes more dominant and harder to remove from the SN spectrum. At day 371.5, the object is highly contaminated by the host background below $\sim$5600 \AA. However, we note that this contamination does not affect our estimation on the progenitor properties since we are only interested in the region redward of $\sim$6000~\AA. 
The comparisons are shown in Figure \ref{fig:nebular_model_comp} and \ref{fig:nebular_model_comp_121d}, while the physical properties of these models are summarized in Table \ref{tab:nebular_model}. 

For the models from \cite{Dessart2021, Dessart2023arXiv230612092D}, we found that the [\ion{O}{1}] intensity of SN~2022crv is between those of the He6.0 and He8.0 models. Other features of the observed spectrum, such as \ion{Na}{1}~$\lambda\lambda$5896,5890, [\ion{N}{2}]~$\lambda\lambda$~6548,6583 and [\ion{Ca}{2}]~$\lambda\lambda$7291,7323, can also be reproduced by these two models. This indicates that the mass of oxygen in SN~2022crv is about 1.0 -- 1.7 \msun{}, and the $\rm M_{preSN}$ is about 4.5 -- 5.6 \msun{}. The corresponding ZAMS star mass of the He6.0 and He8.0 models are 23.3\msun{} and 27.9\msun{}, respectively. 
For the models from \cite{Jerkstrand2015}, we found that the best-fit models are m13G and m17A, while the [\ion{N}{2}]~$\lambda\lambda$6548,6583 is stronger in model m13G than in the observed spectrum. This suggests that the oxygen mass produced in SN~2022crv is about 0.5 -- 1.3\msun{}, and the $\rm M_{preSN}$ is about 3.5 -- 5 \msun{}. The corresponding ZAMS star mass of the m13G and the m17A models are 13\msun{} and 17\msun{}, respectively.


The oxygen mass and the $\rm M_{preSN}$ derived from the models of \cite{Dessart2021, Dessart2023arXiv230612092D} are consistent with the oxygen mass derived from the models of \cite{Jerkstrand2015}, and they are also consistent with the oxygen mass we estimated using the flux of [\ion{O}{1}] emission. 
Since the models in \cite{Woosley2019} used by \cite{Dessart2021, Dessart2023arXiv230612092D} employed a more updated progenitor evolution treatment than the models in \cite{Woosley2007} used by \cite{Jerkstrand2015}, we will adopt an oxygen mass of 1.0 -- 1.7 \msun\ and a $\rm M_{preSN}$ of 4.5 -- 5.6 \msun\ derived from models of \cite{Dessart2021, Dessart2023arXiv230612092D}.

The models from \cite{Dessart2021, Dessart2023arXiv230612092D} give a systematically larger $\rm M_{ZAMS}$ than the models from \cite{Jerkstrand2015}. 
This is mainly due to the different mass loss assumptions of these two model sets.
The models in \cite{Jerkstrand2015} are Type II models with most of the hydrogen envelope artificially removed. This is equivalent to the progenitor promptly losing most of the hydrogen envelope right before the explosion.
Thus, the He core mass would continue to grow until the core collapse due to the hydrogen shell burning. 
While for the models in \cite{Dessart2021, Dessart2023arXiv230612092D}, the hydrogen envelope of the progenitor star is fully stripped by its binary companion close to the time of central He core ignition when the star first becomes a supergiant.
After that, the He star still experiences mass loss due to the winds until core collapse.

\cite{Dessart2021, Dessart2023arXiv230612092D} found that the oxygen yields of CCSNe is closely related to the $\rm M_{preSN}$ of the He stars or the final He core mass of the single stars.
Therefore, in order to synthesize the same amount of oxygen or produce progenitors with the same $\rm M_{preSN}$, the models from \cite{Jerkstrand2015} will always have a smaller $\rm M_{ZAMS}$ than the models from \cite{Dessart2021, Dessart2023arXiv230612092D} (see also Figure 3 in \citealt{Dessart2021}). 

The progenitor of SN~2022crv still retained a tiny amount of hydrogen right before the explosion.
If the progenitor of SN~2022crv was a single star, it likely experienced strong stellar wind mass loss throughout its life.
If the progenitor of SN~2022crv was in a binary system, it likely lost a part of its hydrogen envelope when it first became a supergiant \citep{Yoon2015, Woosley2019, Ertl2020}. After that, the progenitor experienced continuous mass loss via stellar winds until explosion. 
In either case, since the hydrogen envelope was never fully stripped, during the mass loss, the hydrogen envelope mass would decrease, but the He core mass would grow continuously due to H shell burning. 
In order to produce the same amount of oxygen in the core or the same $\rm M_{preSN}$, the progenitor of SN~2022crv must have a $\rm M_{ZAMS}$ smaller than those of the He star models evolved with mass loss used by \cite{Dessart2021, Dessart2023arXiv230612092D} but larger than the trimmed Type II SNe models used by \cite{Jerkstrand2015}. 
Therefore, the $\rm M_{ZAMS}$ of the progenitor of SN~2022crv has to be less than about 23 -- 29 \msun\ and larger than about 13 -- 17 \msun.

\cite{Dessart2023arXiv230612092D} explored another set of models without mass loss after the stripping of the hydrogen envelope. These models are probably more suitable to the case of SN~2022crv since they also have intact He cores. However, since these models lose their hydrogen envelope promptly close to the time of He core ignition, they would still overestimate the $\rm M_{ZAMS}$ of the progenitor of SN~2022crv. The largest He star mass explored by this model set is 4.5 \msun{}, which has a $\rm M_{preSN}$ of 4.5 \msun{} and a $\rm M_{ZAMS}$ of 19.5 \msun{}. The spectral features produced by this model is similar to the He6.0 model evolved with mass loss \cite{Dessart2023arXiv230612092D}, which has a $\rm M_{preSN}$ of 4.44. 
The zero mass loss models with larger masses are not explored in \cite{Dessart2023arXiv230612092D}, but based on equation 4 in \cite{Woosley2019}, we can find that a zero mass loss model that has the same $\rm M_{preSN}$ with the He8.0 model would have a $\rm M_{ZAMS}$ of 22.4 \msun{}.
Therefore, we can further constrain the $\rm M_{ZAMS}$ of the progenitor of SN~2022crv to be less than $\sim$20 -- 22 \msun{}.

Another useful constraint can be placed by using the single star Type II progenitor models that evolved with their hydrogen envelopes. As suggested by \cite{Dessart2021}, the final oxygen yields from the single stars would be similar to those from the stripped He stars as long as the He core mass of the former is equal to the $\rm M_{preSN}$ of the latter. A single star has a much denser hydrogen envelope than the progenitor of SN~2022crv, which means its He core mass would grow faster than that of the progenitor of SN~2022crv. Thus, in order to produce the same He core mass, a single star would have a smaller $\rm M_{ZAMS}$ than the progenitor of SN~2022crv. A single star that finally has a He core mass of 4.44 \msun{} -- 5.63 \msun{} would have a $\rm M_{ZAMS}$ of 15.5 \msun{} -- 18.5 \msun{} \citep{Sukhbold2016ApJ...821...38S}. Therefore, we can further constrain the $\rm M_{ZAMS}$ of the progenitor of SN~2022crv to be larger than $\sim$16 \msun{} -- 19 \msun{}.


In Section \ref{sec:host_properties}, we found that the metallicity at the site of SN~2022crv is slightly larger than solar metallicity, implying that the progenitor experienced strong stellar wind mass loss. However, the progenitor of SN~2022crv ($\rm M_{ZAMS} \simeq$16\msun{} -- 22\msun{}) is still not massive enough to strip most of its hydrogen envelope by itself \citep{Crowther2007ARA&A..45..177C, Sukhbold2016ApJ...821...38S}, so the progenitor has to be in a binary system.

Binary interaction usually can not fully strip the hydrogen envelope from the progenitor star \citep{Yoon2017a, Gotberg2017A&A...608A..11G}. Whether the progenitor can shed all the rest of hydrogen depends on its wind mass-loss rate \citep{Yoon2017a, Gotberg2023arXiv230700074G}.
Since the progenitor of SN~2022crv likely had a high mass-loss rate, to prevent the hydrogen envelope from being fully stripped by wind and binary interaction, the orbital separation of the binary system needs to be large \citep{Yoon2017a, Gotberg2017A&A...608A..11G, Gotberg2023arXiv230700074G}. \cite{Yoon2017a} studied a grid of Type Ib/IIb models in binary systems, and they found that at metallicity Z = 0.02, assuming a mass ratio of $q$ = 0.9, an orbital period of above $2000$ days (or an initial orbital separation larger than $\sim$2200~$\rm R_{\odot}$) is required for a progenitor with $\rm M_{ZAMS}$ = 18 \msun\ in a binary system in order to have some remaining hydrogen envelope left. While for a progenitor with $\rm M_{ZAMS}$ = 16 \msun, the orbital period would be larger than $\sim1700$ days, which is equivalent to an initial orbital separation of $\sim$1900~$\rm R_{\odot}$. If the initial orbital separation is on the order of $\sim$100~$\rm R_{\odot}$, the progenitor would lose all the hydrogen envelope with such a high metallicity. Therefore, the progenitor of SN~2022crv was likely in a binary system with an initial orbital separation larger than $\sim$1000 $\rm R_{\odot}$.


In conclusion, based on the flux of the [\ion{O}{1}] emission line, we found an oxygen mass of 0.9 -- 3.6\msun{}. 
The [\ion{O}{1}]/[\ion{Ca}{2}] ratio implies the progenitor of SN~2022crv has a $\rm M_{ZAMS}$ less than about 25 -- 35 \msun, which implies the envelope of the progenitor was stripped in a binary system via interaction. 
By comparing with hydrodynamic models, we found that the oxygen synthesized in the progenitor is about 1.0 -- 1.7 \msun, and the progenitor is likely a He star with a final mass of $\sim$4.5 -- 5.6 \msun{} that has evolved from a 16 -- 22\msun{} ZAMS star. 
Given the high metallicity at the SN site and the very low-mass hydrogen envelope, the progenitor of SN~2022crv is likely in a binary system with a large initial orbital separation. 


\subsubsection{\texorpdfstring{$\rm ^{56}Ni$}\ \ Model Fit} \label{sec:arnett_model}
In order to estimate the physical parameters of the object, we applied a simple analytical model to the bolometric light curve of SN~2022crv obtained in section \ref{sec:bolo}. The analytical model was proposed for Type Ia SNe \citep{Arnett1982,Sutherland1984,Cappellaro1997}, and can also be used for SESNe \citep{Clocchiatti1997,Valenti2008,Valenti2011,Lyman2016}. For initially pure $\rm^{56}Ni$, the energy production rate of the nickel and cobalt decay can be expressed as \citep{Nadyozhin1994}:
\begin{equation} \label{eqn: Pni}
    P_{\text{Ni}}(t) = \frac{M_{\text{Ni,0}}}{56m_{u}} \frac{Q_{\text{Ni}}}~{\tau_{\text{Ni}}}~e^{-t/\tau_{\text{Ni}}} = M_{\text{Ni,0}}~\epsilon_{\text{Ni}}~e^{-t/\tau_{\text{Ni}}}
\end{equation}
\begin{align} \label{eqn: Pco}
    P_{\text{Co}}(t) &= \frac{M_{\text{Ni,0}}}{56m_{u}}~\frac{Q_{\text{Co}}}{\tau_{\text{Co}} - \tau_{\text{Ni}}}(e^{-t/\tau_{\text{Co}}} - e^{-t/\tau_{\text{Ni}}}) \nonumber \\
    &= M_{\text{Ni,0}}~\epsilon_{\text{Co}}~(e^{-t/\tau_{\text{Co}}} - e^{-t/\tau_{\text{Ni}}}),
\end{align}
where $Q_{\rm Ni} = 1.75$~MeV is the energy released per $^{56}$Ni decay via gamma-ray photons, $Q_{\rm Co} = 3.61$~MeV is the average energy released per $^{56}$Co decay via gamma-ray photons (assuming all the positrons produced in $^{56}$Co are annihilated), $\tau_{\rm Ni} = 8.8$ days is the lifetime of $^{56}$Ni, $\tau_{\rm Co} = 111.3$ days is the lifetime of $^{56}$Co, $\rm m_{u}$ = 1.66$\times10^{-24}$~g is the atomic mass,
$\rm \epsilon_{Ni} = 3.97\times10^{10}~ erg\,s^{-1}\, g^{-1}$ and $\rm \epsilon_{Co} = 7.03\times10^{9}~erg\,s^{-1}\,g^{-1}$. 
The exact $^{56}$Ni and $^{56}$Co decay data presented in \cite{Nadyozhin1994} have been used for calculations.

Following \cite{Valenti2008}, we divided the light curve into photospheric phase and nebular phase. Based on the Arnett model \citep{Arnett1982}, the bolometric light curve in the photospheric phase can be written as \citep{Valenti2008}:
\begin{align} \label{eqn:arnett_lc}
     L_{\rm bol, phot}(t) =~&M_{\rm Ni,0}~e^{-x^2} 
     \nonumber\\
     &\times[(\epsilon_{\rm Ni} - \epsilon_{\rm Co})\int^{x}_{0}A(z) \Gamma(z) dz 
     \nonumber\\
     &~~~~~+ \epsilon_{\rm Co}\int^{x}_{0}B(z) \Gamma(z) dz],
\end{align}
where $A(z) = 2ze^{-2zy+z^2}$ and $B(z)=2ze^{-2zy+2zs+z^2}$, with $x = \frac{t}{\tau_{m}}$, $y = \frac{\tau_{\rm Co}}{2\tau_{\rm Ni}}$ and $s = \frac{\tau_{m}(\tau_{\rm Co}-\tau_{\rm Ni})}{2\tau_{\rm Co}\tau_{\rm Ni}}$. $\Gamma(z) = 1 - e^{-A(z\tau_m)^{-2}}$ is the gamma-ray leakage term, with $A=\frac{3\kappa_{\gamma}M_{ej}}{4\pi v^2}$. $\tau_{m}$ is the effective time scale of the light curve:
\begin{equation}
    \tau_{m} = \frac{\kappa_{opt}}{\beta c}^{1/2}(\frac{6M_{ej}^{3}}{5E_{k}})^{1/4}
\end{equation}
, where $M_{\rm ej}$ is the SN ejecta mass, $E_{k}$ is the kinetic energy of the ejecta, $c$ is the speed of light. We have assumed the optical opacity $\kappa_{opt}$ = 0.07 $\rm cm^{2}\,g^{-1}$ is a constant and $\beta$ = 13.8 is an integration constant.

We note that the gamma-ray leakage term $\Gamma$ we use in Equation \ref{eqn:arnett_lc} is slightly different from what some studies have used in the literature (i.e., \citealt{Chatzopoulos2012}). Specifically, the gamma-ray leakage term here is a function of time, so it should be included within the integral. 
In \cite{Valenti2008}, although not explicitly mentioned in the paper, the gamma-ray leakage term was actually within the integral when calculating the model light curves, which is the same way we do it here.

\begin{figure}
\includegraphics[width=1.\linewidth]{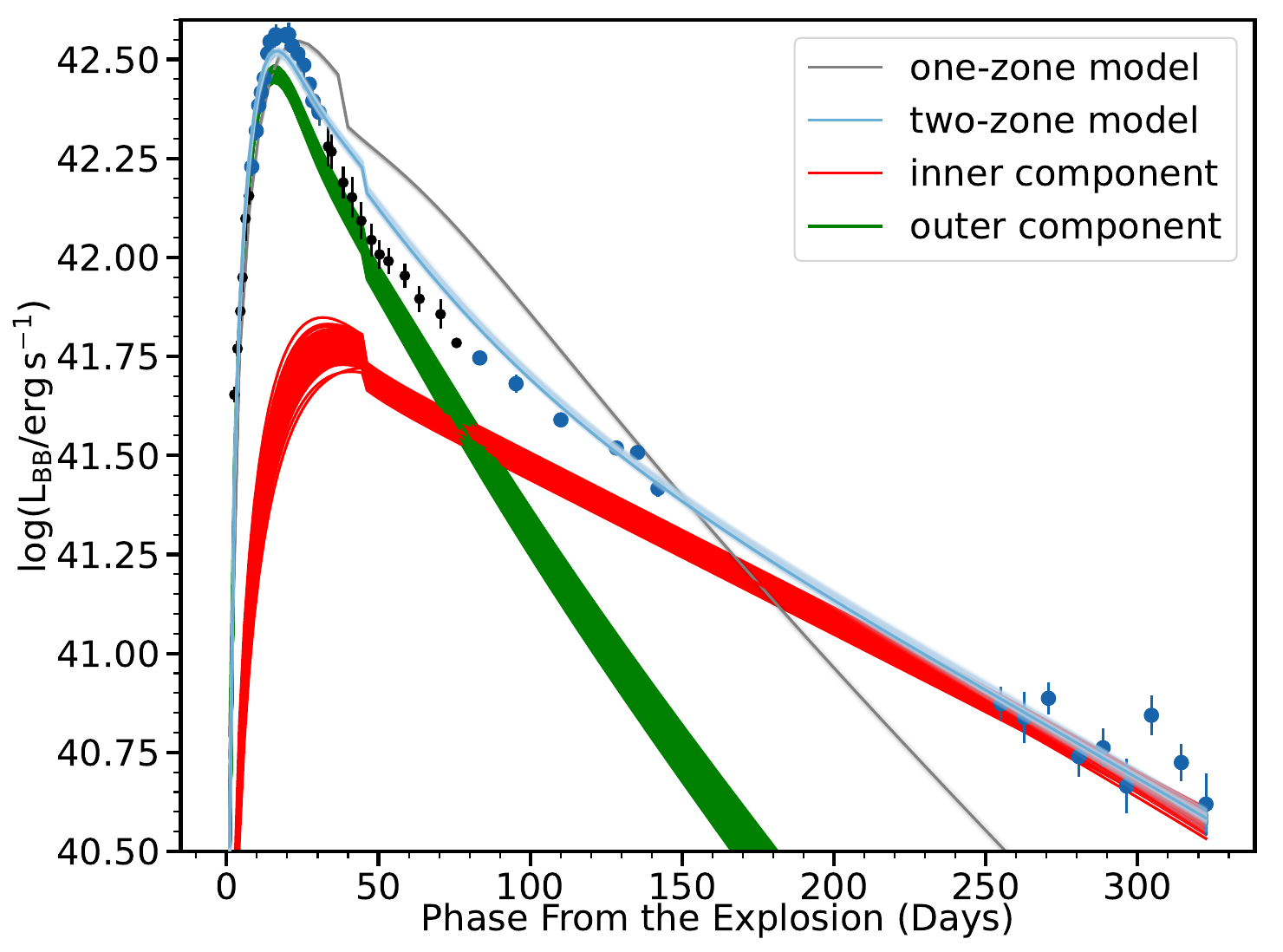}
\caption{Arnett model fit of the bolometric light curve. The fit is done with a MCMC routine. The best fitting results are represented by the one hundred randomly chosen models drawn from the MCMC chain.
The blue lines are the two-component fit and the grey lines are the one-component fit. The red lines and the green lines are the contributions from the inner region and the outer region, respectively. Only the blue points are used for the fit. The phase is shown with respect to the explosion epoch.
\label{fig:arnett_two_component}}
\end{figure}

During the nebular phase, the light curve is powered by the energy deposition produced by the radioactive decay of $\rm{{}^{56}Ni\rightarrow{}^{56}Co\rightarrow{}^{56}Fe}$. Thus, the nebular light curve can be obtained from equation \ref{eqn: Pni} and \ref{eqn: Pco} by adding the incomplete trapping term of $\gamma$-rays:
\begin{align}
    L_{\rm bol, neb} =&~M_{\rm Ni,0}~[\epsilon_{\rm Ni}~e^{-t/\tau_{Ni}} \nonumber\\ 
    &+ \epsilon_{\rm Co}~(e^{-t/\tau_{\rm Co}} - e^{-t/\tau_{\rm Ni}})](1-e^{-\tau_{\gamma}(t)}) \nonumber\\ 
    &+ 0.032\,M_{\rm Ni,0}~\epsilon_{\rm Co}~(e^{-t/\tau_{\rm Co}} - e^{-t/\tau_{\rm Ni}})
\end{align}
where $\tau_{\gamma}=\frac{3\kappa_{\gamma}M_{ej}}{4\pi v^2}$ is the optical depth to $\gamma$-rays \citep{Chatzopoulos2009,Chatzopoulos2012}. The last term takes the kinetic energy of the positrons produced by $\rm ^{56}Co$ decay into account.

The fit has been done with a MCMC method. The free parameters are $t_{0}$, $M_{\rm ej}$ and $M_{\rm Ni,0}$ with uniform priors, where $t_{0}$ is the explosion epoch. The upper and lower bounds of $t_{0}$ are set to be the first detection and the last non-detection, respectively. 
The fitting range has been chosen to be from -10 to 15 days from mamximum and 60 days post maximum following \cite{Valenti2008} and \cite{Lyman2016}.
The best-fitting parameters are constrained to be $M_{\rm Ni,0} = 0.20^{+0.03}_{-0.02}$\,\msun{}, $M_{\rm ej}=4.90^{+0.04}_{-0.04}$\,\msun, $E_{\rm kin}$=2.46$^{+0.02}_{-0.03}$$\times 10^{51}$\,erg. The best-fitting model is shown in Figure \ref{fig:arnett_two_component} (black line).

The model we use here cannot simultaneously reproduce the photospheric and nebular phase of the light curve of SN~2022crv. If we fit these two phases separately, the nickel mass derived from the photospheric phase will be larger than the nickel mass derived from the nebular phase. The ejecta mass derived from the photospheric phase is too small, resulting in a low gamma-ray trapping rate at late phases, leading to a too steep tail. 
This contradiction can be easily solved if a two-zone model is considered, with one of which dominating the photospheric phase and another one contributing to the nebular phase \citep{Valenti2008}.
In the model we adopted above, the nickel is assumed to be concentrated in the center of the progenitor. 
In order to explain the observational properties of SESNe, nickel needs to be mixed in the ejecta to some extent \citep{Woosley1997,Dessart2012,Bersten2013,Dessart2015,Dessart2016, Yoon2019}. Therefore, we adopted the two-zone model initially proposed in \cite{Maeda2003} that has been used for some SESNe \citep{Valenti2008,Valenti2011,Cano2014}. However, we note that since dust likely started to form at late phases in SN~2022crv (Rho et al., in prep), the dust attenuation could also contribute to this contradiction.

In this model, the ejecta of the SN is divided into two separate zones: a high density inner region and a low density outer region. The nickel is assumed to be homogeneously distributed in the two regions, respectively. In this two-zone model, the free parameters are the time of explosion $t_{0}$, the total nickel mass $M_{\rm Ni, total}$, the total ejecta mass $M_{\rm ej, total}$, the kinetic energy of the inner region $E_{\rm kin, inner}$, and the mass fraction of the inner region to the whole ejecta $\rm F_{inner}$. The final best fitting parameters are constrained to be $M_{\rm Ni, total} = 0.18^{+0.06}_{-0.04}$\,\msun{}, $M_{\rm ej, total}=2.78^{+0.05}_{-0.04}$\,\msun, $E_{\rm kin, total}$=1.01$^{+0.02}_{-0.02} \times 10^{51}$erg, $\rm F_{inner}$=0.28$^{+0.01}_{-0.01}$. The two-component model does give a better fit and the best-fitting model is shown in Figure \ref{fig:arnett_two_component}. 

\cite{Lyman2016} analysed a group of bolometric light curves of SESNe by fitting the Arnett model. They found that the average explosion parameters for SNe IIb are $M_{\rm Ni}=0.11(0.04)$\,\msun, $M_{\rm ej}=2.2(0.8)$\,\msun and $E_{\rm kin}$ = 1.0(0.6)$\times 10^{51}$erg. For SNe Ib, these values are $M_{\rm Ni}=0.17(0.16)$\,\msun, $M_{\rm ej}=2.6(1.1)$\,\msun and $E_{\rm kin}$ = 1.6(0.9)$\times 10^{51}$erg. Similar values have also been found for SESNe in more recent studies \cite[e.g.,][]{Taddia2018}. The parameters we derived for SN~2022crv are more similar to those of SNe Ib; but considering the uncertainties, they are consistent with those of both SNe IIb and Ib.


\subsubsection{\texorpdfstring{$\rm ^{56}Ni$}\ \ + Magnetar fit}
Magnetars are thought to be a possible energy source for SESNe \citep{Maeda2007, Kasen2010}. We use a hybrid model in which the object is powered by both $\rm ^{56}Ni$ decay and magnetar spin down. The luminosity of a SN powered by a central magnetar can be written as \citep{Wang2015}:
\begin{equation}
    L_{\rm bol, mag} = e^{-x^2}\times \frac{E_{NS}}{\tau_{NS}}\times \int^{x}_{0}C(z)dz
\end{equation}
where $C(z) = 2ze^{z^2}/(1+z\frac{\tau_{m}}{\tau_{NS}})^2(1-e^{-A(z\tau_{m})^{-2}})$, with $A=\frac{3\kappa_{\gamma}M_{ej}}{4\pi v^2}$. 
\begin{equation}
    E_{NS}=2\times 10^{52}\frac{M_{NS}}{1.4M_{\odot}}\left(\frac{P_0}{1~\rm ms}\right)^{-2}\left(\frac{R_{NS}}{10~\rm km}\right)^2~\rm erg
\end{equation}
\begin{equation}
    \tau_{NS}=1.3\left(\frac{B}{10^{14}~\rm G}\right)^{-2}\left(\frac{P_0}{10~\rm ms}\right)^2
\end{equation}
are the rotational energy and the spin-down timescale of the magnetar, respectively.

We found that the best-fitting model is dominated by the magnetar, and only a small amount of nickel is present in the ejecta. However, for SESNe, a certain amount of nickel should be in the ejecta and power the early part of the light curve \citep[e.g.,][]{Woosley2019, Woosley2021}. Therefore, a magnetar dominated model is likely not suitable in the case of SN~2022crv.


\begin{figure}
\includegraphics[width=1.\linewidth]{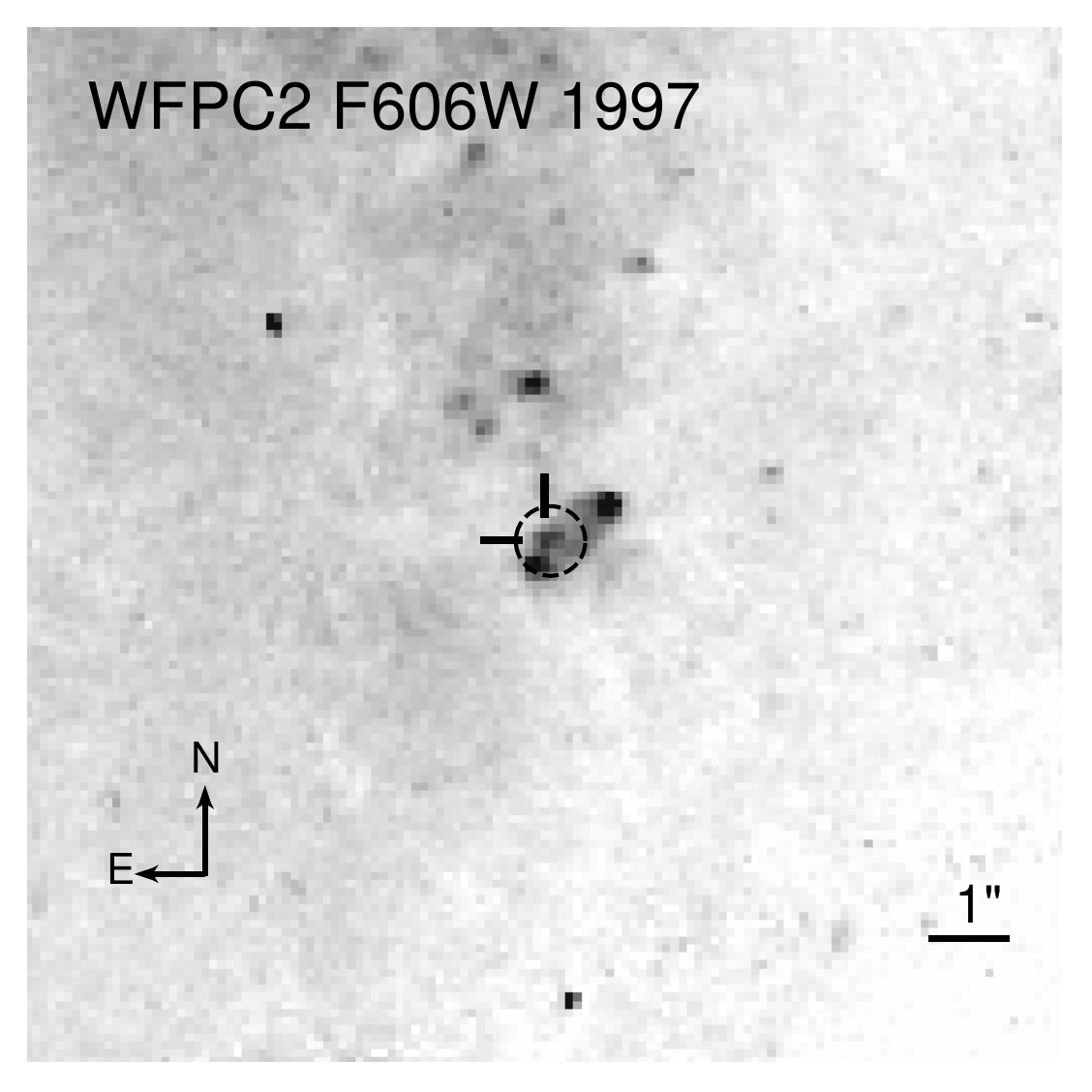}
\caption{A portion of the {\sl HST\/} WFPC2 F606W image mosaic from March 1997, with the location of the SN site indicated by a dashed circle (with radius corresponding to the $3\sigma$ uncertainty in the astrometric registration using a LBT $r$-band image of the SN from March 2022). A stellar-like object with $M_{\rm F606W}  \approx -8.2$ mag is indicated by the black tick markers. We tentatively identify this object as the progenitor candidate. North is up, and east is to the left.
\label{fig:prog_image}}
\end{figure}

\subsubsection{Search for a Progenitor Candidate}
A pre-explosion image obtained with the {\sl Hubble Space Telescope\/} ({\sl HST}) Wide Field Planetary Camera (WFPC2) in band F606W (PI Stiavelli, SNAP-6359)  on 1997 March 27, serendipitously contained the site of SN 2022crv, was identified in the Mikulski Archive for Space Telescope (MAST). We employed a 60-s $r$-band MODS acquisition image of the SN obtained with the LBT on 2022 March 10 to isolate the SN position in the \textit{HST} image mosaic. We were able to identify 17 stars in common between the two image datasets and, using Photutils centroiding and the PyRAF tasks geomap and geotran, we were able to locate the position with a $1\sigma$ astrometric uncertainty of 1.4 WFPC2 pixels ($0{\farcs}14$) (see Figure \ref{fig:prog_image}). The SN site appears to be in a luminous complex of stars or star clusters. We subsequently ran Dolphot \citep{Dolphin2016} with PSF fitting photometry on the individual WFPC2 frames and found an object, indicated to be stellar by the routine, with brightness $m_{\rm F606W} = 24.07 \pm 0.05$ mag. Corrected by our assumed distance and extinction to SN 2022crv, this corresponds to an absolute brightness of $M_{\rm F606W} \approx -8.2$ mag. We tentatively identify this object as the progenitor candidate, although some caution should be applied, given the comparatively low {\sl HST\/} resolution with WFPC2. Additionally, with only the single {\sl HST\/} band, we have no knowledge of the object's color. Nevertheless, if we compare its inferred luminosity to, e.g., the progenitor of SN 2011dh, for which \citet{Maund2011} found $M_{\rm F555W} \approx -7.5$ mag (at the assumed distance to M51 of 7.1 Mpc; adjusted to the recent Cepheid distance from \citealt{Csornyei2023}, this is $\approx -7.6$ mag) and \citet{VanDyk2011} estimated as $M_V^0 \approx -7.7$ mag (at an assumed 7.6 Mpc), this progenitor candidate for SN 2022crv appears to be plausible, albeit somewhat more luminous than the SN 2011dh progenitor. We note, however, that, e.g., \citet{Eldridge2015} found that the absolute brightness of the iPTF13bvn progenitor was at most $M_{\rm F555W} \approx -6.5$ mag, so the progenitor candidate for SN 2022crv is significantly more luminous than that. Whether this object is truly the SN progenitor will require confirmation, with follow-up {\sl HST\/}  observations when the SN has faded significantly (well below mag 24 in F606W).

\section{Summary}\label{sec:summary}
We present optical and NIR observations of SN~2022crv. In general, the optical photometric and spectroscopic evolution of SN~2022crv resembles those of both SNe IIb and Ib. The object showed conspicuous high-velocity H lines at early phase around 6200 \AA, which rapidly disappeared shortly after maximum. The 6200 \AA\ line in SN~2022crv is likely due to H$\alpha$ at early phases and is dominated by \ion{Si}{2}~$\lambda$6355 after the maximum.
The evolution of the H$\alpha$ line pEW in SN~2022crv is similar to those of SNe IIb at early phases, but falls into the Type Ib category shortly after maximum. 
In addition, we applied a SVM classification method to the spectra ranging from $-$15 days to 15 days relative to the maximum, which classifies SN~2022crv as a Type IIb before $-$10 days and a Type Ib afterwards.
This makes SN~2022crv a transitional object on the continuum between Type Ib SNe and Type IIb SNe, and suggests that there is a continuum between these two SN subtypes.

We found that a hydrogen envelope mass of $\sim$10$^{-3}$ \msun{} in the progenitor can reproduce the behaviours of the H lines in SN~2022crv, and the progenitor is constrained to be a He star with a final mass of $\sim$4.5 -- 5.6 \msun{} evolving from a $\sim$16 -- 22 \msun{} ZAMS star in a binary system. The metallicity at the SN site is measured to be slightly higher than the solar metallicity, so the progenitor likely experienced a strong stellar wind mass loss. In this case, an initial orbital separation of the binary system larger than $\sim$1000~$\rm R_{\odot}$ is needed in order to retain a small amount of the hydrogen envelope. 
We found that the bolometric light curve of SN~2022crv can be best fitted by a model with a nickel mass of $0.18^{+0.06}_{-0.04}$\,\msun{}, $M_{\rm ej}$ of $2.78^{+0.05}_{-0.04}$\,\msun, and $E_{\rm kin}$ of 1.01$^{+0.02}_{-0.02}\times 10^{51}$\,erg.

The NIR spectroscopic evolution of SN~2022crv is generally similar to those of other SNe IIb/Ib. However, an extra absorption feature is observed in the NIR spectra, near the blue side of the \ion{He}{1}~$\lambda$1.083~$\mu$m line, referred to as feature A in this paper. To the best of our knowledge, feature A has never been observed in other SNe IIb/Ib. We found that this line is most likely from \ion{Sr}{2}~$\lambda$1.0327~$\mu$m, but we couldn't safely exclude \ion{Fe}{1}~$\lambda$1.05~$\mu$m and \ion{S}{1}~$\lambda$1.0457~$\mu$m. Future detailed modeling is required to further investigate the origin of feature~A.

The peculiar features observed in the optical and NIR spectra of SN~2022crv illustrate that SESNe still have many unsolved mysteries. This emphasizes the importance of obtaining NIR spectra and the early discovery of SESNe. 
As the number of SESNe characterized with detailed datasets increases, the gap between SNe IIb and SNe Ib could be filled, giving us a comprehensive picture of the evolutionary channel and history of the SESNe progenitors.

\section*{Acknowledgements}
We thank Luc Dessart for providing the model spectra and beneficial discussions.
We would like to thank J. Craig Wheeler, Emmanouil Chatzopoulos and Jozsef Vink{\'o} for beneficial discussions. 
We would like to thank Stanford Woosley for generating and providing the model spectra.
We would like to thank Sung-Chul Yoon for providing the model light curves.
We thank the U.C. Berkeley undergraduate students Ivan Altunin, Kate Bostow, Kingsley Ehrich, Nachiket Girish, Neil Pichay and James Sunseri for their effort in taking Lick/Nickel data. YD would like to thank the hospitality and support of LZ in Philadelphia during completion of this paper.

Research by Y.D., S.V., N.M.R, E.H., and D.M. is supported by NSF grant AST-2008108. 

Time-domain research by the University of Arizona team and D.J.S.\ is supported by NSF grants AST-1821987, 1813466, 1908972, 2108032, and 2308181, and by the Heising-Simons Foundation under grant \#2020-1864.

M.M. acknowledges support in part from ADAP program grant No. 80NSSC22K0486, from the NSF grant AST-2206657 and from the HST GO program HST-GO-16656.

This work makes use of data from the Las Cumbres Observatory global telescope network. The Las Cumbres Observatory group is supported by NSF grants AST-1911151 and AST-1911225.

K.A.B. is supported by an LSSTC Catalyst Fellowship; this publication was thus made possible through the support of Grant 62192 from the John Templeton Foundation to LSSTC. The opinions expressed in this publication are those of the authors and do not necessarily reflect the views of LSSTC or the John Templeton Foundation.

L.A.K. acknowledges support by NASA FINESST fellowship 80NSSC22K1599.

S.M. acknowledges support from the Magnus Ehrnrooth Foundation and the Vilho, Yrj\"{o}, and Kalle V\"{a}is\"{a}l\"{a} Foundation.

L.G. acknowledges financial support from the Spanish Ministerio de Ciencia e Innovaci\'on (MCIN), the Agencia Estatal de Investigaci\'on (AEI) 10.13039/501100011033, and the European Social Fund (ESF) "Investing in your future" under the 2019 Ram\'on y Cajal program RYC2019-027683-I and the PID2020-115253GA-I00 HOSTFLOWS project, from Centro Superior de Investigaciones Cient\'ificas (CSIC) under the PIE project 20215AT016, and the program Unidad de Excelencia Mar\'ia de Maeztu CEX2020-001058-M.

M.D.S. is funded by the Independent Research Fund Denmark (IRFD) via Project 2 grant 10.46540/2032-00022B.

The SALT data presented here were obtained through Rutgers University programs 2021-1-MLT-007 and 2022-1-MLT-004 (PI: S.W.J.).  L.A.K. acknowledges support by NASA FINESST fellowship 80NSSC22K1599.

Based on observations obtained at the international Gemini Observatory, a program of NSF's NOIRLab, which is managed by the Association of Universities for Research in Astronomy (AURA) under a cooperative agreement with the National Science Foundation. On behalf of the Gemini Observatory partnership: the National Science Foundation (United States), National Research Council (Canada), Agencia Nacional de Investigaci\'{o}n y Desarrollo (Chile), Ministerio de Ciencia, Tecnolog\'{i}a e Innovaci\'{o}n (Argentina), Minist\'{e}rio da Ci\^{e}ncia, Tecnologia, Inova\c{c}\~{o}es e Comunica\c{c}\~{o}es (Brazil), and Korea Astronomy and Space Science Institute (Republic of Korea).

This work was enabled by observations made from the Gemini North telescope, located within the Maunakea Science Reserve and adjacent to the summit of Maunakea. We are grateful for the privilege of observing the Universe from a place that is unique in both its astronomical quality and its cultural significance.

Observations reported here were obtained at the MMT Observatory, a joint facility of the University of Arizona and the Smithsonian Institution.

This paper uses data gathered with the 6.5 m Magellan telescopes at Las Campanas Observatory, Chile.

The LBT is an international collaboration among institutions in the United States, Italy and Germany. LBT Corporation Members are: The University of Arizona on behalf of the Arizona Board of Regents; Istituto Nazionale di Astrofisica, Italy; LBT Beteiligungsgesellschaft, Germany, representing the Max-Planck Society, The Leibniz Institute for Astrophysics Potsdam, and Heidelberg University; The Ohio State University, and The Research Corporation, on behalf of The University of Notre Dame, University of Minnesota and University of Virginia.
This research is based in part on observations made with the NASA/ESA Hubble Space Telescope obtained from the Space Telescope Science Institute, which is operated by the Association of Universities for Research in Astronomy, Inc., under NASA contract NAS 5-26555.

Based on observations made with the Gran Telescopio Canarias (GTC), installed at the Spanish Observatorio del Roque de los Muchachos of the Instituto de Astrof{\'i}sica de Canarias, on the island of La Palma.

This work is (partly) based on data obtained with the instrument OSIRIS, built by a Consortium led by the Instituto de Astrof{\'i}sica de Canarias in collaboration with the Instituto de Astronom{\'i}a of the Universidad Aut{\'o}noma de M{\'e}xico. OSIRIS was funded by GRANTECAN and the National Plan of Astronomy and Astrophysics of the Spanish Government.

This research has made use of the NASA/IPAC Extragalactic Database (NED), which is funded by the National Aeronautics and Space Administration and operated by the California Institute of Technology.

This research made use of Photutils, an Astropy package for detection and photometry of astronomical sources \citep{larry_bradley_2022_6825092}.

%

\facilities{ADS, DLT40 (Prompt5, Prompt-MO), Lick (KAIT, Nickel), ATLAS, LCOGT (SBIG, Sinistro,
FLOYDS), Gemini:North (GMOS), Keck:I (LRIS), Keck:II (DEIMOS, NIRES), LCOGT (Sinistro), MMT (Binospec, MMIRS), NED, SALT (RSS), Shane (Kast), SOAR (Goodman), Swift (UVOT), Baade (IMACS, FIRE), Clay (LDSS3), UH88 (SNIFS), Bok (B\&C), LBT (MODS)
}


\software{Astropy \citep{astropy13,astropy18}, 
          HOTPANTS \citep{Becker2015},
          Matplotlib \citep{Hunter2007},
          NumPy \citep{2020Natur.585..357H},
          PYRAF \citep{2012ascl.soft07011S},
          Pandas \citep{mckinney-proc-scipy-2010},
          SciPy \citep{2020NatMe..17..261V},
          SWarp \citep{Bertin2002},
          HOTPANTS \citep{Becker2015},
          LCOGTSNpipe \citep{Valenti2016}, 
          Light Curve Fitting \citep{hosseinzadeh_light_2020},
          PypeIt \citep{pypeit:zenodo}
          }



\appendix

\section{Photometric Data Reduction}\label{sec:phot_reduction}

The photometric data from the Las Cumbres Observatory were reduced using the PyRAF-based photometric reduction pipeline {\sc LCOGTSNPIPE} \citep{Valenti2016}. This pipeline uses a low-order polynomial fit to remove the background and calculates instrumental magnitudes using a standard point-spread function (PSF) fitting technique. Apparent magnitudes were calibrated using the APASS ($g, r, i$) and Landolt ($U, B, V$) catalogs.

Unfiltered ($Open$) DLT40 images were processed with a PyRAF-based pipeline. Background contamination was removed by subtracting a reference image, and the aperture photometry was extracted from the subtracted images. The final photometry is calibrated to the $r$ band using the APASS catalog.

The ATLAS survey is carried out primarily in two filters, cyan and orange, roughly equivalent to Pan-STARRS filters g+r and r+i, respectively. A quad of 4 x 30 second exposures are typically obtained for each field over a 1 hour window. The images are processed and difference imaging is performed in real-time to enable rapid discovery of transients in the data stream \citep{Smith2020}. We obtained the forced photometry at the supernova position from the ATLAS forced photometry server \citep{Shingles2021}. We stacked the individual flux measurements for each nightly quad into a single measurement in order to improve signal to noise, and to obtain deeper upper limits for the pre-discovery non-detections.

The images obtained with KAIT and Nickel at Lick Observatory were reduced using a custom
pipeline\footnote{https://github.com/benstahl92/LOSSPhotPypeline}
detailed in \cite{Stahl2019}.
Point-spread function photometry was obtained using DAOPHOT \citep{Stetson1987}
from the IDL Astronomy User's Library\footnote{http://idlastro.gsfc.nasa.gov/}.
Several nearby stars were chosen from the
Pan-STARRS1\footnote{http://archive.stsci.edu/panstarrs/search.php} catalog for calibration.
 Their magnitudes were first transformed into Landolt magnitudes
using the empirical prescription presented by \cite{Tonry2012},
then transformed to the KAIT/Nickel natural system.
The final result were transformed to the standard system using local
calibrators and color terms for KAIT4 and Nickel2 \citep{Stahl2019}.

Swift images were reduced using the High-Energy Astrophysics software (HEA-Soft). The background is measured from a region away from any stars. 
Zero-points were chosen from \cite{Breeveld2011} with time-dependent sensitivity corrections updated in 2020.

\section{Spectroscopic Data Reduction} \label{sec:spec_reduction}
FLOYDS spectra were reduced following standard procedures using the FLOYDS pipeline \citep{Valenti2014}.

The IMACS Baade spectra were reduced using standard methods and {\sc IRAF} routines \footnote{{\sc IRAF} was distributed by the National Optical Astronomy Observatory which was operated by the Association of Universities for Research in Astronomy (AURA) under cooperative agreement with the National Science Foundation.}, as described in \cite{Hamuy2006PASP..118....2H}.

The UH88 data were obtained with SNIFS and reduced using the method outlined in \cite{Tucker2022PASP..134l4502T}.

The SALT spectra were reduced using a custom longslit pipeline based on the PySALT package \citep{PySALT}. 

The spectrum taken with the Gemini Multi-Object Spectrographs (GMOS; \citealp{Hook2004,gimeno16})  as part of our program GN-22A-Q-135 used the B600 grating and a 1.$\arcsec$5 slit. Data were reduced using the {\tt DRAGONS} (Data Reduction for Astronomy from Gemini Observatory North and South) reduction package \citep{Labrie2019}, using the recipe for GMOS long-slit reductions. This includes bias correction, flatfielding, wavelength calibration, and flux calibration.

The spectrum obtained with the Low Dispersion Survey Spectrograph 3 (LDSS-3) 
 was reduced  using
{\sc iraf}
 including bias subtraction, flat fielding, cosmic ray rejection, local sky subtraction and extraction of one-dimensional spectra. The slit was aligned along the parallactic angle to minimize differential light losses, and flux calibration was done using a spectrophotometric standard taken that night at similar airmass.

The Keck NIRES data were reduced following standard procedures described in the IDL package Spextools version 5.0.2 for NIRES \citep{Cushing2004PASP..116..362C}. The extracted 1D spectrum was flux calibrated and also corrected for telluric features with Xtellcorr \citep{Vacca2003} version 5.0.2 for NIRES, making use of an A0V standard star close in time and at similar airmass to the science target.

The two spectra taken with the Kast double spectrograph \citep{miller1994} mounted on the Shane 3~m telescope at Lick Observatory utilized the 2" slit, 600/4310 grism, and 300/7500 grating. This instrument configuration has a combined wavelength range of ~ 3500-10,500 A, and a spectral resolving power of R $\approx$ 800. To minimize slit losses caused by atmospheric dispersion \citep{Filippenko1982}, the slit was oriented at or near the parallactic angle. The data were reduced following standard techniques for CCD processing and spectrum extraction \citep{Silverman2012} utilizing IRAF \citep{Tody1986} routines and custom Python and IDL codes\footnote{https://github.com/ishivvers/TheKastShiv}.  Low-order polynomial fits to comparison-lamp spectra were used to calibrate the wavelength scale, and small adjustments derived from night-sky lines in the target frames were applied. The spectra were flux calibrated using observations of appropriate spectrophotometric standard stars observed on the same night, at similar airmasses, and with an identical instrument configuration.

The NIR spectrum obtained using the Folded port InfraRed Echellette \cite[FIRE;][]{Simcoe2013} spectrograph mounted on the 6.5-m Magellan Baade telescope at Las Campanas Observatory, Chile, was taken in the high-throughput prism mode with a 0$\farcs$6 slit. For telluric correction, an A0V star was observed close in time and at similar airmass to the science target. The spectra were reduced using the IDL pipeline firehose \citep{Simcoe2013}. Details of the observation setup and reduction were outlined in \cite{Hsiao2019PASP..131a4002H}.

NIR spectra were obtained with the MMT and Magellan Infrared Spectrograph (MMIRS; \citealp{McLeod2012}) on the 6.5~m MMT Observatory telescope on Mt.\ Hopkins at the Smithsonian's Fred Lawrence Whipple Observatory using a 1$\farcs$0 slit in both the zJ and HK (with the high-throughput HK3 filter) spectroscopic modes. The data were reduced using the automated MMIRS pipeline \citep{Chilingarian2015}. Telluric corrections and absolute flux calibrations were performed using observations of the A0V star HD~72033 close in time and at similar airmass to the science target. We employed the method of \citet{Vacca2003} implemented in the IDL tool \textsc{xtellcor\_general} developed by \citet{Cushing2004PASP..116..362C} as part of the Spextool package.

The Keck LRIS spectra were reduced in the standard manner with {\sc PypeIt} \citep{pypeit:joss_arXiv,pypeit:joss_pub,pypeit:zenodo}.

The GTC spectrum was reduced following standard procedures with {\sc pyraf} routines via the graphical user interface {\sc FOSCGUI}\footnote{{\sc FOSCGUI} is a graphic user interface aimed at extracting SN spectroscopy and photometry obtained with FOSC-like instruments. It was developed by E. Cappellaro. A package description can be found at \url{http://sngroup.oapd.inaf.it/foscgui.html.}}. The two-dimensional frames were corrected for bias and flat-fielded before the one-dimensional spectra extraction. We wavelength-calibrated the spectra via comparison with arc-lamp spectra and calibrated the flux using spectrophotometric standard stars. These also helped in removing the strongest telluric absorption bands present in the spectrum. Finally, the absolute flux calibration of the spectrum was cross-checked against the broadband photometry. 



\bibliography{sn2022crv}{}

\begin{thebibliography}{}
\expandafter\ifx\csname natexlab\endcsname\relax\def\natexlab#1{#1}\fi
\providecommand{\url}[1]{\href{#1}{#1}}
\providecommand{\dodoi}[1]{doi:~\href{http://doi.org/#1}{\nolinkurl{#1}}}
\providecommand{\doeprint}[1]{\href{http://ascl.net/#1}{\nolinkurl{http://ascl.net/#1}}}
\providecommand{\doarXiv}[1]{\href{https://arxiv.org/abs/#1}{\nolinkurl{https://arxiv.org/abs/#1}}}

\bibitem[{{Aldering} {et~al.}(1994){Aldering}, {Humphreys}, \&
  {Richmond}}]{Aldering1994}
{Aldering}, G., {Humphreys}, R.~M., \& {Richmond}, M. 1994, \aj, 107, 662,
  \dodoi{10.1086/116886}

\bibitem[{{Allende Prieto} {et~al.}(2001){Allende Prieto}, {Lambert}, \&
  {Asplund}}]{Allende_Prieto2001ApJ...556L..63A}
{Allende Prieto}, C., {Lambert}, D.~L., \& {Asplund}, M. 2001, \apjl, 556, L63,
  \dodoi{10.1086/322874}

\bibitem[{{Allington-Smith} {et~al.}(1994){Allington-Smith}, {Breare}, {Ellis},
  {Gellatly}, {Glazebrook}, {Jorden}, {Maclean}, {Oates}, {Shaw}, {Tanvir},
  {Taylor}, {Taylor}, {Webster}, \& {Worswick}}]{Allington-Smith1994}
{Allington-Smith}, J., {Breare}, M., {Ellis}, R., {et~al.} 1994, \pasp, 106,
  983, \dodoi{10.1086/133471}

\bibitem[{{Andrews} {et~al.}(2022){Andrews}, {Lundquist}, {Sand},
  {Hosseinzadeh}, {Dong}, {Janzen}, {Bostroem}, {Valenti}, {Jencson}, {Wyatt},
  {Meza}, \& {Pearson}}]{Andrews2022}
{Andrews}, J.~E., {Lundquist}, M., {Sand}, D.~J., {et~al.} 2022, Transient Name
  Server Classification Report, 2022-454, 1

\bibitem[{{Arnett}(1982)}]{Arnett1982}
{Arnett}, W.~D. 1982, \apj, 253, 785, \dodoi{10.1086/159681}

\bibitem[{{Asplund} {et~al.}(2009){Asplund}, {Grevesse}, {Sauval}, \&
  {Scott}}]{Asplund2009ARA&A..47..481A}
{Asplund}, M., {Grevesse}, N., {Sauval}, A.~J., \& {Scott}, P. 2009, \araa, 47,
  481, \dodoi{10.1146/annurev.astro.46.060407.145222}

\bibitem[{{Astropy Collaboration} {et~al.}(2013){Astropy Collaboration},
  {Robitaille}, {Tollerud}, {Greenfield}, {Droettboom}, {Bray}, {Aldcroft},
  {Davis}, {Ginsburg}, {Price-Whelan}, {Kerzendorf}, {Conley}, {Crighton},
  {Barbary}, {Muna}, {Ferguson}, {Grollier}, {Parikh}, {Nair}, {Unther},
  {Deil}, {Woillez}, {Conseil}, {Kramer}, {Turner}, {Singer}, {Fox}, {Weaver},
  {Zabalza}, {Edwards}, {Azalee Bostroem}, {Burke}, {Casey}, {Crawford},
  {Dencheva}, {Ely}, {Jenness}, {Labrie}, {Lim}, {Pierfederici}, {Pontzen},
  {Ptak}, {Refsdal}, {Servillat}, \& {Streicher}}]{astropy13}
{Astropy Collaboration}, {Robitaille}, T.~P., {Tollerud}, E.~J., {et~al.} 2013,
  \aap, 558, A33, \dodoi{10.1051/0004-6361/201322068}

\bibitem[{{Astropy Collaboration} {et~al.}(2018){Astropy Collaboration},
  {Price-Whelan}, {Sip{\H o}cz}, {G{\"u}nther}, {Lim}, {Crawford}, {Conseil},
  {Shupe}, {Craig}, {Dencheva}, {Ginsburg}, {VanderPlas}, {Bradley},
  {P{\'e}rez-Su{\'a}rez}, {de Val-Borro}, {Aldcroft}, {Cruz}, {Robitaille},
  {Tollerud}, {Ardelean}, {Babej}, {Bach}, {Bachetti}, {Bakanov}, {Bamford},
  {Barentsen}, {Barmby}, {Baumbach}, {Berry}, {Biscani}, {Boquien}, {Bostroem},
  {Bouma}, {Brammer}, {Bray}, {Breytenbach}, {Buddelmeijer}, {Burke},
  {Calderone}, {Cano Rodr{\'{\i}}guez}, {Cara}, {Cardoso}, {Cheedella},
  {Copin}, {Corrales}, {Crichton}, {D'Avella}, {Deil}, {Depagne}, {Dietrich},
  {Donath}, {Droettboom}, {Earl}, {Erben}, {Fabbro}, {Ferreira}, {Finethy},
  {Fox}, {Garrison}, {Gibbons}, {Goldstein}, {Gommers}, {Greco}, {Greenfield},
  {Groener}, {Grollier}, {Hagen}, {Hirst}, {Homeier}, {Horton}, {Hosseinzadeh},
  {Hu}, {Hunkeler}, {Ivezi{\'c}}, {Jain}, {Jenness}, {Kanarek}, {Kendrew},
  {Kern}, {Kerzendorf}, {Khvalko}, {King}, {Kirkby}, {Kulkarni}, {Kumar},
  {Lee}, {Lenz}, {Littlefair}, {Ma}, {Macleod}, {Mastropietro}, {McCully},
  {Montagnac}, {Morris}, {Mueller}, {Mumford}, {Muna}, {Murphy}, {Nelson},
  {Nguyen}, {Ninan}, {N{\"o}the}, {Ogaz}, {Oh}, {Parejko}, {Parley}, {Pascual},
  {Patil}, {Patil}, {Plunkett}, {Prochaska}, {Rastogi}, {Reddy Janga},
  {Sabater}, {Sakurikar}, {Seifert}, {Sherbert}, {Sherwood-Taylor}, {Shih},
  {Sick}, {Silbiger}, {Singanamalla}, {Singer}, {Sladen}, {Sooley},
  {Sornarajah}, {Streicher}, {Teuben}, {Thomas}, {Tremblay}, {Turner},
  {Terr{\'o}n}, {van Kerkwijk}, {de la Vega}, {Watkins}, {Weaver}, {Whitmore},
  {Woillez}, {Zabalza}, \& {Astropy Contributors}}]{astropy18}
{Astropy Collaboration}, {Price-Whelan}, A.~M., {Sip{\H o}cz}, B.~M., {et~al.}
  2018, \aj, 156, 123, \dodoi{10.3847/1538-3881/aabc4f}

\bibitem[{{Becker}(2015)}]{Becker2015}
{Becker}, A. 2015, {HOTPANTS: High Order Transform of PSF ANd Template
  Subtraction}.
\newblock \doeprint{1504.004}

\bibitem[{{Bersten} {et~al.}(2013){Bersten}, {Tanaka}, {Tominaga}, {Benvenuto},
  \& {Nomoto}}]{Bersten2013}
{Bersten}, M.~C., {Tanaka}, M., {Tominaga}, N., {Benvenuto}, O.~G., \&
  {Nomoto}, K. 2013, \apj, 767, 143, \dodoi{10.1088/0004-637X/767/2/143}

\bibitem[{{Bertin} {et~al.}(2002){Bertin}, {Mellier}, {Radovich}, {Missonnier},
  {Didelon}, \& {Morin}}]{Bertin2002}
{Bertin}, E., {Mellier}, Y., {Radovich}, M., {et~al.} 2002, in Astronomical
  Society of the Pacific Conference Series, Vol. 281, Astronomical Data
  Analysis Software and Systems XI, ed. D.~A. {Bohlender}, D.~{Durand}, \&
  T.~H. {Handley}, 228

\bibitem[{{Blondin} \& {Tonry}(2007)}]{Blondin2007}
{Blondin}, S., \& {Tonry}, J.~L. 2007, \apj, 666, 1024, \dodoi{10.1086/520494}

\bibitem[{Bradley {et~al.}(2022)Bradley, Sip{\H{o}}cz, Robitaille, Tollerud,
  Vinícius, Deil, Barbary, Wilson, Busko, Donath, Günther, Cara, Lim,
  Meßlinger, Conseil, Bostroem, Droettboom, Bray, Bratholm, Barentsen, Craig,
  Rathi, Pascual, Perren, Georgiev, de~Val-Borro, Kerzendorf, Bach, Quint, \&
  Souchereau}]{larry_bradley_2022_6825092}
Bradley, L., Sip{\H{o}}cz, B., Robitaille, T., {et~al.} 2022,
  astropy/photutils: 1.5.0, 1.5.0,  Zenodo, \dodoi{10.5281/zenodo.6825092}

\bibitem[{{Branch} {et~al.}(2002){Branch}, {Benetti}, {Kasen}, {Baron},
  {Jeffery}, {Hatano}, {Stathakis}, {Filippenko}, {Matheson}, {Pastorello},
  {Altavilla}, {Cappellaro}, {Rizzi}, {Turatto}, {Li}, {Leonard}, \&
  {Shields}}]{Branch2002}
{Branch}, D., {Benetti}, S., {Kasen}, D., {et~al.} 2002, \apj, 566, 1005,
  \dodoi{10.1086/338127}

\bibitem[{{Breeveld} {et~al.}(2011){Breeveld}, {Landsman}, {Holland}, {Roming},
  {Kuin}, \& {Page}}]{Breeveld2011}
{Breeveld}, A.~A., {Landsman}, W., {Holland}, S.~T., {et~al.} 2011, in American
  Institute of Physics Conference Series, Vol. 1358, Gamma Ray Bursts 2010, ed.
  J.~E. {McEnery}, J.~L. {Racusin}, \& N.~{Gehrels}, 373--376,
  \dodoi{10.1063/1.3621807}

\bibitem[{{Brethauer} {et~al.}(2022){Brethauer}, {Margutti}, {Milisavljevic},
  {Bietenholz}, {Chornock}, {Coppejans}, {De Colle}, {Hajela}, {Terreran},
  {Vargas}, {DeMarchi}, {Harris}, {Jacobson-Gal{\'a}n}, {Kamble}, {Patnaude},
  \& {Stroh}}]{Brethauer2022}
{Brethauer}, D., {Margutti}, R., {Milisavljevic}, D., {et~al.} 2022, \apj, 939,
  105, \dodoi{10.3847/1538-4357/ac8b14}

\bibitem[{{Brown} {et~al.}(2013){Brown}, {Baliber}, {Bianco}, {Bowman},
  {Burleson}, {Conway}, {Crellin}, {Depagne}, {De Vera}, {Dilday}, {Dragomir},
  {Dubberley}, {Eastman}, {Elphick}, {Falarski}, {Foale}, {Ford}, {Fulton},
  {Garza}, {Gomez}, {Graham}, {Greene}, {Haldeman}, {Hawkins}, {Haworth},
  {Haynes}, {Hidas}, {Hjelstrom}, {Howell}, {Hygelund}, {Lister}, {Lobdill},
  {Martinez}, {Mullins}, {Norbury}, {Parrent}, {Paulson}, {Petry}, {Pickles},
  {Posner}, {Rosing}, {Ross}, {Sand}, {Saunders}, {Shobbrook}, {Shporer},
  {Street}, {Thomas}, {Tsapras}, {Tufts}, {Valenti}, {Vander Horst}, {Walker},
  {White}, \& {Willis}}]{Brown2013}
{Brown}, T.~M., {Baliber}, N., {Bianco}, F.~B., {et~al.} 2013, \pasp, 125,
  1031, \dodoi{10.1086/673168}

\bibitem[{{Burns} {et~al.}(2021){Burns}, {Hsiao}, {Suntzeff}, {Baron},
  {Shappee}, {Aldoroty}, {Anderson}, {Ashall}, {Bersten}, {Brown}, {Burrow},
  {Clochiatti}, {Davis}, {DerKacy}, {Do}, {Folatelli}, {Forster Buron},
  {Galbany}, {Hoeflich}, {Holmbo}, {Karamehmetoglu}, {Krisciunas}, {Kumar},
  {Lu}, {Mazzali}, {Morrell}, {Pessi}, {Phillips}, {Pignata}, {Piro}, {Polin},
  {Shahbandeh}, {Stangl}, {Stritzinger}, {Teffs}, {Tonry}, {Tucker}, {Uddin},
  \& {Yang}}]{Burns2021ATel14441....1B}
{Burns}, C., {Hsiao}, E., {Suntzeff}, N., {et~al.} 2021, The Astronomer's
  Telegram, 14441, 1

\bibitem[{{Cano} {et~al.}(2014){Cano}, {Maeda}, \& {Schulze}}]{Cano2014}
{Cano}, Z., {Maeda}, K., \& {Schulze}, S. 2014, \mnras, 438, 2924,
  \dodoi{10.1093/mnras/stt2400}

\bibitem[{{Cappellaro} {et~al.}(1997){Cappellaro}, {Mazzali}, {Benetti},
  {Danziger}, {Turatto}, {della Valle}, \& {Patat}}]{Cappellaro1997}
{Cappellaro}, E., {Mazzali}, P.~A., {Benetti}, S., {et~al.} 1997, \aap, 328,
  203, \dodoi{10.48550/arXiv.astro-ph/9707016}

\bibitem[{Cepa {et~al.}(2000)Cepa, Aguiar-Gonzalez, Gonzalez-Escalera,
  Gonzalez-Serrano, Joven-Alvarez, Cano, Rasilla, Rodriguez-Ramos, Gonzalez,
  Duenas, Sanchez, Tejada, Bland-Hawthorn, Militello, \&
  Rosa}]{Cepa10.1117/12.395520}
Cepa, J., Aguiar-Gonzalez, M., Gonzalez-Escalera, V., {et~al.} 2000, in Optical
  and IR Telescope Instrumentation and Detectors, ed. M.~Iye \& A.~F.~M.
  Moorwood, Vol. 4008, International Society for Optics and Photonics (SPIE),
  623 -- 631, \dodoi{10.1117/12.395520}

\bibitem[{{Chatzopoulos} {et~al.}(2009){Chatzopoulos}, {Wheeler}, \&
  {Vinko}}]{Chatzopoulos2009}
{Chatzopoulos}, E., {Wheeler}, J.~C., \& {Vinko}, J. 2009, \apj, 704, 1251,
  \dodoi{10.1088/0004-637X/704/2/1251}

\bibitem[{{Chatzopoulos} {et~al.}(2012){Chatzopoulos}, {Wheeler}, \&
  {Vinko}}]{Chatzopoulos2012}
---. 2012, \apj, 746, 121, \dodoi{10.1088/0004-637X/746/2/121}

\bibitem[{{Chevalier} \& {Soderberg}(2010)}]{Chevalier2010ApJ...711L..40C}
{Chevalier}, R.~A., \& {Soderberg}, A.~M. 2010, \apjl, 711, L40,
  \dodoi{10.1088/2041-8205/711/1/L40}

\bibitem[{{Chilingarian} {et~al.}(2015){Chilingarian}, {Beletsky}, {Moran},
  {Brown}, {McLeod}, \& {Fabricant}}]{Chilingarian2015}
{Chilingarian}, I., {Beletsky}, Y., {Moran}, S., {et~al.} 2015, \pasp, 127,
  406, \dodoi{10.1086/680598}

\bibitem[{{Clemens} {et~al.}(2004){Clemens}, {Crain}, \&
  {Anderson}}]{clemens2004}
{Clemens}, J.~C., {Crain}, J.~A., \& {Anderson}, R. 2004, in Society of
  Photo-Optical Instrumentation Engineers (SPIE) Conference Series, Vol. 5492,
  Ground-based Instrumentation for Astronomy, ed. A.~F.~M. {Moorwood} \&
  M.~{Iye}, 331--340, \dodoi{10.1117/12.550069}

\bibitem[{{Clocchiatti} \& {Wheeler}(1997)}]{Clocchiatti1997}
{Clocchiatti}, A., \& {Wheeler}, J.~C. 1997, \apj, 491, 375,
  \dodoi{10.1086/304961}

\bibitem[{{Crawford} {et~al.}(2010){Crawford}, {Still}, {Schellart}, {Balona},
  {Buckley}, {Dugmore}, {Gulbis}, {Kniazev}, {Kotze}, {Loaring}, {Nordsieck},
  {Pickering}, {Potter}, {Romero Colmenero}, {Vaisanen}, {Williams}, \&
  {Zietsman}}]{PySALT}
{Crawford}, S.~M., {Still}, M., {Schellart}, P., {et~al.} 2010, in Society of
  Photo-Optical Instrumentation Engineers (SPIE) Conference Series, Society of
  Photo-Optical Instrumentation Engineers (SPIE) Conference Series, 25,
  \dodoi{10.1117/12.857000}

\bibitem[{{Crowther}(2007)}]{Crowther2007ARA&A..45..177C}
{Crowther}, P.~A. 2007, \araa, 45, 177,
  \dodoi{10.1146/annurev.astro.45.051806.110615}

\bibitem[{{Cs{\"o}rnyei} {et~al.}(2023){Cs{\"o}rnyei}, {Anderson}, {Vogl},
  {Taubenberger}, {Blondin}, {Leibundgut}, \& {Hillebrandt}}]{Csornyei2023}
{Cs{\"o}rnyei}, G., {Anderson}, R.~I., {Vogl}, C., {et~al.} 2023, arXiv
  e-prints, arXiv:2305.13943, \dodoi{10.48550/arXiv.2305.13943}

\bibitem[{{Curti} {et~al.}(2020){Curti}, {Mannucci}, {Cresci}, \&
  {Maiolino}}]{Curti2020MNRAS.491..944C}
{Curti}, M., {Mannucci}, F., {Cresci}, G., \& {Maiolino}, R. 2020, \mnras, 491,
  944, \dodoi{10.1093/mnras/stz2910}

\bibitem[{{Cushing} {et~al.}(2004){Cushing}, {Vacca}, \&
  {Rayner}}]{Cushing2004PASP..116..362C}
{Cushing}, M.~C., {Vacca}, W.~D., \& {Rayner}, J.~T. 2004, \pasp, 116, 362,
  \dodoi{10.1086/382907}

\bibitem[{{Davis} {et~al.}(2019){Davis}, {Hsiao}, {Ashall}, {Hoeflich},
  {Phillips}, {Marion}, {Kirshner}, {Morrell}, {Sand}, {Burns}, {Contreras},
  {Stritzinger}, {Anderson}, {Baron}, {Diamond}, {Guti{\'e}rrez}, {Hamuy},
  {Holmbo}, {Kasliwal}, {Krisciunas}, {Kumar}, {Lu}, {Pessi}, {Piro}, {Prieto},
  {Shahbandeh}, \& {Suntzeff}}]{Davis2019ApJ...887....4D}
{Davis}, S., {Hsiao}, E.~Y., {Ashall}, C., {et~al.} 2019, \apj, 887, 4,
  \dodoi{10.3847/1538-4357/ab4c40}

\bibitem[{{Deng} {et~al.}(2000){Deng}, {Qiu}, {Hu}, {Hatano}, \&
  {Branch}}]{Deng2000}
{Deng}, J.~S., {Qiu}, Y.~L., {Hu}, J.~Y., {Hatano}, K., \& {Branch}, D. 2000,
  \apj, 540, 452, \dodoi{10.1086/309335}

\bibitem[{{Dessart} {et~al.}(2012){Dessart}, {Hillier}, {Li}, \&
  {Woosley}}]{Dessart2012}
{Dessart}, L., {Hillier}, D.~J., {Li}, C., \& {Woosley}, S. 2012, \mnras, 424,
  2139, \dodoi{10.1111/j.1365-2966.2012.21374.x}

\bibitem[{{Dessart} {et~al.}(2011){Dessart}, {Hillier}, {Livne}, {Yoon},
  {Woosley}, {Waldman}, \& {Langer}}]{Dessart2011MNRAS.414.2985D}
{Dessart}, L., {Hillier}, D.~J., {Livne}, E., {et~al.} 2011, \mnras, 414, 2985,
  \dodoi{10.1111/j.1365-2966.2011.18598.x}

\bibitem[{{Dessart} {et~al.}(2021){Dessart}, {Hillier}, {Sukhbold}, {Woosley},
  \& {Janka}}]{Dessart2021}
{Dessart}, L., {Hillier}, D.~J., {Sukhbold}, T., {Woosley}, S.~E., \& {Janka},
  H.~T. 2021, \aap, 656, A61, \dodoi{10.1051/0004-6361/202141927}

\bibitem[{{Dessart} {et~al.}(2015){Dessart}, {Hillier}, {Woosley}, {Livne},
  {Waldman}, {Yoon}, \& {Langer}}]{Dessart2015}
{Dessart}, L., {Hillier}, D.~J., {Woosley}, S., {et~al.} 2015, \mnras, 453,
  2189, \dodoi{10.1093/mnras/stv1747}

\bibitem[{{Dessart} {et~al.}(2016){Dessart}, {Hillier}, {Woosley}, {Livne},
  {Waldman}, {Yoon}, \& {Langer}}]{Dessart2016}
---. 2016, \mnras, 458, 1618, \dodoi{10.1093/mnras/stw418}

\bibitem[{{Dessart} {et~al.}(2023){Dessart}, {Hillier}, {Woosley}, \&
  {Kuncarayakti}}]{Dessart2023arXiv230612092D}
{Dessart}, L., {Hillier}, D.~J., {Woosley}, S.~E., \& {Kuncarayakti}, H. 2023,
  arXiv e-prints, arXiv:2306.12092, \dodoi{10.48550/arXiv.2306.12092}

\bibitem[{{Dolphin}(2016)}]{Dolphin2016}
{Dolphin}, A. 2016, {DOLPHOT: Stellar photometry}, Astrophysics Source Code
  Library, record ascl:1608.013.
\newblock \doeprint{1608.013}

\bibitem[{{Dong} {et~al.}(2022){Dong}, {Valenti}, {Sand}, {Janzen},
  {Hosseinzadeh}, {Jencson}, {Bostroem}, {Wyatt}, {Meza}, {Lundquist},
  {Pearson}, \& {Andrews}}]{Dong2022TNS}
{Dong}, Y., {Valenti}, S., {Sand}, D.~J., {et~al.} 2022, Transient Name Server
  Discovery Report, 2022-448, 1

\bibitem[{{Dressler} {et~al.}(2011){Dressler}, {Bigelow}, {Hare}, {Sutin},
  {Thompson}, {Burley}, {Epps}, {Oemler}, {Bagish}, {Birk}, {Clardy},
  {Gunnels}, {Kelson}, {Shectman}, \& {Osip}}]{Dressler2011}
{Dressler}, A., {Bigelow}, B., {Hare}, T., {et~al.} 2011, \pasp, 123, 288,
  \dodoi{10.1086/658908}

\bibitem[{{Drout} {et~al.}(2011){Drout}, {Soderberg}, {Gal-Yam}, {Cenko},
  {Fox}, {Leonard}, {Sand}, {Moon}, {Arcavi}, \& {Green}}]{Drout2011}
{Drout}, M.~R., {Soderberg}, A.~M., {Gal-Yam}, A., {et~al.} 2011, \apj, 741,
  97, \dodoi{10.1088/0004-637X/741/2/97}

\bibitem[{{Drout} {et~al.}(2016){Drout}, {Milisavljevic}, {Parrent},
  {Margutti}, {Kamble}, {Soderberg}, {Challis}, {Chornock}, {Fong}, {Frank},
  {Gehrels}, {Graham}, {Hsiao}, {Itagaki}, {Kasliwal}, {Kirshner}, {Macomb},
  {Marion}, {Norris}, \& {Phillips}}]{Drout2016}
{Drout}, M.~R., {Milisavljevic}, D., {Parrent}, J., {et~al.} 2016, \apj, 821,
  57, \dodoi{10.3847/0004-637X/821/1/57}

\bibitem[{{Eldridge} {et~al.}(2015){Eldridge}, {Fraser}, {Maund}, \&
  {Smartt}}]{Eldridge2015}
{Eldridge}, J.~J., {Fraser}, M., {Maund}, J.~R., \& {Smartt}, S.~J. 2015,
  \mnras, 446, 2689, \dodoi{10.1093/mnras/stu2197}

\bibitem[{{Eldridge} {et~al.}(2013){Eldridge}, {Fraser}, {Smartt}, {Maund}, \&
  {Crockett}}]{Eldridge2013}
{Eldridge}, J.~J., {Fraser}, M., {Smartt}, S.~J., {Maund}, J.~R., \&
  {Crockett}, R.~M. 2013, \mnras, 436, 774, \dodoi{10.1093/mnras/stt1612}

\bibitem[{{Eldridge} \& {Tout}(2004)}]{Eldridge2004}
{Eldridge}, J.~J., \& {Tout}, C.~A. 2004, \mnras, 353, 87,
  \dodoi{10.1111/j.1365-2966.2004.08041.x}

\bibitem[{{Elmhamdi} {et~al.}(2006){Elmhamdi}, {Danziger}, {Branch},
  {Leibundgut}, {Baron}, \& {Kirshner}}]{Elmhamdi2006}
{Elmhamdi}, A., {Danziger}, I.~J., {Branch}, D., {et~al.} 2006, \aap, 450, 305,
  \dodoi{10.1051/0004-6361:20054366}

\bibitem[{{Elmhamdi} {et~al.}(2004){Elmhamdi}, {Danziger}, {Cappellaro}, {Della
  Valle}, {Gouiffes}, {Phillips}, \& {Turatto}}]{Elmhamdi2004}
{Elmhamdi}, A., {Danziger}, I.~J., {Cappellaro}, E., {et~al.} 2004, \aap, 426,
  963, \dodoi{10.1051/0004-6361:20041318}

\bibitem[{{Ergon} {et~al.}(2014){Ergon}, {Sollerman}, {Fraser}, {Pastorello},
  {Taubenberger}, {Elias-Rosa}, {Bersten}, {Jerkstrand}, {Benetti},
  {Botticella}, {Fransson}, {Harutyunyan}, {Kotak}, {Smartt}, {Valenti},
  {Bufano}, {Cappellaro}, {Fiaschi}, {Howell}, {Kankare}, {Magill}, {Mattila},
  {Maund}, {Naves}, {Ochner}, {Ruiz}, {Smith}, {Tomasella}, \&
  {Turatto}}]{Ergon2014A&A...562A..17E}
{Ergon}, M., {Sollerman}, J., {Fraser}, M., {et~al.} 2014, \aap, 562, A17,
  \dodoi{10.1051/0004-6361/201321850}

\bibitem[{{Ertl} {et~al.}(2020){Ertl}, {Woosley}, {Sukhbold}, \&
  {Janka}}]{Ertl2020}
{Ertl}, T., {Woosley}, S.~E., {Sukhbold}, T., \& {Janka}, H.~T. 2020, \apj,
  890, 51, \dodoi{10.3847/1538-4357/ab6458}

\bibitem[{{Fabricant} {et~al.}(2019){Fabricant}, {Fata}, {Epps}, {Gauron},
  {Mueller}, {Zajac}, {Amato}, {Barberis}, {Bergner}, {Brennan}, {Brown},
  {Chilingarian}, {Geary}, {Kradinov}, {McLeod}, {Smith}, \&
  {Woods}}]{Binospec}
{Fabricant}, D., {Fata}, R., {Epps}, H., {et~al.} 2019, \pasp, 131, 075004,
  \dodoi{10.1088/1538-3873/ab1d78}

\bibitem[{{Fang} {et~al.}(2022){Fang}, {Maeda}, {Kuncarayakti}, {Tanaka},
  {Kawabata}, {Hattori}, {Aoki}, {Moriya}, \&
  {Yamanaka}}]{Fang2022ApJ...928..151F}
{Fang}, Q., {Maeda}, K., {Kuncarayakti}, H., {et~al.} 2022, \apj, 928, 151,
  \dodoi{10.3847/1538-4357/ac4f60}

\bibitem[{{Filippenko}(1982)}]{Filippenko1982}
{Filippenko}, A.~V. 1982, \pasp, 94, 715, \dodoi{10.1086/131052}

\bibitem[{{Filippenko}(1988)}]{Filippenk1988}
---. 1988, \aj, 96, 1941, \dodoi{10.1086/114940}

\bibitem[{{Filippenko}(1997)}]{Filippenko1997}
---. 1997, \araa, 35, 309, \dodoi{10.1146/annurev.astro.35.1.309}

\bibitem[{{Filippenko} {et~al.}(2001){Filippenko}, {Li}, {Treffers}, \&
  {Modjaz}}]{Filippenko2001}
{Filippenko}, A.~V., {Li}, W.~D., {Treffers}, R.~R., \& {Modjaz}, M. 2001, in
  Astronomical Society of the Pacific Conference Series, Vol. 246, IAU Colloq.
  183: Small Telescope Astronomy on Global Scales, ed. B.~{Paczynski}, W.-P.
  {Chen}, \& C.~{Lemme}, 121

\bibitem[{{Filippenko} {et~al.}(1993){Filippenko}, {Matheson}, \&
  {Ho}}]{Filippenko1993}
{Filippenko}, A.~V., {Matheson}, T., \& {Ho}, L.~C. 1993, \apjl, 415, L103,
  \dodoi{10.1086/187043}

\bibitem[{{Folatelli} {et~al.}(2014){Folatelli}, {Bersten}, {Kuncarayakti},
  {Olivares Estay}, {Anderson}, {Holmbo}, {Maeda}, {Morrell}, {Nomoto},
  {Pignata}, {Stritzinger}, {Contreras}, {F{\"o}rster}, {Hamuy}, {Phillips},
  {Prieto}, {Valenti}, {Afonso}, {Altenm{\"u}ller}, {Elliott}, {Greiner},
  {Updike}, {Haislip}, {LaCluyze}, {Moore}, \&
  {Reichart}}]{Folatelli2014ApJ...792....7F}
{Folatelli}, G., {Bersten}, M.~C., {Kuncarayakti}, H., {et~al.} 2014, \apj,
  792, 7, \dodoi{10.1088/0004-637X/792/1/7}

\bibitem[{{Folatelli} {et~al.}(2016){Folatelli}, {Van Dyk}, {Kuncarayakti},
  {Maeda}, {Bersten}, {Nomoto}, {Pignata}, {Hamuy}, {Quimby}, {Zheng},
  {Filippenko}, {Clubb}, {Smith}, {Elias-Rosa}, {Foley}, \&
  {Miller}}]{Folatelli2016}
{Folatelli}, G., {Van Dyk}, S.~D., {Kuncarayakti}, H., {et~al.} 2016, \apjl,
  825, L22, \dodoi{10.3847/2041-8205/825/2/L22}

\bibitem[{{Fremling} {et~al.}(2014){Fremling}, {Sollerman}, {Taddia}, {Ergon},
  {Valenti}, {Arcavi}, {Ben-Ami}, {Cao}, {Cenko}, {Filippenko}, {Gal-Yam}, \&
  {Howell}}]{Fremling2014}
{Fremling}, C., {Sollerman}, J., {Taddia}, F., {et~al.} 2014, \aap, 565, A114,
  \dodoi{10.1051/0004-6361/201423884}

\bibitem[{{Fremling} {et~al.}(2018){Fremling}, {Sollerman}, {Kasliwal},
  {Kulkarni}, {Barbarino}, {Ergon}, {Karamehmetoglu}, {Taddia}, {Arcavi},
  {Cenko}, {Clubb}, {De Cia}, {Duggan}, {Filippenko}, {Gal-Yam}, {Graham},
  {Horesh}, {Hosseinzadeh}, {Howell}, {Kuesters}, {Lunnan}, {Matheson},
  {Nugent}, {Perley}, {Quimby}, \& {Saunders}}]{Fremling2018A&A...618A..37F}
{Fremling}, C., {Sollerman}, J., {Kasliwal}, M.~M., {et~al.} 2018, \aap, 618,
  A37, \dodoi{10.1051/0004-6361/201731701}

\bibitem[{{Fryer} {et~al.}(2007){Fryer}, {Mazzali}, {Prochaska}, {Cappellaro},
  {Panaitescu}, {Berger}, {van Putten}, {van den Heuvel}, {Young},
  {Hungerford}, {Rockefeller}, {Yoon}, {Podsiadlowski}, {Nomoto}, {Chevalier},
  {Schmidt}, \& {Kulkarni}}]{Fryer2007}
{Fryer}, C.~L., {Mazzali}, P.~A., {Prochaska}, J., {et~al.} 2007, \pasp, 119,
  1211, \dodoi{10.1086/523768}

\bibitem[{Gal-Yam(2017)}]{Gal-Yam2017}
Gal-Yam, A. 2017, in Handbook of Supernovae (Springer International
  Publishing), 195--237, \dodoi{10.1007/978-3-319-21846-5_35}

\bibitem[{{Galbany} {et~al.}(2018){Galbany}, {Anderson}, {S{\'a}nchez},
  {Kuncarayakti}, {Pedraz}, {Gonz{\'a}lez-Gait{\'a}n}, {Stanishev},
  {Dom{\'\i}nguez}, {Moreno-Raya}, {Wood-Vasey}, {Mour{\~a}o}, {Ponder},
  {Badenes}, {Moll{\'a}}, {L{\'o}pez-S{\'a}nchez}, {Rosales-Ortega},
  {V{\'\i}lchez}, {Garc{\'\i}a-Benito}, \&
  {Marino}}]{Galbany2018ApJ...855..107G}
{Galbany}, L., {Anderson}, J.~P., {S{\'a}nchez}, S.~F., {et~al.} 2018, \apj,
  855, 107, \dodoi{10.3847/1538-4357/aaaf20}

\bibitem[{{Gangopadhyay} {et~al.}(2023){Gangopadhyay}, {Maeda}, {Singh},
  {Nayana}, {Nakaoka}, {Kawabata}, {Taguchi}, {Singh}, {Chandra}, {Ryder},
  {Dastidar}, {Yamanaka}, {Kawabata}, {Alsaberi}, {Dukiya}, {Teja},
  {Ailawadhi}, {Dutta}, {Sahu}, {Moriya}, {Misra}, {Tanaka}, {Chevalier},
  {Tominaga}, {Uno}, {Imazawa}, {Hamada}, {Hori}, \&
  {Isogai}}]{gangopadhyay2023bridging}
{Gangopadhyay}, A., {Maeda}, K., {Singh}, A., {et~al.} 2023, \apj, 957, 100,
  \dodoi{10.3847/1538-4357/acfa94}

\bibitem[{{Gehrels} {et~al.}(2004){Gehrels}, {Chincarini}, {Giommi}, {Mason},
  {Nousek}, {Wells}, {White}, {Barthelmy}, {Burrows}, {Cominsky}, {Hurley},
  {Marshall}, {M{\'e}sz{\'a}ros}, {Roming}, {Angelini}, {Barbier}, {Belloni},
  {Campana}, {Caraveo}, {Chester}, {Citterio}, {Cline}, {Cropper}, {Cummings},
  {Dean}, {Feigelson}, {Fenimore}, {Frail}, {Fruchter}, {Garmire}, {Gendreau},
  {Ghisellini}, {Greiner}, {Hill}, {Hunsberger}, {Krimm}, {Kulkarni}, {Kumar},
  {Lebrun}, {Lloyd-Ronning}, {Markwardt}, {Mattson}, {Mushotzky}, {Norris},
  {Osborne}, {Paczynski}, {Palmer}, {Park}, {Parsons}, {Paul}, {Rees},
  {Reynolds}, {Rhoads}, {Sasseen}, {Schaefer}, {Short}, {Smale}, {Smith},
  {Stella}, {Tagliaferri}, {Takahashi}, {Tashiro}, {Townsley}, {Tueller},
  {Turner}, {Vietri}, {Voges}, {Ward}, {Willingale}, {Zerbi}, \&
  {Zhang}}]{Gehrels2004}
{Gehrels}, N., {Chincarini}, G., {Giommi}, P., {et~al.} 2004, \apj, 611, 1005,
  \dodoi{10.1086/422091}

\bibitem[{{Gilkis} \& {Arcavi}(2022)}]{Gilkis2022}
{Gilkis}, A., \& {Arcavi}, I. 2022, \mnras, 511, 691,
  \dodoi{10.1093/mnras/stac088}

\bibitem[{{Gimeno} {et~al.}(2016){Gimeno}, {Roth}, {Chiboucas}, {Hibon},
  {Boucher}, {White}, {Rippa}, {Labrie}, {Turner}, {Hanna}, {Lazo},
  {P{\'e}rez}, {Rogers}, {Rojas}, {Placco}, \& {Murowinski}}]{gimeno16}
{Gimeno}, G., {Roth}, K., {Chiboucas}, K., {et~al.} 2016, in Society of
  Photo-Optical Instrumentation Engineers (SPIE) Conference Series, Vol. 9908,
  \procspie, 99082S, \dodoi{10.1117/12.2233883}

\bibitem[{{G{\"o}tberg} {et~al.}(2017){G{\"o}tberg}, {de Mink}, \&
  {Groh}}]{Gotberg2017A&A...608A..11G}
{G{\"o}tberg}, Y., {de Mink}, S.~E., \& {Groh}, J.~H. 2017, \aap, 608, A11,
  \dodoi{10.1051/0004-6361/201730472}

\bibitem[{{G{\"o}tberg} {et~al.}(2018){G{\"o}tberg}, {de Mink}, {Groh},
  {Kupfer}, {Crowther}, {Zapartas}, \& {Renzo}}]{Gotberg2018A&A...615A..78G}
{G{\"o}tberg}, Y., {de Mink}, S.~E., {Groh}, J.~H., {et~al.} 2018, \aap, 615,
  A78, \dodoi{10.1051/0004-6361/201732274}

\bibitem[{{Gotberg} {et~al.}(2023){Gotberg}, {Drout}, {Ji}, {Groh}, {Ludwig},
  {Crowther}, {Smith}, {de Koter}, \& {de Mink}}]{Gotberg2023arXiv230700074G}
{Gotberg}, Y., {Drout}, M.~R., {Ji}, A.~P., {et~al.} 2023, arXiv e-prints,
  arXiv:2307.00074, \dodoi{10.48550/arXiv.2307.00074}

\bibitem[{Green {et~al.}(1995)Green, Schmidt, Oey, Wittman, \&
  Hall}]{green1995}
Green, R., Schmidt, G., Oey, S., Wittman, D., \& Hall, P. 1995, Steward
  {{Observatory}} 2.3-m {{Boller}} and {{Chivens Spectrograph Manual}}.
\newblock
  \url{http://james.as.arizona.edu/~psmith/90inch/bcman/html/bcman.html}

\bibitem[{{Hachinger} {et~al.}(2012){Hachinger}, {Mazzali}, {Taubenberger},
  {Hillebrandt}, {Nomoto}, \& {Sauer}}]{Hachinger2012}
{Hachinger}, S., {Mazzali}, P.~A., {Taubenberger}, S., {et~al.} 2012, \mnras,
  422, 70, \dodoi{10.1111/j.1365-2966.2012.20464.x}

\bibitem[{{Hamuy} {et~al.}(2002){Hamuy}, {Maza}, {Pinto}, {Phillips},
  {Suntzeff}, {Blum}, {Olsen}, {Pinfield}, {Ivanov}, {Augusteijn}, {Brillant},
  {Chadid}, {Cuby}, {Doublier}, {Hainaut}, {Le Floc'h}, {Lidman},
  {Petr-Gotzens}, {Pompei}, \& {Vanzi}}]{Hamuy2002}
{Hamuy}, M., {Maza}, J., {Pinto}, P.~A., {et~al.} 2002, \aj, 124, 417,
  \dodoi{10.1086/340968}

\bibitem[{{Hamuy} {et~al.}(2006){Hamuy}, {Folatelli}, {Morrell}, {Phillips},
  {Suntzeff}, {Persson}, {Roth}, {Gonzalez}, {Krzeminski}, {Contreras},
  {Freedman}, {Murphy}, {Madore}, {Wyatt}, {Maza}, {Filippenko}, {Li}, \&
  {Pinto}}]{Hamuy2006PASP..118....2H}
{Hamuy}, M., {Folatelli}, G., {Morrell}, N.~I., {et~al.} 2006, \pasp, 118, 2,
  \dodoi{10.1086/500228}

\bibitem[{{Harkness} {et~al.}(1987){Harkness}, {Wheeler}, {Margon}, {Downes},
  {Kirshner}, {Uomoto}, {Barker}, {Cochran}, {Dinerstein}, {Garnett}, \&
  {Levreault}}]{Harkness1987}
{Harkness}, R.~P., {Wheeler}, J.~C., {Margon}, B., {et~al.} 1987, \apj, 317,
  355, \dodoi{10.1086/165283}

\bibitem[{{Harris} {et~al.}(2020){Harris}, {Millman}, {van der Walt},
  {Gommers}, {Virtanen}, {Cournapeau}, {Wieser}, {Taylor}, {Berg}, {Smith},
  {Kern}, {Picus}, {Hoyer}, {van Kerkwijk}, {Brett}, {Haldane}, {del R{\'\i}o},
  {Wiebe}, {Peterson}, {G{\'e}rard-Marchant}, {Sheppard}, {Reddy}, {Weckesser},
  {Abbasi}, {Gohlke}, \& {Oliphant}}]{2020Natur.585..357H}
{Harris}, C.~R., {Millman}, K.~J., {van der Walt}, S.~J., {et~al.} 2020, at,
  585, 357, \dodoi{10.1038/s41586-020-2649-2}

\bibitem[{{Hiramatsu} {et~al.}(2021){Hiramatsu}, {Howell}, {Moriya},
  {Goldberg}, {Hosseinzadeh}, {Arcavi}, {Anderson}, {Guti{\'e}rrez}, {Burke},
  {McCully}, {Valenti}, {Galbany}, {Fang}, {Maeda}, {Folatelli}, {Hsiao},
  {Morrell}, {Phillips}, {Stritzinger}, {Suntzeff}, {Gromadzki}, {Maguire},
  {M{\"u}ller-Bravo}, \& {Young}}]{Hiramatsu2021ApJ...913...55H}
{Hiramatsu}, D., {Howell}, D.~A., {Moriya}, T.~J., {et~al.} 2021, \apj, 913,
  55, \dodoi{10.3847/1538-4357/abf6d6}

\bibitem[{{Holmbo} {et~al.}(2023){Holmbo}, {Stritzinger}, {Karamehmetoglu},
  {Burns}, {Morrell}, {Ashall}, {Hsiao}, {Galbany}, {Folatelli}, {Phillips},
  {Baron}, {Guti{\'e}rrez}, {Leloudas}, {M{\"u}ller-Bravo}, {Hoeflich},
  {Taddia}, \& {Suntzeff}}]{Holmbo2023A&A...675A..83H}
{Holmbo}, S., {Stritzinger}, M.~D., {Karamehmetoglu}, E., {et~al.} 2023, \aap,
  675, A83, \dodoi{10.1051/0004-6361/202245334}

\bibitem[{{Hook} {et~al.}(2004){Hook}, {J{\o}rgensen}, {Allington-Smith},
  {Davies}, {Metcalfe}, {Murowinski}, \& {Crampton}}]{Hook2004}
{Hook}, I.~M., {J{\o}rgensen}, I., {Allington-Smith}, J.~R., {et~al.} 2004,
  \pasp, 116, 425, \dodoi{10.1086/383624}

\bibitem[{Hosseinzadeh \& Gomez(2020)}]{hosseinzadeh_light_2020}
Hosseinzadeh, G., \& Gomez, S. 2020, Light {{Curve Fitting}}, Zenodo,
  \dodoi{10.5281/zenodo.4312178}

\bibitem[{{Hsiao} {et~al.}(2019){Hsiao}, {Phillips}, {Marion}, {Kirshner},
  {Morrell}, {Sand}, {Burns}, {Contreras}, {Hoeflich}, {Stritzinger},
  {Valenti}, {Anderson}, {Ashall}, {Baltay}, {Baron}, {Banerjee}, {Davis},
  {Diamond}, {Folatelli}, {Freedman}, {F{\"o}rster}, {Galbany}, {Gall},
  {Gonz{\'a}lez-Gait{\'a}n}, {Goobar}, {Hamuy}, {Holmbo}, {Kasliwal},
  {Krisciunas}, {Kumar}, {Lidman}, {Lu}, {Nugent}, {Perlmutter}, {Persson},
  {Piro}, {Rabinowitz}, {Roth}, {Ryder}, {Schmidt}, {Shahbandeh}, {Suntzeff},
  {Taddia}, {Uddin}, \& {Wang}}]{Hsiao2019PASP..131a4002H}
{Hsiao}, E.~Y., {Phillips}, M.~M., {Marion}, G.~H., {et~al.} 2019, \pasp, 131,
  014002, \dodoi{10.1088/1538-3873/aae961}

\bibitem[{{Hunter} {et~al.}(2009){Hunter}, {Valenti}, {Kotak}, {Meikle},
  {Taubenberger}, {Pastorello}, {Benetti}, {Stanishev}, {Smartt}, {Trundle},
  {Arkharov}, {Bufano}, {Cappellaro}, {Di Carlo}, {Dolci}, {Elias-Rosa},
  {Frandsen}, {Fynbo}, {Hopp}, {Larionov}, {Laursen}, {Mazzali}, {Navasardyan},
  {Ries}, {Riffeser}, {Rizzi}, {Tsvetkov}, {Turatto}, \&
  {Wilke}}]{Hunter2009A&A...508..371H}
{Hunter}, D.~J., {Valenti}, S., {Kotak}, R., {et~al.} 2009, \aap, 508, 371,
  \dodoi{10.1051/0004-6361/200912896}

\bibitem[{{Hunter}(2007)}]{Hunter2007}
{Hunter}, J.~D. 2007, Computing in Science and Engineering, 9, 90,
  \dodoi{10.1109/MCSE.2007.55}

\bibitem[{{James} \& {Baron}(2010)}]{James2010ApJ...718..957J}
{James}, S., \& {Baron}, E. 2010, \apj, 718, 957,
  \dodoi{10.1088/0004-637X/718/2/957}

\bibitem[{{Jerkstrand} {et~al.}(2015){Jerkstrand}, {Ergon}, {Smartt},
  {Fransson}, {Sollerman}, {Taubenberger}, {Bersten}, \&
  {Spyromilio}}]{Jerkstrand2015}
{Jerkstrand}, A., {Ergon}, M., {Smartt}, S.~J., {et~al.} 2015, \aap, 573, A12,
  \dodoi{10.1051/0004-6361/201423983}

\bibitem[{{Jerkstrand} {et~al.}(2012){Jerkstrand}, {Fransson}, {Maguire},
  {Smartt}, {Ergon}, \& {Spyromilio}}]{Jerkstrand2012}
{Jerkstrand}, A., {Fransson}, C., {Maguire}, K., {et~al.} 2012, \aap, 546, A28,
  \dodoi{10.1051/0004-6361/201219528}

\bibitem[{{Kasen} \& {Bildsten}(2010)}]{Kasen2010}
{Kasen}, D., \& {Bildsten}, L. 2010, \apj, 717, 245,
  \dodoi{10.1088/0004-637X/717/1/245}

\bibitem[{{Kilpatrick} {et~al.}(2017){Kilpatrick}, {Foley}, {Abramson}, {Pan},
  {Lu}, {Williams}, {Treu}, {Siebert}, {Fassnacht}, \& {Max}}]{Kilpatrick2017}
{Kilpatrick}, C.~D., {Foley}, R.~J., {Abramson}, L.~E., {et~al.} 2017, \mnras,
  465, 4650, \dodoi{10.1093/mnras/stw3082}

\bibitem[{{Kilpatrick} {et~al.}(2021){Kilpatrick}, {Drout}, {Auchettl},
  {Dimitriadis}, {Foley}, {Jones}, {DeMarchi}, {French}, {Gall}, {Hjorth},
  {Jacobson-Gal{\'a}n}, {Margutti}, {Piro}, {Ramirez-Ruiz}, {Rest}, \&
  {Rojas-Bravo}}]{Kilpatrick2021}
{Kilpatrick}, C.~D., {Drout}, M.~R., {Auchettl}, K., {et~al.} 2021, \mnras,
  504, 2073, \dodoi{10.1093/mnras/stab838}

\bibitem[{{Kinney} {et~al.}(1996){Kinney}, {Calzetti}, {Bohlin}, {McQuade},
  {Storchi-Bergmann}, \& {Schmitt}}]{Kinney1996ApJ...467...38K}
{Kinney}, A.~L., {Calzetti}, D., {Bohlin}, R.~C., {et~al.} 1996, \apj, 467, 38,
  \dodoi{10.1086/177583}

\bibitem[{{Kuncarayakti} {et~al.}(2015){Kuncarayakti}, {Maeda}, {Bersten},
  {Folatelli}, {Morrell}, {Hsiao}, {Gonz{\'a}lez-Gait{\'a}n}, {Anderson},
  {Hamuy}, {de Jaeger}, {Guti{\'e}rrez}, \& {Kawabata}}]{Kuncarayakti2015}
{Kuncarayakti}, H., {Maeda}, K., {Bersten}, M.~C., {et~al.} 2015, \aap, 579,
  A95, \dodoi{10.1051/0004-6361/201425604}

\bibitem[{{Labrie} {et~al.}(2019){Labrie}, {Anderson}, {C{\'a}rdenes},
  {Simpson}, \& {Turner}}]{Labrie2019}
{Labrie}, K., {Anderson}, K., {C{\'a}rdenes}, R., {Simpson}, C., \& {Turner},
  J. E.~H. 2019, in Astronomical Society of the Pacific Conference Series, Vol.
  523, Astronomical Data Analysis Software and Systems XXVII, ed. P.~J.
  {Teuben}, M.~W. {Pound}, B.~A. {Thomas}, \& E.~M. {Warner}, 321

\bibitem[{{Lantz} {et~al.}(2004){Lantz}, {Aldering}, {Antilogus}, {Bonnaud},
  {Capoani}, {Castera}, {Copin}, {Dubet}, {Gangler}, {Henault}, {Lemonnier},
  {Pain}, {Pecontal}, {Pecontal}, \& {Smadja}}]{Lantz2004}
{Lantz}, B., {Aldering}, G., {Antilogus}, P., {et~al.} 2004, in Society of
  Photo-Optical Instrumentation Engineers (SPIE) Conference Series, Vol. 5249,
  Optical Design and Engineering, ed. L.~{Mazuray}, P.~J. {Rogers}, \&
  R.~{Wartmann}, 146--155, \dodoi{10.1117/12.512493}

\bibitem[{{Li} {et~al.}(2003){Li}, {Filippenko}, {Chornock}, \& {Jha}}]{Li2003}
{Li}, W., {Filippenko}, A.~V., {Chornock}, R., \& {Jha}, S. 2003, \pasp, 115,
  844, \dodoi{10.1086/376432}

\bibitem[{{Liu} {et~al.}(2016){Liu}, {Modjaz}, {Bianco}, \& {Graur}}]{Liu2016}
{Liu}, Y.-Q., {Modjaz}, M., {Bianco}, F.~B., \& {Graur}, O. 2016, \apj, 827,
  90, \dodoi{10.3847/0004-637X/827/2/90}

\bibitem[{{Lyman} {et~al.}(2014){Lyman}, {Bersier}, \& {James}}]{Lyman2014}
{Lyman}, J.~D., {Bersier}, D., \& {James}, P.~A. 2014, \mnras, 437, 3848,
  \dodoi{10.1093/mnras/stt2187}

\bibitem[{{Lyman} {et~al.}(2016){Lyman}, {Bersier}, {James}, {Mazzali},
  {Eldridge}, {Fraser}, \& {Pian}}]{Lyman2016}
{Lyman}, J.~D., {Bersier}, D., {James}, P.~A., {et~al.} 2016, \mnras, 457, 328,
  \dodoi{10.1093/mnras/stv2983}

\bibitem[{{Maeda} {et~al.}(2003){Maeda}, {Mazzali}, {Deng}, {Nomoto}, {Yoshii},
  {Tomita}, \& {Kobayashi}}]{Maeda2003}
{Maeda}, K., {Mazzali}, P.~A., {Deng}, J., {et~al.} 2003, \apj, 593, 931,
  \dodoi{10.1086/376591}

\bibitem[{{Maeda} {et~al.}(2007){Maeda}, {Tanaka}, {Nomoto}, {Tominaga},
  {Kawabata}, {Mazzali}, {Umeda}, {Suzuki}, \& {Hattori}}]{Maeda2007}
{Maeda}, K., {Tanaka}, M., {Nomoto}, K., {et~al.} 2007, \apj, 666, 1069,
  \dodoi{10.1086/520054}

\bibitem[{{Maeda} {et~al.}(2008){Maeda}, {Kawabata}, {Mazzali}, {Tanaka},
  {Valenti}, {Nomoto}, {Hattori}, {Deng}, {Pian}, {Taubenberger}, {Iye},
  {Matheson}, {Filippenko}, {Aoki}, {Kosugi}, {Ohyama}, {Sasaki}, \&
  {Takata}}]{Maeda2008}
{Maeda}, K., {Kawabata}, K., {Mazzali}, P.~A., {et~al.} 2008, Science, 319,
  1220, \dodoi{10.1126/science.1149437}

\bibitem[{{Marion} {et~al.}(2014){Marion}, {Vinko}, {Kirshner}, {Foley},
  {Berlind}, {Bieryla}, {Bloom}, {Calkins}, {Challis}, {Chevalier}, {Chornock},
  {Culliton}, {Curtis}, {Esquerdo}, {Everett}, {Falco}, {France}, {Fransson},
  {Friedman}, {Garnavich}, {Leibundgut}, {Meyer}, {Smith}, {Soderberg},
  {Sollerman}, {Starr}, {Szklenar}, {Takats}, \& {Wheeler}}]{Marion2014}
{Marion}, G.~H., {Vinko}, J., {Kirshner}, R.~P., {et~al.} 2014, \apj, 781, 69,
  \dodoi{10.1088/0004-637X/781/2/69}

\bibitem[{{Maund} {et~al.}(2004){Maund}, {Smartt}, {Kudritzki},
  {Podsiadlowski}, \& {Gilmore}}]{Maund2004}
{Maund}, J.~R., {Smartt}, S.~J., {Kudritzki}, R.~P., {Podsiadlowski}, P., \&
  {Gilmore}, G.~F. 2004, \nat, 427, 129, \dodoi{10.1038/nature02161}

\bibitem[{{Maund} {et~al.}(2011){Maund}, {Fraser}, {Ergon}, {Pastorello},
  {Smartt}, {Sollerman}, {Benetti}, {Botticella}, {Bufano}, {Danziger},
  {Kotak}, {Magill}, {Stephens}, \& {Valenti}}]{Maund2011}
{Maund}, J.~R., {Fraser}, M., {Ergon}, M., {et~al.} 2011, \apjl, 739, L37,
  \dodoi{10.1088/2041-8205/739/2/L37}

\bibitem[{{Maurer} {et~al.}(2010){Maurer}, {Mazzali}, {Taubenberger}, \&
  {Hachinger}}]{Maurer2010}
{Maurer}, I., {Mazzali}, P.~A., {Taubenberger}, S., \& {Hachinger}, S. 2010,
  \mnras, 409, 1441, \dodoi{10.1111/j.1365-2966.2010.17186.x}

\bibitem[{{Mazzali} {et~al.}(2005){Mazzali}, {Kawabata}, {Maeda}, {Nomoto},
  {Filippenko}, {Ramirez-Ruiz}, {Benetti}, {Pian}, {Deng}, {Tominaga},
  {Ohyama}, {Iye}, {Foley}, {Matheson}, {Wang}, \& {Gal-Yam}}]{Mazzali2005}
{Mazzali}, P.~A., {Kawabata}, K.~S., {Maeda}, K., {et~al.} 2005, Science, 308,
  1284, \dodoi{10.1126/science.1111384}

\bibitem[{{McLeod} {et~al.}(2012){McLeod}, {Fabricant}, {Nystrom}, {McCracken},
  {Amato}, {Bergner}, {Brown}, {Burke}, {Chilingarian}, {Conroy}, {Curley},
  {Furesz}, {Geary}, {Hertz}, {Holwell}, {Matthews}, {Norton}, {Park}, {Roll},
  {Zajac}, {Epps}, \& {Martini}}]{McLeod2012}
{McLeod}, B., {Fabricant}, D., {Nystrom}, G., {et~al.} 2012, \pasp, 124, 1318,
  \dodoi{10.1086/669044}

\bibitem[{{Meynet} \& {Maeder}(2005)}]{Meynet2005}
{Meynet}, G., \& {Maeder}, A. 2005, \aap, 429, 581,
  \dodoi{10.1051/0004-6361:20047106}

\bibitem[{{Milisavljevic} {et~al.}(2010){Milisavljevic}, {Fesen}, {Gerardy},
  {Kirshner}, \& {Challis}}]{Milisavljevic2010}
{Milisavljevic}, D., {Fesen}, R.~A., {Gerardy}, C.~L., {Kirshner}, R.~P., \&
  {Challis}, P. 2010, \apj, 709, 1343, \dodoi{10.1088/0004-637X/709/2/1343}

\bibitem[{{Milisavljevic} {et~al.}(2013){Milisavljevic}, {Margutti},
  {Soderberg}, {Pignata}, {Chomiuk}, {Fesen}, {Bufano}, {Sanders}, {Parrent},
  {Parker}, {Mazzali}, {Pian}, {Pickering}, {Buckley}, {Crawford}, {Gulbis},
  {Hettlage}, {Hooper}, {Nordsieck}, {O'Donoghue}, {Husser}, {Potter},
  {Kniazev}, {Kotze}, {Romero-Colmenero}, {Vaisanen}, {Wolf}, {Bietenholz},
  {Bartel}, {Fransson}, {Walker}, {Brunthaler}, {Chakraborti}, {Levesque},
  {MacFadyen}, {Drescher}, {Bock}, {Marples}, {Anderson}, {Benetti},
  {Reichart}, \& {Ivarsen}}]{Milisavljevic2013}
{Milisavljevic}, D., {Margutti}, R., {Soderberg}, A.~M., {et~al.} 2013, \apj,
  767, 71, \dodoi{10.1088/0004-637X/767/1/71}

\bibitem[{Miller \& Stone(1994)}]{miller1994}
Miller, J.~S., \& Stone, R. P.~S. 1994, The {{Kast}} Double Spectograph, Lick
  {{Observatory Tech}}. {{Rep}}. No.~66 ({Santa Cruz}: {Lick Observatory})

\bibitem[{{Modjaz} {et~al.}(2019){Modjaz}, {Guti{\'e}rrez}, \&
  {Arcavi}}]{Modjaz2019NatAs...3..717M}
{Modjaz}, M., {Guti{\'e}rrez}, C.~P., \& {Arcavi}, I. 2019, Nature Astronomy,
  3, 717, \dodoi{10.1038/s41550-019-0856-2}

\bibitem[{{Modjaz} {et~al.}(2008){Modjaz}, {Kirshner}, {Blondin}, {Challis}, \&
  {Matheson}}]{Modjaz2008}
{Modjaz}, M., {Kirshner}, R.~P., {Blondin}, S., {Challis}, P., \& {Matheson},
  T. 2008, \apjl, 687, L9, \dodoi{10.1086/593135}

\bibitem[{{Modjaz} {et~al.}(2009){Modjaz}, {Li}, {Butler}, {Chornock},
  {Perley}, {Blondin}, {Bloom}, {Filippenko}, {Kirshner}, {Kocevski},
  {Poznanski}, {Hicken}, {Foley}, {Stringfellow}, {Berlind}, {Barrado y
  Navascues}, {Blake}, {Bouy}, {Brown}, {Challis}, {Chen}, {de Vries},
  {Dufour}, {Falco}, {Friedman}, {Ganeshalingam}, {Garnavich}, {Holden},
  {Illingworth}, {Lee}, {Liebert}, {Marion}, {Olivier}, {Prochaska},
  {Silverman}, {Smith}, {Starr}, {Steele}, {Stockton}, {Williams}, \&
  {Wood-Vasey}}]{Modjaz2009}
{Modjaz}, M., {Li}, W., {Butler}, N., {et~al.} 2009, \apj, 702, 226,
  \dodoi{10.1088/0004-637X/702/1/226}

\bibitem[{{Mokiem} {et~al.}(2007){Mokiem}, {de Koter}, {Vink}, {Puls}, {Evans},
  {Smartt}, {Crowther}, {Herrero}, {Langer}, {Lennon}, {Najarro}, \&
  {Villamariz}}]{Mokiem2007A&A...473..603M}
{Mokiem}, M.~R., {de Koter}, A., {Vink}, J.~S., {et~al.} 2007, \aap, 473, 603,
  \dodoi{10.1051/0004-6361:20077545}

\bibitem[{{Monard}(2006)}]{Monard2006}
{Monard}, L.~A.~G. 2006, \iaucirc, 8666, 2

\bibitem[{{Munari} \& {Zwitter}(1997)}]{Munari1997}
{Munari}, U., \& {Zwitter}, T. 1997, \aap, 318, 269

\bibitem[{{Nadyozhin}(1994)}]{Nadyozhin1994}
{Nadyozhin}, D.~K. 1994, \apjs, 92, 527, \dodoi{10.1086/192008}

\bibitem[{{Nomoto} {et~al.}(2006){Nomoto}, {Tominaga}, {Umeda}, {Kobayashi}, \&
  {Maeda}}]{Nomoto2006}
{Nomoto}, K., {Tominaga}, N., {Umeda}, H., {Kobayashi}, C., \& {Maeda}, K.
  2006, \nphysa, 777, 424, \dodoi{10.1016/j.nuclphysa.2006.05.008}

\bibitem[{{Oke} {et~al.}(1995){Oke}, {Cohen}, {Carr}, {Cromer}, {Dingizian},
  {Harris}, {Labrecque}, {Lucinio}, {Schaal}, {Epps}, \& {Miller}}]{Oke1995}
{Oke}, J.~B., {Cohen}, J.~G., {Carr}, M., {et~al.} 1995, \pasp, 107, 375,
  \dodoi{10.1086/133562}

\bibitem[{{Parrent} {et~al.}(2007){Parrent}, {Branch}, {Troxel}, {Casebeer},
  {Jeffery}, {Ketchum}, {Baron}, {Serduke}, \& {Filippenko}}]{Parrent2007}
{Parrent}, J., {Branch}, D., {Troxel}, M.~A., {et~al.} 2007, \pasp, 119, 135,
  \dodoi{10.1086/512494}

\bibitem[{{Pastorello} {et~al.}(2008){Pastorello}, {Kasliwal}, {Crockett},
  {Valenti}, {Arbour}, {Itagaki}, {Kaspi}, {Gal-Yam}, {Smartt}, {Griffith},
  {Maguire}, {Ofek}, {Seymour}, {Stern}, \& {Wiethoff}}]{Pastorello2008}
{Pastorello}, A., {Kasliwal}, M.~M., {Crockett}, R.~M., {et~al.} 2008, \mnras,
  389, 955, \dodoi{10.1111/j.1365-2966.2008.13618.x}

\bibitem[{{Pettini} \& {Pagel}(2004)}]{Pettini2004MNRAS.348L..59P}
{Pettini}, M., \& {Pagel}, B. E.~J. 2004, \mnras, 348, L59,
  \dodoi{10.1111/j.1365-2966.2004.07591.x}

\bibitem[{{Podsiadlowski} {et~al.}(1992){Podsiadlowski}, {Joss}, \&
  {Hsu}}]{Podsiadlowski1992}
{Podsiadlowski}, P., {Joss}, P.~C., \& {Hsu}, J.~J.~L. 1992, \apj, 391, 246,
  \dodoi{10.1086/171341}

\bibitem[{{Pogge} {et~al.}(2010){Pogge}, {Atwood}, {Brewer}, {Byard},
  {Derwent}, {Gonzalez}, {Martini}, {Mason}, {O'Brien}, {Osmer}, {Pappalardo},
  {Steinbrecher}, {Teiga}, \& {Zhelem}}]{MODS}
{Pogge}, R.~W., {Atwood}, B., {Brewer}, D.~F., {et~al.} 2010, in Society of
  Photo-Optical Instrumentation Engineers (SPIE) Conference Series, Vol. 7735,
  Ground-based and Airborne Instrumentation for Astronomy III, ed. I.~S.
  {McLean}, S.~K. {Ramsay}, \& H.~{Takami}, 77350A, \dodoi{10.1117/12.857215}

\bibitem[{{Poznanski} {et~al.}(2012){Poznanski}, {Prochaska}, \&
  {Bloom}}]{Poznanski2012}
{Poznanski}, D., {Prochaska}, J.~X., \& {Bloom}, J.~S. 2012, \mnras, 426, 1465,
  \dodoi{10.1111/j.1365-2966.2012.21796.x}

\bibitem[{{Prentice} \& {Mazzali}(2017)}]{Prentice2017}
{Prentice}, S.~J., \& {Mazzali}, P.~A. 2017, \mnras, 469, 2672,
  \dodoi{10.1093/mnras/stx980}

\bibitem[{{Prentice} {et~al.}(2019){Prentice}, {Ashall}, {James}, {Short},
  {Mazzali}, {Bersier}, {Crowther}, {Barbarino}, {Chen}, {Copperwheat},
  {Darnley}, {Denneau}, {Elias-Rosa}, {Fraser}, {Galbany}, {Gal-Yam},
  {Harmanen}, {Howell}, {Hosseinzadeh}, {Inserra}, {Kankare}, {Karamehmetoglu},
  {Lamb}, {Limongi}, {Maguire}, {McCully}, {Olivares E}, {Piascik}, {Pignata},
  {Reichart}, {Rest}, {Reynolds}, {Rodr{\'\i}guez}, {Saario}, {Schulze},
  {Smartt}, {Smith}, {Sollerman}, {Stalder}, {Sullivan}, {Taddia}, {Valenti},
  {Vergani}, {Williams}, \& {Young}}]{Prentice2019MNRAS.485.1559P}
{Prentice}, S.~J., {Ashall}, C., {James}, P.~A., {et~al.} 2019, \mnras, 485,
  1559, \dodoi{10.1093/mnras/sty3399}

\bibitem[{{Prochaska} {et~al.}(2020{\natexlab{a}}){Prochaska}, {Hennawi},
  {Westfall}, {Cooke}, {Wang}, {Hsyu}, {Davies}, \&
  {Farina}}]{pypeit:joss_arXiv}
{Prochaska}, J.~X., {Hennawi}, J.~F., {Westfall}, K.~B., {et~al.}
  2020{\natexlab{a}}, arXiv e-prints, arXiv:2005.06505.
\newblock \doarXiv{2005.06505}

\bibitem[{{Prochaska} {et~al.}(2020{\natexlab{b}}){Prochaska}, {Hennawi},
  {Cooke}, {Westfall}, {Wang}, {EmAstro}, {Tiffanyhsyu}, {Wasserman},
  {Villaume}, {Marijana777}, {Schindler}, {Young}, {Simha}, {Wilde}, {Tejos},
  {Isbell}, {Fl{\"o}rs}, {Sandford}, {Vasovi{\'c}}, {Betts}, \&
  {Holden}}]{pypeit:zenodo}
{Prochaska}, J.~X., {Hennawi}, J., {Cooke}, R., {et~al.} 2020{\natexlab{b}},
  {pypeit/PypeIt: Release 1.0.0}, v1.0.0,  Zenodo,
  \dodoi{10.5281/zenodo.3743493}

\bibitem[{Prochaska {et~al.}(2020)Prochaska, Hennawi, Westfall, Cooke, Wang,
  Hsyu, Davies, Farina, \& Pelliccia}]{pypeit:joss_pub}
Prochaska, J.~X., Hennawi, J.~F., Westfall, K.~B., {et~al.} 2020, Journal of
  Open Source Software, 5, 2308, \dodoi{10.21105/joss.02308}

\bibitem[{{Reichart} {et~al.}(2005){Reichart}, {Nysewander}, {Moran},
  {Bartelme}, {Bayliss}, {Foster}, {Clemens}, {Price}, {Evans}, {Salmonson},
  {Trammell}, {Carney}, {Keohane}, \& {Gotwals}}]{Reichart2005}
{Reichart}, D., {Nysewander}, M., {Moran}, J., {et~al.} 2005, Nuovo Cimento C
  Geophysics Space Physics C, 28, 767, \dodoi{10.1393/ncc/i2005-10149-6}

\bibitem[{{Sahu} {et~al.}(2011){Sahu}, {Gurugubelli}, {Anupama}, \&
  {Nomoto}}]{Sahu2011}
{Sahu}, D.~K., {Gurugubelli}, U.~K., {Anupama}, G.~C., \& {Nomoto}, K. 2011,
  \mnras, 413, 2583, \dodoi{10.1111/j.1365-2966.2011.18326.x}

\bibitem[{{Salpeter}(1955)}]{Salpeter1955ApJ...121..161S}
{Salpeter}, E.~E. 1955, \apj, 121, 161, \dodoi{10.1086/145971}

\bibitem[{{Schlafly} \& {Finkbeiner}(2011)}]{Schlafly2011}
{Schlafly}, E.~F., \& {Finkbeiner}, D.~P. 2011, \apj, 737, 103,
  \dodoi{10.1088/0004-637X/737/2/103}

\bibitem[{Schlafly {et~al.}(2010)Schlafly, Finkbeiner, Schlegel, Juri{\'c},
  Ivezi{\'c}, Gibson, Knapp, \& Weaver}]{schlafly_blue_2010}
Schlafly, E.~F., Finkbeiner, D.~P., Schlegel, D.~J., {et~al.} 2010, ApJ, 725,
  1175, \dodoi{10.1088/0004-637X/725/1/1175}

\bibitem[{{Science Software Branch at STScI}(2012)}]{2012ascl.soft07011S}
{Science Software Branch at STScI}. 2012, {PyRAF: Python alternative for IRAF},
  Astrophysics Source Code Library, record ascl:1207.011.
\newblock \doeprint{1207.011}

\bibitem[{{Shahbandeh} {et~al.}(2022){Shahbandeh}, {Hsiao}, {Ashall}, {Teffs},
  {Hoeflich}, {Morrell}, {Phillips}, {Anderson}, {Baron}, {Burns}, {Contreras},
  {Davis}, {Diamond}, {Folatelli}, {Galbany}, {Gall}, {Hachinger}, {Holmbo},
  {Karamehmetoglu}, {Kasliwal}, {Kirshner}, {Krisciunas}, {Kumar}, {Lu},
  {Marion}, {Mazzali}, {Piro}, {Sand}, {Stritzinger}, {Suntzeff}, {Taddia}, \&
  {Uddin}}]{Shahbandeh2022}
{Shahbandeh}, M., {Hsiao}, E.~Y., {Ashall}, C., {et~al.} 2022, \apj, 925, 175,
  \dodoi{10.3847/1538-4357/ac4030}

\bibitem[{{Shigeyama} {et~al.}(1990){Shigeyama}, {Nomoto}, {Tsujimoto}, \&
  {Hashimoto}}]{Shigeyama1990}
{Shigeyama}, T., {Nomoto}, K., {Tsujimoto}, T., \& {Hashimoto}, M.-A. 1990,
  \apjl, 361, L23, \dodoi{10.1086/185818}

\bibitem[{{Shingles} {et~al.}(2021){Shingles}, {Smith}, {Young}, {Smartt},
  {Tonry}, {Denneau}, {Heinze}, {Weiland}, {Flewelling}, {Stalder},
  {Clocchiatti}, {F{\"o}rster}, {Pignata}, {Rest}, {Anderson}, {Stubbs}, \&
  {Erasmus}}]{Shingles2021}
{Shingles}, L., {Smith}, K.~W., {Young}, D.~R., {et~al.} 2021, Transient Name
  Server AstroNote, 7, 1

\bibitem[{{Silverman} {et~al.}(2012){Silverman}, {Foley}, {Filippenko},
  {Ganeshalingam}, {Barth}, {Chornock}, {Griffith}, {Kong}, {Lee}, {Leonard},
  {Matheson}, {Miller}, {Steele}, {Barris}, {Bloom}, {Cobb}, {Coil},
  {Desroches}, {Gates}, {Ho}, {Jha}, {Kandrashoff}, {Li}, {Mandel}, {Modjaz},
  {Moore}, {Mostardi}, {Papenkova}, {Park}, {Perley}, {Poznanski}, {Reuter},
  {Scala}, {Serduke}, {Shields}, {Swift}, {Tonry}, {Van Dyk}, {Wang}, \&
  {Wong}}]{Silverman2012}
{Silverman}, J.~M., {Foley}, R.~J., {Filippenko}, A.~V., {et~al.} 2012, \mnras,
  425, 1789, \dodoi{10.1111/j.1365-2966.2012.21270.x}

\bibitem[{{Simcoe} {et~al.}(2013){Simcoe}, {Burgasser}, {Schechter}, {Fishner},
  {Bernstein}, {Bigelow}, {Pipher}, {Forrest}, {McMurtry}, {Smith}, \&
  {Bochanski}}]{Simcoe2013}
{Simcoe}, R.~A., {Burgasser}, A.~J., {Schechter}, P.~L., {et~al.} 2013, \pasp,
  125, 270, \dodoi{10.1086/670241}

\bibitem[{{Smith} {et~al.}(2020){Smith}, {Smartt}, {Young}, {Tonry}, {Denneau},
  {Flewelling}, {Heinze}, {Weiland}, {Stalder}, {Rest}, {Stubbs}, {Anderson},
  {Chen}, {Clark}, {Do}, {F{\"o}rster}, {Fulton}, {Gillanders}, {McBrien},
  {O'Neill}, {Srivastav}, \& {Wright}}]{Smith2020}
{Smith}, K.~W., {Smartt}, S.~J., {Young}, D.~R., {et~al.} 2020, \pasp, 132,
  085002, \dodoi{10.1088/1538-3873/ab936e}

\bibitem[{{Smith} {et~al.}(2006){Smith}, {Nordsieck}, {Burgh}, {Percival},
  {Williams}, {O'Donohue}, {O'Connor}, \& {Schier}}]{smith2006}
{Smith}, M.~P., {Nordsieck}, K.~H., {Burgh}, E.~B., {et~al.} 2006, in Society
  of Photo-Optical Instrumentation Engineers (SPIE) Conference Series, Vol.
  6269, Society of Photo-Optical Instrumentation Engineers (SPIE) Conference
  Series, ed. I.~S. {McLean} \& M.~{Iye}, 62692A, \dodoi{10.1117/12.672415}

\bibitem[{{Smith} {et~al.}(2011){Smith}, {Li}, {Filippenko}, \&
  {Chornock}}]{Smith2011}
{Smith}, N., {Li}, W., {Filippenko}, A.~V., \& {Chornock}, R. 2011, \mnras,
  412, 1522, \dodoi{10.1111/j.1365-2966.2011.17229.x}

\bibitem[{{Stahl} {et~al.}(2019){Stahl}, {Zheng}, {de Jaeger}, {Filippenko},
  {Bigley}, {Blanchard}, {Blanchard}, {Brink}, {Cargill}, {Casper}, {Channa},
  {Choi}, {Choksi}, {Chu}, {Clubb}, {Cohen}, {Ellison}, {Falcon}, {Fazeli},
  {Fuller}, {Ganeshalingam}, {Gates}, {Gould}, {Halevi}, {Hayakawa},
  {Hestenes}, {Jeffers}, {Joubert}, {Kandrashoff}, {Kim}, {Kim}, {Kislak},
  {Kleiser}, {Kong}, {de Kouchkovsky}, {Krishnan}, {Kumar}, {Leja}, {Leonard},
  {Li}, {Li}, {Lu}, {Mason}, {Molloy}, {Pina}, {Rex}, {Ross}, {Stegman},
  {Tang}, {Thrasher}, {Wang}, {Wilkins}, {Yuk}, {Yunus}, \&
  {Zhang}}]{Stahl2019}
{Stahl}, B.~E., {Zheng}, W., {de Jaeger}, T., {et~al.} 2019, \mnras, 490, 3882,
  \dodoi{10.1093/mnras/stz2742}

\bibitem[{{Stetson}(1987)}]{Stetson1987}
{Stetson}, P.~B. 1987, \pasp, 99, 191, \dodoi{10.1086/131977}

\bibitem[{{Stritzinger} {et~al.}(2009){Stritzinger}, {Mazzali}, {Phillips},
  {Immler}, {Soderberg}, {Sollerman}, {Boldt}, {Braithwaite}, {Brown}, {Burns},
  {Contreras}, {Covarrubias}, {Folatelli}, {Freedman}, {Gonz{\'a}lez}, {Hamuy},
  {Krzeminski}, {Madore}, {Milne}, {Morrell}, {Persson}, {Roth}, {Smith}, \&
  {Suntzeff}}]{Stritzinger2009}
{Stritzinger}, M., {Mazzali}, P., {Phillips}, M.~M., {et~al.} 2009, \apj, 696,
  713, \dodoi{10.1088/0004-637X/696/1/713}

\bibitem[{{Stritzinger} {et~al.}(2018){Stritzinger}, {Taddia}, {Burns},
  {Phillips}, {Bersten}, {Contreras}, {Folatelli}, {Holmbo}, {Hsiao},
  {Hoeflich}, {Leloudas}, {Morrell}, {Sollerman}, \&
  {Suntzeff}}]{Stritzinger2018}
{Stritzinger}, M.~D., {Taddia}, F., {Burns}, C.~R., {et~al.} 2018, \aap, 609,
  A135, \dodoi{10.1051/0004-6361/201730843}

\bibitem[{{Stritzinger} {et~al.}(2020){Stritzinger}, {Taddia}, {Holmbo},
  {Baron}, {Contreras}, {Karamehmetoglu}, {Phillips}, {Sollerman}, {Suntzeff},
  {Vinko}, {Ashall}, {Avila}, {Burns}, {Campillay}, {Castellon}, {Folatelli},
  {Galbany}, {Hoeflich}, {Hsiao}, {Marion}, {Morrell}, \&
  {Wheeler}}]{Stritzinger2020}
{Stritzinger}, M.~D., {Taddia}, F., {Holmbo}, S., {et~al.} 2020, \aap, 634,
  A21, \dodoi{10.1051/0004-6361/201936619}

\bibitem[{{Sukhbold} {et~al.}(2016){Sukhbold}, {Ertl}, {Woosley}, {Brown}, \&
  {Janka}}]{Sukhbold2016ApJ...821...38S}
{Sukhbold}, T., {Ertl}, T., {Woosley}, S.~E., {Brown}, J.~M., \& {Janka}, H.~T.
  2016, \apj, 821, 38, \dodoi{10.3847/0004-637X/821/1/38}

\bibitem[{{Sutherland} \& {Wheeler}(1984)}]{Sutherland1984}
{Sutherland}, P.~G., \& {Wheeler}, J.~C. 1984, \apj, 280, 282,
  \dodoi{10.1086/161995}

\bibitem[{{Taddia} {et~al.}(2015){Taddia}, {Sollerman}, {Leloudas},
  {Stritzinger}, {Valenti}, {Galbany}, {Kessler}, {Schneider}, \&
  {Wheeler}}]{Taddia2015}
{Taddia}, F., {Sollerman}, J., {Leloudas}, G., {et~al.} 2015, \aap, 574, A60,
  \dodoi{10.1051/0004-6361/201423915}

\bibitem[{{Taddia} {et~al.}(2018){Taddia}, {Stritzinger}, {Bersten}, {Baron},
  {Burns}, {Contreras}, {Holmbo}, {Hsiao}, {Morrell}, {Phillips}, {Sollerman},
  \& {Suntzeff}}]{Taddia2018}
{Taddia}, F., {Stritzinger}, M.~D., {Bersten}, M., {et~al.} 2018, \aap, 609,
  A136, \dodoi{10.1051/0004-6361/201730844}

\bibitem[{{Tanaka} {et~al.}(2009){Tanaka}, {Tominaga}, {Nomoto}, {Valenti},
  {Sahu}, {Minezaki}, {Yoshii}, {Yoshida}, {Anupama}, {Benetti}, {Chincarini},
  {Della Valle}, {Mazzali}, \& {Pian}}]{Tanaka2009}
{Tanaka}, M., {Tominaga}, N., {Nomoto}, K., {et~al.} 2009, \apj, 692, 1131,
  \dodoi{10.1088/0004-637X/692/2/1131}

\bibitem[{{Tartaglia} {et~al.}(2017){Tartaglia}, {Fraser}, {Sand}, {Valenti},
  {Smartt}, {McCully}, {Anderson}, {Arcavi}, {Elias-Rosa}, {Galbany},
  {Gal-Yam}, {Haislip}, {Hosseinzadeh}, {Howell}, {Inserra}, {Jha}, {Kankare},
  {Lundqvist}, {Maguire}, {Mattila}, {Reichart}, {Smith}, {Smith},
  {Stritzinger}, {Sullivan}, {Taddia}, \& {Tomasella}}]{Tartaglia2017}
{Tartaglia}, L., {Fraser}, M., {Sand}, D.~J., {et~al.} 2017, \apjl, 836, L12,
  \dodoi{10.3847/2041-8213/aa5c7f}

\bibitem[{{Tartaglia} {et~al.}(2018){Tartaglia}, {Sand}, {Valenti}, {Wyatt},
  {Anderson}, {Arcavi}, {Ashall}, {Botticella}, {Cartier}, {Chen}, {Cikota},
  {Coulter}, {Della Valle}, {Foley}, {Gal-Yam}, {Galbany}, {Gall}, {Haislip},
  {Harmanen}, {Hosseinzadeh}, {Howell}, {Hsiao}, {Inserra}, {Jha}, {Kankare},
  {Kilpatrick}, {Kouprianov}, {Kuncarayakti}, {Maccarone}, {Maguire},
  {Mattila}, {Mazzali}, {McCully}, {Meland ri}, {Morrell}, {Phillips},
  {Pignata}, {Piro}, {Prentice}, {Reichart}, {Rojas-Bravo}, {Smartt}, {Smith},
  {Sollerman}, {Stritzinger}, {Sullivan}, {Taddia}, \& {Young}}]{Tartaglia2018}
{Tartaglia}, L., {Sand}, D.~J., {Valenti}, S., {et~al.} 2018, \apj, 853, 62,
  \dodoi{10.3847/1538-4357/aaa014}

\bibitem[{{Taubenberger} {et~al.}(2009){Taubenberger}, {Valenti}, {Benetti},
  {Cappellaro}, {Della Valle}, {Elias-Rosa}, {Hachinger}, {Hillebrandt},
  {Maeda}, {Mazzali}, {Pastorello}, {Patat}, {Sim}, \&
  {Turatto}}]{Taubenberger2009}
{Taubenberger}, S., {Valenti}, S., {Benetti}, S., {et~al.} 2009, \mnras, 397,
  677, \dodoi{10.1111/j.1365-2966.2009.15003.x}

\bibitem[{{Teffs} {et~al.}(2020){Teffs}, {Ertl}, {Mazzali}, {Hachinger}, \&
  {Janka}}]{Teffs2020}
{Teffs}, J., {Ertl}, T., {Mazzali}, P., {Hachinger}, S., \& {Janka}, T. 2020,
  \mnras, 492, 4369, \dodoi{10.1093/mnras/staa123}

\bibitem[{{Tody}(1986)}]{Tody1986}
{Tody}, D. 1986, in Society of Photo-Optical Instrumentation Engineers (SPIE)
  Conference Series, Vol. 627, Instrumentation in astronomy VI, ed. D.~L.
  {Crawford}, 733, \dodoi{10.1117/12.968154}

\bibitem[{{Tonry}(2011)}]{Tonry2011}
{Tonry}, J.~L. 2011, \pasp, 123, 58, \dodoi{10.1086/657997}

\bibitem[{{Tonry} {et~al.}(2012){Tonry}, {Stubbs}, {Lykke}, {Doherty},
  {Shivvers}, {Burgett}, {Chambers}, {Hodapp}, {Kaiser}, {Kudritzki},
  {Magnier}, {Morgan}, {Price}, \& {Wainscoat}}]{Tonry2012}
{Tonry}, J.~L., {Stubbs}, C.~W., {Lykke}, K.~R., {et~al.} 2012, \apj, 750, 99,
  \dodoi{10.1088/0004-637X/750/2/99}

\bibitem[{{Tonry} {et~al.}(2018){Tonry}, {Denneau}, {Heinze}, {Stalder},
  {Smith}, {Smartt}, {Stubbs}, {Weiland}, \& {Rest}}]{Tonry2018}
{Tonry}, J.~L., {Denneau}, L., {Heinze}, A.~N., {et~al.} 2018, \pasp, 130,
  064505, \dodoi{10.1088/1538-3873/aabadf}

\bibitem[{{Tucker} {et~al.}(2022){Tucker}, {Shappee}, {Huber}, {Payne}, {Do},
  {Hinkle}, {de Jaeger}, {Ashall}, {Desai}, {Hoogendam}, {Aldering},
  {Auchettl}, {Baranec}, {Bulger}, {Chambers}, {Chun}, {Hodapp}, {Lowe},
  {McKay}, {Rampy}, {Rubin}, \& {Tonry}}]{Tucker2022PASP..134l4502T}
{Tucker}, M.~A., {Shappee}, B.~J., {Huber}, M.~E., {et~al.} 2022, \pasp, 134,
  124502, \dodoi{10.1088/1538-3873/aca719}

\bibitem[{{Tully} {et~al.}(2009){Tully}, {Rizzi}, {Shaya}, {Courtois},
  {Makarov}, \& {Jacobs}}]{Tully2009}
{Tully}, R.~B., {Rizzi}, L., {Shaya}, E.~J., {et~al.} 2009, \aj, 138, 323,
  \dodoi{10.1088/0004-6256/138/2/323}

\bibitem[{{Uomoto}(1986)}]{Uomoto1986}
{Uomoto}, A. 1986, \apjl, 310, L35, \dodoi{10.1086/184777}

\bibitem[{{Vacca} {et~al.}(2003){Vacca}, {Cushing}, \& {Rayner}}]{Vacca2003}
{Vacca}, W.~D., {Cushing}, M.~C., \& {Rayner}, J.~T. 2003, \pasp, 115, 389,
  \dodoi{10.1086/346193}

\bibitem[{{Valenti} {et~al.}(2008){Valenti}, {Benetti}, {Cappellaro}, {Patat},
  {Mazzali}, {Turatto}, {Hurley}, {Maeda}, {Gal-Yam}, {Foley}, {Filippenko},
  {Pastorello}, {Challis}, {Frontera}, {Harutyunyan}, {Iye}, {Kawabata},
  {Kirshner}, {Li}, {Lipkin}, {Matheson}, {Nomoto}, {Ofek}, {Ohyama}, {Pian},
  {Poznanski}, {Salvo}, {Sauer}, {Schmidt}, {Soderberg}, \&
  {Zampieri}}]{Valenti2008}
{Valenti}, S., {Benetti}, S., {Cappellaro}, E., {et~al.} 2008, \mnras, 383,
  1485, \dodoi{10.1111/j.1365-2966.2007.12647.x}

\bibitem[{{Valenti} {et~al.}(2011){Valenti}, {Fraser}, {Benetti}, {Pignata},
  {Sollerman}, {Inserra}, {Cappellaro}, {Pastorello}, {Smartt}, {Ergon},
  {Botticella}, {Brimacombe}, {Bufano}, {Crockett}, {Eder}, {Fugazza},
  {Haislip}, {Hamuy}, {Harutyunyan}, {Ivarsen}, {Kankare}, {Kotak}, {Lacluyze},
  {Magill}, {Mattila}, {Maza}, {Mazzali}, {Reichart}, {Taubenberger},
  {Turatto}, \& {Zampieri}}]{Valenti2011}
{Valenti}, S., {Fraser}, M., {Benetti}, S., {et~al.} 2011, \mnras, 416, 3138,
  \dodoi{10.1111/j.1365-2966.2011.19262.x}

\bibitem[{{Valenti} {et~al.}(2014){Valenti}, {Sand}, {Pastorello}, {Graham},
  {Howell}, {Parrent}, {Tomasella}, {Ochner}, {Fraser}, {Benetti}, {Yuan},
  {Smartt}, {Maund}, {Arcavi}, {Gal-Yam}, {Inserra}, \& {Young}}]{Valenti2014}
{Valenti}, S., {Sand}, D., {Pastorello}, A., {et~al.} 2014, \mnras, 438, L101,
  \dodoi{10.1093/mnrasl/slt171}

\bibitem[{{Valenti} {et~al.}(2016){Valenti}, {Howell}, {Stritzinger}, {Graham},
  {Hosseinzadeh}, {Arcavi}, {Bildsten}, {Jerkstrand}, {McCully}, {Pastorello},
  {Piro}, {Sand}, {Smartt}, {Terreran}, {Baltay}, {Benetti}, {Brown},
  {Filippenko}, {Fraser}, {Rabinowitz}, {Sullivan}, \& {Yuan}}]{Valenti2016}
{Valenti}, S., {Howell}, D.~A., {Stritzinger}, M.~D., {et~al.} 2016, \mnras,
  459, 3939, \dodoi{10.1093/mnras/stw870}

\bibitem[{{Van Dyk} {et~al.}(2011){Van Dyk}, {Li}, {Cenko}, {Kasliwal},
  {Horesh}, {Ofek}, {Kraus}, {Silverman}, {Arcavi}, {Filippenko}, {Gal-Yam},
  {Quimby}, {Kulkarni}, {Yaron}, \& {Polishook}}]{VanDyk2011}
{Van Dyk}, S.~D., {Li}, W., {Cenko}, S.~B., {et~al.} 2011, \apjl, 741, L28,
  \dodoi{10.1088/2041-8205/741/2/L28}

\bibitem[{{Van Dyk} {et~al.}(2014){Van Dyk}, {Zheng}, {Fox}, {Cenko}, {Clubb},
  {Filippenko}, {Foley}, {Miller}, {Smith}, {Kelly}, {Lee}, {Ben-Ami}, \&
  {Gal-Yam}}]{VanDyk2014}
{Van Dyk}, S.~D., {Zheng}, W., {Fox}, O.~D., {et~al.} 2014, \aj, 147, 37,
  \dodoi{10.1088/0004-6256/147/2/37}

\bibitem[{{Vink} {et~al.}(2001){Vink}, {de Koter}, \&
  {Lamers}}]{Vink2001A&A...369..574V}
{Vink}, J.~S., {de Koter}, A., \& {Lamers}, H.~J.~G.~L.~M. 2001, \aap, 369,
  574, \dodoi{10.1051/0004-6361:20010127}

\bibitem[{{Virtanen} {et~al.}(2020){Virtanen}, {Gommers}, {Oliphant},
  {Haberland}, {Reddy}, {Cournapeau}, {Burovski}, {Peterson}, {Weckesser},
  {Bright}, {van der Walt}, {Brett}, {Wilson}, {Millman}, {Mayorov}, {Nelson},
  {Jones}, {Kern}, {Larson}, {Carey}, {Polat}, {Feng}, {Moore}, {VanderPlas},
  {Laxalde}, {Perktold}, {Cimrman}, {Henriksen}, {Quintero}, {Harris},
  {Archibald}, {Ribeiro}, {Pedregosa}, {van Mulbregt}, \& {SciPy 1. 0
  Contributors}}]{2020NatMe..17..261V}
{Virtanen}, P., {Gommers}, R., {Oliphant}, T.~E., {et~al.} 2020, Nature
  Methods, 17, 261, \dodoi{10.1038/s41592-019-0686-2}

\bibitem[{{Wang} {et~al.}(2015){Wang}, {Wang}, {Dai}, \& {Wu}}]{Wang2015}
{Wang}, S.~Q., {Wang}, L.~J., {Dai}, Z.~G., \& {Wu}, X.~F. 2015, \apj, 807,
  147, \dodoi{10.1088/0004-637X/807/2/147}

\bibitem[{{Wellons} {et~al.}(2012){Wellons}, {Soderberg}, \&
  {Chevalier}}]{Wellons2012}
{Wellons}, S., {Soderberg}, A.~M., \& {Chevalier}, R.~A. 2012, \apj, 752, 17,
  \dodoi{10.1088/0004-637X/752/1/17}

\bibitem[{{Wellstein} \& {Langer}(1999)}]{Wellstein1999}
{Wellstein}, S., \& {Langer}, N. 1999, \aap, 350, 148,
  \dodoi{10.48550/arXiv.astro-ph/9904256}

\bibitem[{{W}es {M}c{K}inney(2010)}]{mckinney-proc-scipy-2010}
{W}es {M}c{K}inney. 2010, in {P}roceedings of the 9th {P}ython in {S}cience
  {C}onference, ed. {S}t\'efan van~der {W}alt \& {J}arrod {M}illman, 56 -- 61,
  \dodoi{10.25080/Majora-92bf1922-00a}

\bibitem[{{Wheeler} \& {Harkness}(1990)}]{Wheeler1990}
{Wheeler}, J.~C., \& {Harkness}, R.~P. 1990, Reports on Progress in Physics,
  53, 1467, \dodoi{10.1088/0034-4885/53/12/001}

\bibitem[{{Williamson} {et~al.}(2019){Williamson}, {Modjaz}, \&
  {Bianco}}]{Williamson2019}
{Williamson}, M., {Modjaz}, M., \& {Bianco}, F.~B. 2019, \apjl, 880, L22,
  \dodoi{10.3847/2041-8213/ab2edb}

\bibitem[{{Wilson} {et~al.}(2004){Wilson}, {Henderson}, {Herter}, {Matthews},
  {Skrutskie}, {Adams}, {Moon}, {Smith}, {Gautier}, {Ressler}, {Soifer}, {Lin},
  {Howard}, {LaMarr}, {Stolberg}, \& {Zink}}]{Wilson2004}
{Wilson}, J.~C., {Henderson}, C.~P., {Herter}, T.~L., {et~al.} 2004, in Society
  of Photo-Optical Instrumentation Engineers (SPIE) Conference Series, Vol.
  5492, Ground-based Instrumentation for Astronomy, ed. A.~F.~M. {Moorwood} \&
  M.~{Iye}, 1295--1305, \dodoi{10.1117/12.550925}

\bibitem[{{Woosley}(2019)}]{Woosley2019}
{Woosley}, S.~E. 2019, \apj, 878, 49, \dodoi{10.3847/1538-4357/ab1b41}

\bibitem[{{Woosley} \& {Eastman}(1997)}]{Woosley1997}
{Woosley}, S.~E., \& {Eastman}, R.~G. 1997, in NATO Advanced Study Institute
  (ASI) Series C, Vol. 486, Thermonuclear Supernovae, ed. P.~{Ruiz-Lapuente},
  R.~{Canal}, \& J.~{Isern}, 821, \dodoi{10.1007/978-94-011-5710-0_51}

\bibitem[{{Woosley} \& {Heger}(2007)}]{Woosley2007}
{Woosley}, S.~E., \& {Heger}, A. 2007, \physrep, 442, 269,
  \dodoi{10.1016/j.physrep.2007.02.009}

\bibitem[{{Woosley} {et~al.}(2002){Woosley}, {Heger}, \&
  {Weaver}}]{Woosley2002}
{Woosley}, S.~E., {Heger}, A., \& {Weaver}, T.~A. 2002, Reviews of Modern
  Physics, 74, 1015, \dodoi{10.1103/RevModPhys.74.1015}

\bibitem[{{Woosley} {et~al.}(1993){Woosley}, {Langer}, \&
  {Weaver}}]{Woosley1993}
{Woosley}, S.~E., {Langer}, N., \& {Weaver}, T.~A. 1993, \apj, 411, 823,
  \dodoi{10.1086/172886}

\bibitem[{{Woosley} {et~al.}(1995){Woosley}, {Langer}, \&
  {Weaver}}]{Woosley1995ApJ...448..315W}
---. 1995, \apj, 448, 315, \dodoi{10.1086/175963}

\bibitem[{{Woosley} {et~al.}(2021){Woosley}, {Sukhbold}, \&
  {Kasen}}]{Woosley2021}
{Woosley}, S.~E., {Sukhbold}, T., \& {Kasen}, D.~N. 2021, \apj, 913, 145,
  \dodoi{10.3847/1538-4357/abf3be}

\bibitem[{{Woosley} \& {Weaver}(1995)}]{Woosley1995ApJS..101..181W}
{Woosley}, S.~E., \& {Weaver}, T.~A. 1995, \apjs, 101, 181,
  \dodoi{10.1086/192237}

\bibitem[{{Yoon}(2015)}]{Yoon2015}
{Yoon}, S.-C. 2015, \pasa, 32, e015, \dodoi{10.1017/pasa.2015.16}

\bibitem[{{Yoon}(2017)}]{Yoon2017b}
---. 2017, \mnras, 470, 3970, \dodoi{10.1093/mnras/stx1496}

\bibitem[{{Yoon} {et~al.}(2019){Yoon}, {Chun}, {Tolstov}, {Blinnikov}, \&
  {Dessart}}]{Yoon2019}
{Yoon}, S.-C., {Chun}, W., {Tolstov}, A., {Blinnikov}, S., \& {Dessart}, L.
  2019, \apj, 872, 174, \dodoi{10.3847/1538-4357/ab0020}

\bibitem[{{Yoon} {et~al.}(2017){Yoon}, {Dessart}, \& {Clocchiatti}}]{Yoon2017a}
{Yoon}, S.-C., {Dessart}, L., \& {Clocchiatti}, A. 2017, \apj, 840, 10,
  \dodoi{10.3847/1538-4357/aa6afe}

\bibitem[{{Yoon} {et~al.}(2010){Yoon}, {Woosley}, \& {Langer}}]{Yoon2010}
{Yoon}, S.~C., {Woosley}, S.~E., \& {Langer}, N. 2010, \apj, 725, 940,
  \dodoi{10.1088/0004-637X/725/1/940}

\end{thebibliography}
\bibliographystyle{aasjournal}



\end{document}